\documentclass{JHEP3}

\usepackage{epsfig}
\usepackage{amssymb}
\usepackage{amsmath}
\usepackage{slashbox}
\usepackage{bbm,bm}
\usepackage{cite}
\usepackage{verbatim}
\usepackage{slashed}
\usepackage{multirow}

\newcommand{\bea}{\begin{eqnarray}}
\newcommand{\eea}{\end{eqnarray}}
\newcommand{\bean}{\begin{eqnarray*}}
\newcommand{\eean}{\end{eqnarray*}}
\newcommand{\nn}{\nonumber \\}

\def\W #1{\widetilde{#1}}
\def\WH #1{\widehat{#1}}

\def\eref#1{(\ref{#1})}

\def\a{{\alpha}}

\def\b{{\beta}}

\def\c{{\gamma}}

\def\la{\lambda}
\def\eps{\epsilon}

\def\Spaa #1{ \left\langle #1 \right\rangle}
\def\Spab #1{ \left\langle #1 \right]}

\def\Label#1{\label{#1}%
  \smash{\hbox to0pt{\raise1ex\hbox{\tiny[#1]}\hss}}}

\title{The classification of two-loop integrand basis in
pure four-dimension}

\author{Bo Feng$^{a,b,c}$, Rijun Huang$^{c}$\\$^a$Zhejiang Institute of Modern Physics, Zhejiang
University, Hangzhou, 310027, P. R. China\\$^b$Center of
Mathematical Sciences, Zhejiang University, Hangzhou, 310027, P. R.
China\\$^c$Niels Bohr International Academy and Discovery Center,
The Niels Bohr Institute, \\Copenhagen University, Blegdamsvej 17,
DK-2100 Copenhagen, Denmark}

\date{\today}
\abstract{In this paper, we have made the attempt to classify the
integrand basis of  all two-loop diagrams in pure four-dimensional
space-time. The first step of our classification is to determine all
different topologies of two-loop diagrams, i.e., the structure of
denominators. The second step is to determine the set of independent
numerators for each topology  using Gr\"{o}bner basis method. For
the second step, varieties defined by putting all propagators
on-shell has played an important role.  We discuss the structures of
varieties and how they split to various irreducible  branches under
specific kinematic configurations of external momenta. The
structures of varieties are crucial to determine coefficients of
integrand basis in reduction both numerically or analytically.

}


\begin{document}

\section{Introduction}

In the past few years we have seen tremendous progresses for
one-loop diagram computations\footnote{See reports
\cite{AlcarazMaestre:2012vp, Binoth:2010ra} for references.} using
Passrino-Veltman(PV) reduction method \cite{Passarino:1978jh}.
 The newly developed reduction methods can be sorted into two categories:
(a) the reduction performed at the integral level, such as the
unitarity cut method \cite{Landau:1959fi,Bern:1994cg,Bern:1994zx,
Anastasiou:2006jv,Anastasiou:2006gt} and generalized unitarity cut
method\cite{Britto:2004nc,Britto:2005ha}; (b) the reduction
performed at the integrand level, which was initiated by
Ossola-Papadopoulos-Pittau(OPP) in \cite{Ossola:2006us} and further
generalized in \cite{Forde:2007mi, Ellis:2007br, Kilgore:2007qr,
Giele:2008ve, Ossola:2008xq, Badger:2008cm}. Comparing methods of
these two categories, methods in the first one  focus only on
coefficients having nonzero final contributions while methods in the
second one must also include spurious coefficients. Although more
coefficients must be calculated, methods in the second category are
still very useful because all manipulations are performed purely
algebraically at the integrand level, thus they can be easily
programmed.

Encouraged by  successful computations at one-loop level, it is
natural to generalize these methods to higher loops, partially
because of our theoretical curiosity and partially because of the
precise prediction  for modern collide experiments. However, the
generalization is not so trivial. The first difficulty is that in
general we do not know much about the basis for multi-loop
amplitudes. In fact, now it is clear that we should distinguish the
integral basis and the integrand basis. Unlike the one-loop
amplitude, the number of integrand basis is much larger than the
number of integral basis for multi-loop amplitudes. Thus it is
highly desirable to reduce integrand basis to integral basis
further. One standard method of doing so is the Integrate-by-Part
(IBP) method\cite{IBP}. The IBP can be carried out in a reasonable
short time if the amplitude involves only a few external particles,
but it becomes unpractical with time consuming when the number of
external particles increases. The second difficulty  is how to
extract coefficients of basis. For one-loop amplitudes, finding
coefficients is separated from finding basis, while the
frequently-used IBP reduction method combines these two tasks
together at the same time.

These computation difficulties for multi-loop amplitudes have been
addressed in the past few years by several groups
\cite{Mastrolia:2011pr, Badger:2012dp,Gluza:2010ws , Kosower:2011ty,
Larsen:2012sx, CaronHuot:2012ab,Kleiss:2012yv, Johansson:2012zv,
Zhang:2012ce, Mastrolia:2012an, Badger:2012dv}. The main focus of
study is  the reduction  at integrand level, which includes finding
integrand basis and matching their coefficients. The  step towards
this direction was first taken in \cite{Mastrolia:2011pr}, where
four-dimensional constructive algorithm for integrand has been
applied to two-loop planar and non-planar contributions of four and
five-point Maximal-Helicity-Violating(MHV) amplitudes in
$\mathcal{N}=4$ Super-Yang-Mills theory. Using constraints from Gram
matrix, similar determinant of monomials of numerators was achieved
in \cite{Badger:2012dp}. Besides reduction at the integrand level,
reduction at the integral level is discussed in \cite{Gluza:2010ws,
Kosower:2011ty, Larsen:2012sx, CaronHuot:2012ab,Kleiss:2012yv},
where in order to determine the physical contour for integral basis,
the variety defined by setting all propagators on-shell has been
carefully analyzed.

Among these new developments, the application of computational
algebraic geometry method to multi-loop amplitude calculations is
very intriguing \cite{Zhang:2012ce,Mastrolia:2012an}, where the
Gr\"{o}bner basis plays a central role. It is quite easy to
determine integrand basis by  Gr\"{o}bner basis method, although
different  sets of integrand basis can be obtained with different
 orderings in polynomial division. Besides the integrand basis,
their coefficients  can also be determined by the same method. The
knowledge of variety, including its branch structure and
intersection pattern of branches, is very important in the
application of this method. This  method has been tested by several
examples in  at two and three-loops
\cite{Zhang:2012ce,Badger:2012dv}\footnote{The numeric algebraic
geometry method\cite{Mehta:2012wk} can also be used if we  only want
 the number of irreducible components.}. Encouraged by the
success, in this paper we will use algebraic geometry method to
systematically study all possible topologies of two-loop diagrams in
pure four-dimension for any external momentum configurations, not
only restrict to double-box or penta-triangle studied in various
references.

The paper is organized as follows. In section 2 we classify all
possible topologies for two-loop diagrams. In section 3 the one-loop
topologies are re-examined using algebraic geometry method. Ideas
from the reexamination will be used to analyze two-loop topologies.
 In section 4, as
a warm-up, we present  results of some trivial two-loop topologies.
In section 5, a careful analysis of planar penta-triangle topology
has been given, while in section 6, we give a detailed study of
non-planar crossed double-triangle topology. In section 7, we
summarize  results of all remaining topologies. In the last section,
conclusions and discussions are given. In appendix, we introduce
some mathematical facts that can be used to study the branch
structure of variety.

\section{An overview of general two-loop topologies}

In this section, we give an overview of general two-loop topologies.
Much of the results are scattered in literatures, and we assemble
them here to make the paper self-contained.

\subsection{The two-loop topology }

Since two-loop diagrams can always be reduced to one-loop diagrams
by cutting an inner propagator, we can inversely reconstruct
two-loop ones by sewing two external legs of one-loop diagrams. The
topology of one-loop diagrams is very simple: we just attach various
tree structures along the loop at some vertices $V_i$ (see Figure
\ref{1-loop-gen} ).
\EPSFIGURE[ht]{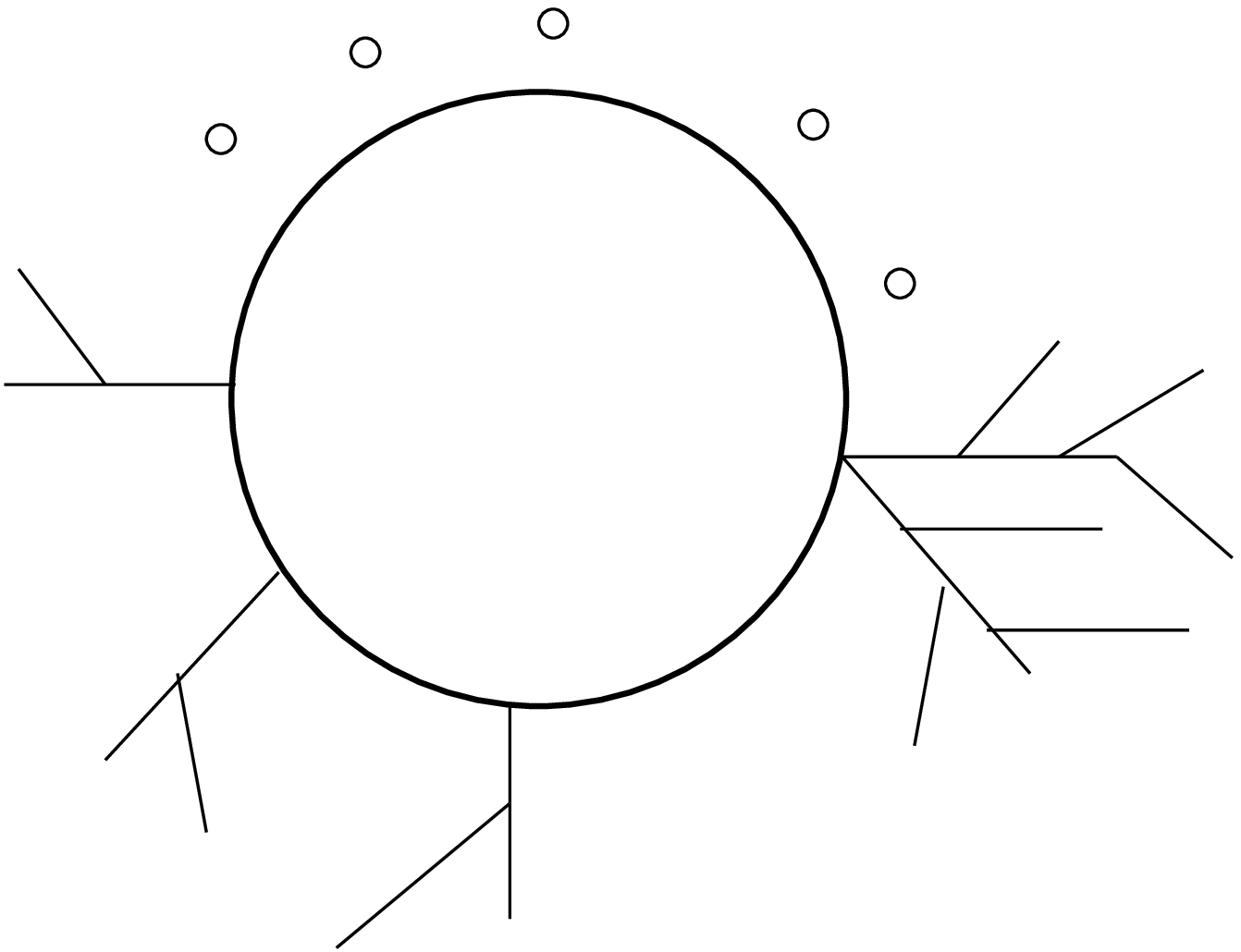,width=11.5cm} {The general topology of
one-loop diagrams where various tree structures are attached along
the loop at some vertices.  \label{1-loop-gen} }

From one-loop topology we can reconstruct two-loop topology by
connecting two external legs, and there are several ways of doing
so, which give different two-loop topologies:
\begin{itemize}
\item (A) If the two to-be-connected external legs are attached
to the same tree structure, we will get two-loop topology as drawn
in Figure \ref{2-loop-A}. Explicit illustration shows that there are
two kinds of connections. In the first kind (A1), two one-loop
sub-topologies do not share the same vertex while in the second kind
(A2), they do share a common vertex.

\EPSFIGURE[ht]{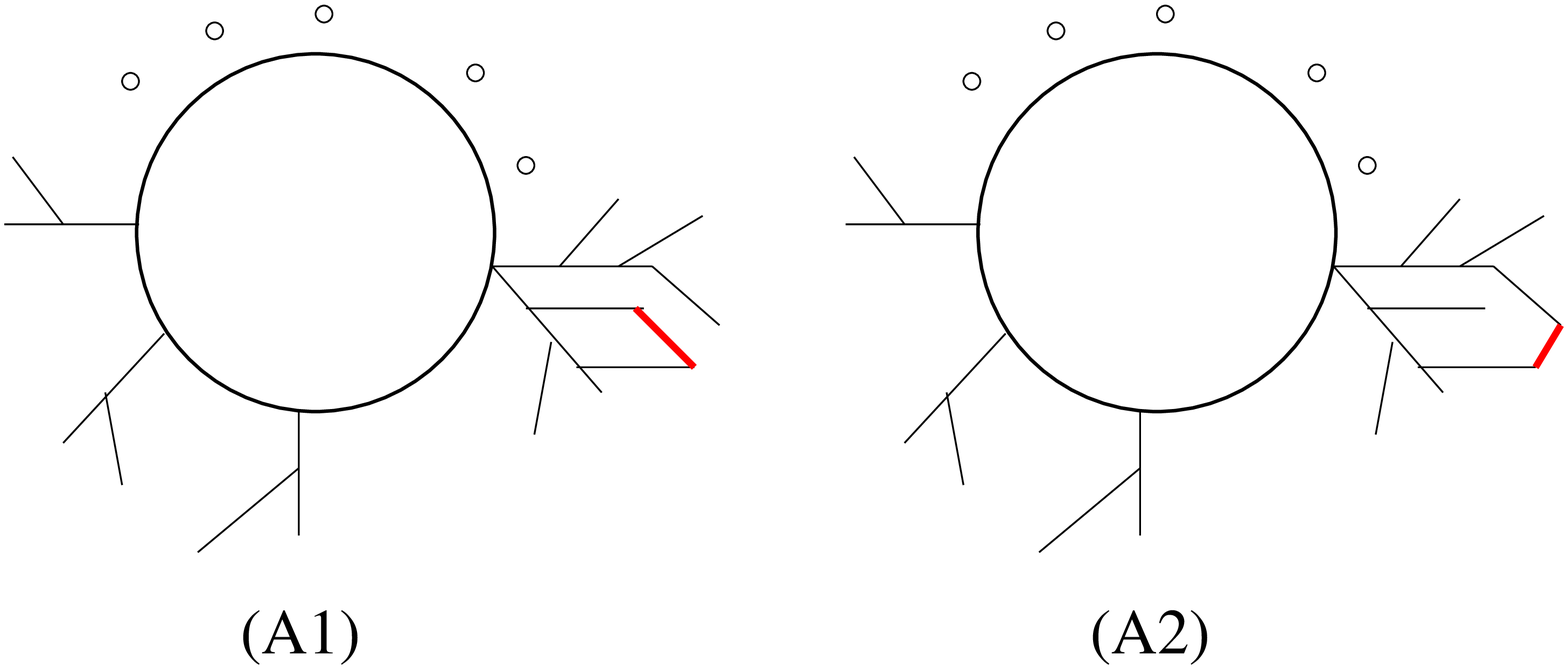,width=12.5cm} {The two-loop
topology  generated from one-loop topology by connecting two
external legs attached to the same tree structure. The connection
has been denoted by red color thick line. In connection (A1), two
one-loop sub-topologies do not share the same vertex while in
connection (A2), they do share a common vertex.
    \label{2-loop-A} }

\item (B) If the two to-be-connected external legs are attached
to two nearby vertices along the loop, we will get two-loop topology
as drawn in (B) of Figure \ref{2-loop-BC}. All two-loop planar
topologies can be generated from this type.

\EPSFIGURE[ht]{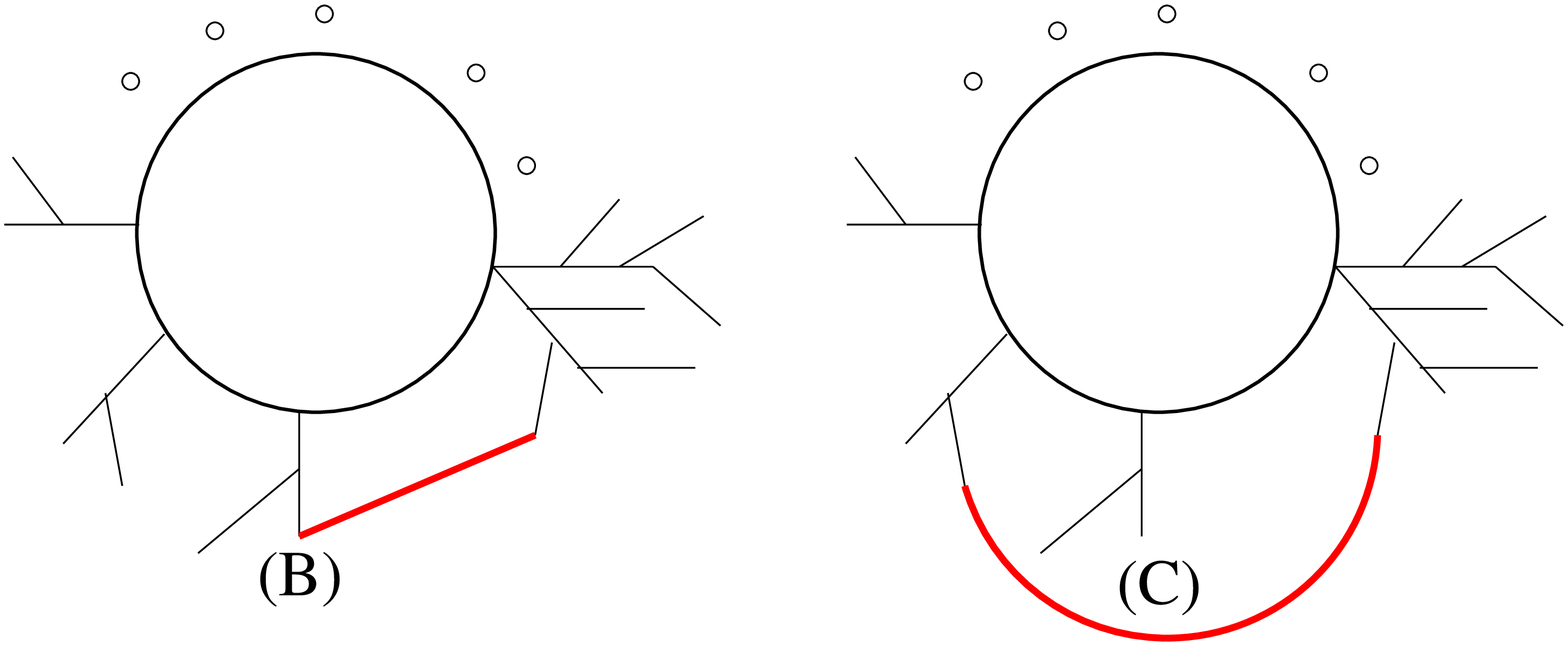,width=12.5cm} {The  two-loop
topologies of case (B) and case (C) obtained by connecting two
external legs attached to two different tree structures. For case
(B), two tree structures are adjacent while for case (C), not
adjacent.
    \label{2-loop-BC} }

\item (C) If the two to-be-connected external legs are attached
to two non-nearby tree structures along the loop, we will get
two-loop topology as drawn in (C) of Figure \ref{2-loop-BC}. All
two-loop non-planar topologies can be generated from this type.

\end{itemize}
%

\subsection{Classification of denominators of two-loop basis}

Having understood the general two-loop topologies, the next step is
to classify the basis  used to expand any two-loop amplitudes. This
is similar to the classification of scalar basis for one-loop
diagrams, which includes box, triangle, bubble and tadpole. However,
it is necessary to distinguish the {\bf integrand basis} and the
{\bf integral basis}. The integrand basis is that used by OPP method
to expand expressions coming from Feynman diagrams at integrand
level. However, after carrying out the integrations, some elements
of integrand basis will vanish, while others may have nontrivial
linear relations. After excluded these redundancies from integrand
basis we obtain the integral basis. The integral basis is also
called the {\bf master integrals}(MIs). The number of integral basis
is much smaller than the number of integrand basis, since after
integration. The difference between these two kinds of basis can be
easily seen in the one-loop box topology: there is only one master
integral ${1\over D_1 D_2 D_3 D_4}$, but there are two integrand
basis ${1\over D_1 D_2 D_3 D_4}$ and
${\epsilon(\ell,K_1,K_2,K_3)\over D_1 D_2 D_3 D_4}$. Numerator of
$\epsilon \cdot \ell$ with odd power will vanish after integration
because of parity.

To find the integrand and integral basis, we can use the procedure
called PV-reduction. Among  manipulations on expressions coming from
Feynman diagrams, some are done at the integrand level, such as
rewriting $2K_1\cdot \ell=-(\ell-K_1)^2 +\ell^2+K_1^2$, while some
manipulations are carried out using  properties of integral, such as
IBP method. Pure algebraic manipulations at the integrand level will
produce the integrand basis, while combining with operations such as
IBP, will reduce integrand basis further to integral basis. Above
reduction has been discussed in many references, for example,
\cite{Kleiss:2012yv} for details and reference.

For two-loop diagrams, denominators of expressions coming from
Feynman diagrams can always be written as products of three kinds of
propagators
\bea \mathcal{D} & = & D \widetilde{D}
\widehat{D}~,~~~\label{BC-den-form-1}\eea
where
\bea D& = & \ell_1^2
(\ell_1-K_{a,1})^2(\ell_1-K_{a,2})^2...(\ell_1-K_{a,n_1-1})^2~,~~~
\nn
\widetilde{D} & = &
\ell_2^2(\ell_2-K_{b,1})^2(\ell_2-K_{b,2})^2...(\ell_2-K_{b,n_2-1})^2~,~~~\nn
\widehat{D} & = & (\ell_1+\ell_2+K_{c,1})^2...
(\ell_1+\ell_2+K_{c,n_3})^2~.~~~\label{BC-den-form-2}\eea
Here $n_1$, $n_2$ are  numbers of propagators  containing only
$\ell_1$ or $\ell_2$, while $n_3$ is the number of propagators
containing both $\ell_1,\ell_2$. By the freedom of relabeling
$\ell_1,\ell_2$, we can always restrict $n_i$ with condition
\bea n_1\geq n_2\geq n_3~.~~~\label{n-rel-1}\eea
The up-bound of $n_1,n_2,n_3$ and their summation depend on the
space-time dimension. For example, if we consider physics in
$(4-2\epsilon)$-dimension, we would have
\bea n_1,~n_2,~n_3\leq 5~,~~~n_1+n_2+n_3\leq
11~.~~\label{42e-cond}\eea
But if we constrain to pure four-dimension, the condition becomes
\bea n_1,~n_2,~n_3~\leq 4~,~~~n_1+n_2+n_3\leq
8~.~~~\label{4-cond}\eea
By combining conditions \eref{n-rel-1} and \eref{4-cond} for
4-dimension case (or \eref{n-rel-1} and \eref{42e-cond} for
$(4-2\epsilon)$-dimension case), we can classify denominators of
integrand and integral.
 For $(4-2\epsilon)$-dimension, conditions \eref{n-rel-1} and
\eref{42e-cond} constrain $n_3\leq 3$. Thus if we arrange all
possible solutions of $(n_1,n_2,n_3)$ by value of $n_3$, we have
following 4 groups of solutions
\bea n_3 =3: &~~~& (5,3,3), (4,4,3), (4,3,3), (3,3,3)~;~~~ \nn
n_3 =2: &~~~& (5,4,2), (5,3,2),(4,4,2), (5,2,2),(4,3,2),(4,2,2),
(3,3,2), (3,2,2),(2,2,2)~;~~~\nn
n_3 =1: &~~~& (5,5,1), (5,4,1),(5,3,1), (4,4,1),(5,2,1),(4,3,1),
(5,1,1),(4,2,1), \nn & & (3,3,1), (4,1,1), (3,2,1),
(3,1,1),(2,2,1),(2,1,1), (1,1,1)~;~~~\nn
n_3 =0: &~~~& (5,5,0),(5,4,0),(5,3,0), (4,4,0), (5,2,0),(4,3,0),
(5,1,0),(4,2,0), (3,3,0), \nn & & (4,1,0),(3,2,0), (3,1,0),(2,2,0),
(2,1,0),(1,1,0)~.~~~\label{42e-n123}\eea
For pure four-dimension, the number of solutions decreases a lot,
since now we have $n_3\leq 2$. The possible solutions of
$(n_1,n_2,n_3)$ for \eref{n-rel-1} and \eref{4-cond} are listed into
3 groups:
\bea n_3 =2: &~~~& (4,2,2), (3,3,2), (3,2,2),(2,2,2)~;~~\label{4-n123}\\
n_3 =1: &~~~&  (4,3,1), (4,2,1), (3,3,1), (4,1,1), (3,2,1),
(3,1,1),(2,2,1),(2,1,1), (1,1,1)~;~~~\nn
n_3 =0: &~~~&  (4,4,0), (4,3,0), (4,2,0), (3,3,0), (4,1,0),(3,2,0),
(3,1,0),(2,2,0), (2,1,0),(1,1,0)~.~~~\nonumber\eea
Solutions with $n_3=0$ contain two-loop topologies coming from
sewing two one-loop topologies at a single vertex as shown in Figure
\ref{4d-A} , while solutions with $n_3=1$ contain planar two-loop
topologies with one common propagator  as shown in Figure
\ref{4d-B}. All two-loop non-planar topologies are included in
solutions $n_3=2$  as shown in Figure \ref{4d-C}.

While two-loop topologies of  basis have been classified
 by $(n_1,n_2,n_3)$,  to get the integrand or integral basis, we
 still need to determine corresponding numerators. For two-loop the so called
 "scalar basis"  is not enough to expand all amplitudes, we also need
terms with numerators containing Lorentz invariant scalar product
having internal momenta. The distinction between integrand and
integral basis becomes important when discussing the classification
of numerators. In this paper, we will focus only on the integrand
basis in pure four-dimension.

\section{The integrand basis of one-loop diagrams in pure four-dimension}

As a warm-up, we take the one-loop integrand basis as a simple
example to demonstrate various ideas that we will meet in later part
of this paper. All results in this section are known in other
references such as \cite{Ossola:2006us,
Zhang:2012ce,Mastrolia:2012an}, however, we recall  them here  since
these results are also related to two-loop integrand basis
 with $n_3=0$.

In pure four-dimension, since each external or internal momentum has
four components, we need four independent momenta to expand all
kinematics. One  construction of  momentum basis is to take two
arbitrary independent momenta $K_1,K_2$ and    construct following
four null momenta $e_i$, $i=1,2,3,4$ (assuming $(K_1+K_2)^2\neq 0$)
\bea e_1 & = & {1\over \gamma_{12}} \left(K_1- { K_1^2+ K_1\cdot
K_2-{\rm sgn} (K_1\cdot K_2)|\sqrt{ (K_1\cdot K_2)^2- K_1^2
K_2^2}|\over (K_1+K_2)^2} K_{12}\right)~,~~~ \nn
 e_2 & = & {1\over \gamma_{12}} \left(K_2- { K_2^2+ K_1\cdot K_2-{\rm sgn}
(K_1\cdot K_2)|\sqrt{ (K_1\cdot K_2)^2- K_1^2 K_2^2}|\over
(K_1+K_2)^2} K_{12}\right)~,~~~\nn
e_3 & = & {\Spab{e_1|\gamma^\mu|e_2}\over
2i}~,~~~e_4={\Spab{e_2|\gamma^\mu|e_1}\over
2i}~,~~~\label{Define-e}\eea
where $\gamma_{12}^2   =  {2 [ (K_1\cdot K_2)^2- K_1^2 K_2^2]\over
(K_1+K_2)^2}$. This momentum basis has following property: among all
inner products
 of $e_i\cdot e_j$, the only non-zero ones are  $e_1\cdot e_2=1$ and
 $e_3\cdot e_4=1$\footnote{In fact,
this property has not determined $e_i$ uniquely, since there is a
freedom to rescale $e_1\to w e_1$ and $e_2\to w^{-1} e_2$ and
similarly for  $e_3,e_4$ pair.}. Definition \eref{Define-e} also
makes massless limit smoothly, {\sl i.e.}, when $K_1^2 \to 0$,
$e_1\to K_1$ and when $K_2^2\to 0$, $e_2\to K_2$. Using above
momentum basis, we can expand any momentum, such as
\bea &&K_i  =  (K_i\cdot e_2) e_1+ (K_i\cdot e_1) e_2+ (K_i\cdot
e_4) e_3+ (K_i\cdot e_3) e_4 ~,~~~ \nn
&&\ell  =  (\ell\cdot e_2) e_1+(\ell\cdot e_1) e_2+(\ell\cdot e_4)
e_3+(\ell\cdot e_3) e_4 \equiv x_2 e_1+x_1 e_2+ x_4 e_3+ x_3
e_4~,~~~\label{ell-exp-by-e}\eea
and the Lorentz invariant scalar products are given by
\bea \ell^2 & = & x_1 x_2+ x_3 x_4~,~~~\ell\cdot K_i= \sum_{j=1}^4
\a_{ij} x_j~.~~~\label{inner-under-e}\eea

The importance of above expansions \eref{ell-exp-by-e} and
\eref{inner-under-e}  is that any integrand  can be written as a
rational function ${f(x_1,x_2,x_3,x_4)\over \prod_t
D_t(x_1,x_2,x_3,x_4)}$, and the PV-reduction procedure is equivalent
to finding following expansion of numerator
\bea f(x_i)=\sum_t c_t(x_i) D_t(x_i)+ r(x_i)~,~~\label{f-by-D}\eea
where the remaining polynomial $r(x_i)$ is nothing but the integrand
basis we are looking for. In a more mathematical language,
propagators $D_t$ generate an ideal $I$ in polynomial ring
$k[x_1,x_2,x_3,x_4]$, and the integrand basis is constructed by
representative elements in the quotient ring $k/I$ under some
physical constraints. One physical constraint  is the total degree
$n_\ell$ of loop momentum $\ell$ in numerator. For renormalizable
theory, we require $n_\ell\leq n_D$ where $n_D$ is the number of
propagators in denominator.

Having these general preparations, we will discuss explicitly
various one-loop integrand basis,  such as box, triangle, bubble and
tadpole \cite{Ossola:2006us, Zhang:2012ce,Mastrolia:2012an}. For
simplicity, we will only consider massless propagators, but the
massive ones can be discussed in a  similar way.

\subsection{One-loop box topology}

For box topology, four propagators are given by
\bea D_0=
\ell^2~,~D_1=(\ell-K_1)^2~,~D_2=(\ell-K_1-K_2)^2~,~D_3=(\ell-K_1-K_2-K_3)^2~.~~~
\label{One-loop-D}\eea
Without loss of generality, we can use $K_1, K_2$ to construct
momentum basis and use it to expand all momenta. There are 4
variables $(x_1,x_2,x_3,x_4)$ coming from loop momentum expansion.
All above propagators can be translated into following polynomials
of $x_i$ variables
\bea D_0 & = & x_1 x_2+x_3 x_4~,~D_1= D_0-2(\a_{11}x_1+\a_{12} x_2)+
2\a_{11}\a_{12}~,~~~ \nn
D_2 & = & D_0-2(\a_{21}x_1+\a_{22} x_2)+ 2\a_{21}\a_{22}~,~~~\nn D_3
&= &D_0-2(\a_{31}x_1+\a_{32} x_2+\a_{33}x_3+\a_{34}x_4)+
2\a_{31}\a_{32}+2\a_{33}\a_{34}~,~~~\eea
where we have used the parametrization $K_1+...+K_i\equiv
\sum_{t=1}^4 \a_{it} e_t$. It is easy to see that
$(D_0-D_1)/2=\a_{11}x_1+\a_{12} x_2-\a_{11}\a_{12} $ belongs to the
ideal generated by $D_0, D_1, D_2, D_3$. However, the linearity of
this equation means that in quotient ring $k[x_1,x_2,x_3,x_4]/I$, we
can always treat variable $x_1$ as combination
$\a_{12}-{\a_{12}\over \a_{11}} x_2$. In other words, we can use
equation $0=\a_{11}x_1+\a_{12} x_2-\a_{11}\a_{12} $ to solve $x_1$
and eliminate variable $x_1$ from the quotient ring
$k[x_1,x_2,x_3,x_4]/I$. Similarly  using other two linear equations
$(D_0-D_2), (D_0-D_3)$ we can solve  variables $x_2, x_3$
\bea {x_1} & = &  \frac{\a_{12}\a_{22} (\a_{11}-\a_{21}) }{\a_{11}
   \a_{22}-\a_{12} \a_{21}}~~,~~
{x_2}  =   \frac{\a_{11} \a_{21}
   (\a_{12}-\a_{22})}{\a_{12} \a_{21}-\a_{11} \a_{22}}~,~~~\nn
{x_3} & = &
   {-{x_4} \a_{34}-\a_{21} \a_{32}+\a_{31} \a_{32}\over \a_{33}}+\frac{\a_{12}
   (\a_{11}-\a_{21}) (\a_{21} \a_{32}-\a_{22} \a_{31})}{\a_{33}(\a_{11} \a_{22}-\a_{12}
   \a_{21})}+\a_{34}~.~~~\label{box-sol}\eea
Since $x_1,x_2,x_3$ have been solved as linear polynomial of $x_4$,
we will call them {\bf reducible scalar products (RSP)}, while the
remaining variable $x_4$,  {\bf irreducible scalar products (ISP)}.

After substituting solution \eref{box-sol} into $D_0$ we get a
quadratic polynomial of single variable $x_4$
\bea D_0(x_4)= -{\a_{34}\over \a_{33}} x_4^2+ c_1
x_4+c_0~,~~~\label{box-x4}\eea
where
\bea c_1 & = & \frac{\a_{12} \a_{21} (\a_{31} (\a_{32}-\a_{22})+\a_{33} \a_{34})+\a_{11} (\a_{12}
   (\a_{22} \a_{31}-\a_{21} \a_{32})+\a_{22} ((\a_{21}-\a_{31}) \a_{32}-\a_{33}
   \a_{34}))}{(\a_{12} \a_{21}-\a_{11} \a_{22}) \a_{33}}~,~~~\nn
c_0 & = & -\frac{\a_{11} \a_{12} \a_{21}\a_{22} (\a_{11}-\a_{21})
(\a_{12}-\a_{22})
   }{(\a_{12} \a_{21}-\a_{11} \a_{22})^2}~.~~~\eea
The problem of finding integrand basis for box topology is then
reduced to finding representative elements in  quotient ring
 $k[x_4]/\Spaa{D_0(x_4)}$. Since \eref{box-x4}
is a quadratic polynomial, the representative elements in quotient
ring can take following two terms: $1$ and $x_4$. It is worth to
notice that although in this example the dimension of quotient ring
is finite, it is not true in general. In fact, if we consider the
quotient ring as linear space, in general the dimension of it will
be infinity, {\sl i.e.}, there are infinite number of representative
elements. Only when some constraints are imposed  we get finite
number of representative elements.

There is another issue regarding to the ideal defined by
\eref{box-x4}. The quadratic polynomial is reducible, {\sl i.e.}, it
can be factorized as product of two factors $a(x_4-z_1)(x_4-z_2)$
where $z_1,z_2$ are two roots. This will split the solution space
into two branches, which are obtained by setting either factor to
zero. The variety \footnote{we call the solution space as {\bf
variety} following the terminology used in algebraic geometry.}
defined by this polynomial is the union of two branches (here is
just two points). Both branches are needed to analytically (or
numerically) determine coefficients of two integrand basis ${1\over
D_0 D_1, D_2 D_3}$ and ${x_4\over D_0 D_1, D_2 D_3}$ at the
integrand level. One of the main focus of this paper is  varieties
determined by setting all propagators of a given topology to zero.
Their branch structures as well as degeneracy for specific kinematic
configurations, such as massless limit of external momenta or some
attached momenta becoming zero, will be studied carefully.

\subsection{One-loop triangle topology}

The three propagators are given by $D_0, D_1, D_2$ as in
\eref{One-loop-D}, thus we can solve
\bea {x_1}=\frac{\a_{12}\a_{22} (\a_{11}-\a_{21}) }{\a_{11}
   \a_{22}-\a_{12} \a_{21}}~,~{x_2}= \frac{\a_{11} \a_{21}
   (\a_{22}-\a_{12})}{\a_{11} \a_{22}-\a_{12} \a_{21}}~.~~~\eea
For triangle, $x_1,x_2$ become RSPs, while  $x_3, x_4$ are left as
ISPs. Putting them back to $D_0$ we obtain
\bea D_0(x_3, x_4)= \frac{\a_{11} \a_{12}  \a_{21}
\a_{22}(\a_{11}-\a_{21}) (\a_{22}-\a_{12})}{(\a_{12}
   \a_{21}-\a_{11} \a_{22})^2}+{x_3} {x_4}~.~~~\label{triangle-x34}\eea
The quotient ring is given by $k[x_3,x_4]/\Spaa{D_0(x_3,x_4)}$. Its
representative elements can be taken as $1, x_3^{n_3}, x_4^{n_4}$
with $n_3,n_4\geq 1$\footnote{ Using \eref{triangle-x34} we can
eliminate any product of $x_3, x_4$. }. Unlike the box topology, the
dimension of this quotient ring will be infinity. To select finite
number of representative elements from quotient ring, we constrain
the power $n_3,n_4$  to be no larger than three. This corresponds to
the condition that the power of $\ell$ in numerator is no more than
three for triangle topology. Under this constraint we get following
seven representative elements $\{1,x_3,x_4, x_3^2, x_4^2,
x_3^3,x_4^3\}$ as given in \cite{Ossola:2006us}.

After getting the integrand basis, we need to find their
coefficients in expansion of amplitudes. For this purpose,
understanding the variety defined by \eref{triangle-x34} becomes
important. Assuming that the equation is given by $x_3 x_4+ d=0$
with $d\neq 0$, we can solve $x_3=-d/ x_4$. Putting $x_3$ back to
integrand basis we get seven monomials of $x_4$ only: $x_4^t$ with
$t=-3,-2,-1,0,1,2,3$. Thus to find  coefficients of integrand basis,
we just need to substitute $x_1, x_2, x_3$ as functions of $x_4$
into {\sl integrand } obtained by Feynman diagrams or sewing three
on-shell amplitudes using unitarity cut method. Having the monomial
of $x_4$, we can identify corresponding coefficients for a each
power of $x_4$. For numerical analysis, we can take seven arbitrary
values of $x_4$ to write down seven linear equations and by solving
them, find the seven unknown coefficients of integrand basis.

There is a technical issue regarding to the method we  just
described. To guarantee that we will get exactly the form
$\sum_{t=-3}^{+3} c_t x_4^t$, we must first subtract all
contributions from box topologies. Similar manipulation should be
taken when finding coefficients for bubble and tadpole at one-loop.
In other words, we should subtract contributions from all other
higher topologies which contain the same set of propagators in the
problem.

The procedure we have just described is called {\sl parametrization
of variety}. For the simple example with $d\neq 0$, there is only
one irreducible branch parameterized by $x_4$. However, with some
specific kinematic configurations, above  branch can split to two
branches. This happens when $K_1^2=0$, so $\a_{12}=0$ or $K_2^2=0$,
so $(\a_{21}-\a_{11})(\a_{22}-\a_{12})=0$, or $K_3^2=(K_1+K_2)^2=0$,
so $\a_{21}\a_{22}=0$. In other words, when at least one leg is
massless, the definition equation of variety is reduced to $x_3
x_4=0$, and we get two irreducible branches. The first branch is
parameterized by setting $x_3=0$ with  $x_4$ as free parameter, and
the second branch,  by setting $x_4=0$ with $x_3$ as free parameter.
Using the parametrization of the first branch,  integrand basis
$x_3^n$ with $n=1,2,3$ will be zero and their coefficients can not
be detected by method described in previous paragraph. It means that
the first branch can only be used to find four coefficients of
integrand basis $1,x_4,x_4^2, x_4^3$. Similarly, the second branch
can only be used to find coefficients of integrand basis $1, x_3,
x_3^2, x_3^3$. These two branches intersect at one point
$x_3=x_4=0$, thus we have $4+4-1=7$, {\sl i.e.}, both branches are
necessary to fully determine coefficients of integrand basis.

\subsection{One-loop bubble topology}

Because of momentum conservation there is only one external
momentum. In this case, we take $K_1$ and another auxiliary momentum
$P$ to construct the momentum basis. With two propagators $D_0, D_1$
we can solve
\bea {x_1}= \a_{12}
   \left(1-\frac{{x_2}}{\a_{11}}\right)~,~~~\label{bubble-x1}\eea
and there are three ISPs $(x_2, x_3, x_4)$.  After eliminating
$x_1$, $D_0$ becomes polynomial of three ISPs
\bea D_0(x_2,x_3,x_4)={x_2} \a_{12}
\left(1-\frac{{x_2}}{\a_{11}}\right)+{x_3}
{x_4}~,~~\label{bubble-x234}\eea
which defines the variety  in polynomial ring $k[x_2,x_3,x_4]$.
Unlike box and triangle topologies, it is hard to find
representative elements in the quotient ring
$$k[x_2,x_3,x_4]/\Spaa{D_0(x_2,x_3,x_4)}$$ and we need a systematic
way to do so. A good way  is to use the Gr\"{o}bner basis of ideal.
Firstly we write down all possible monomials $x_2^{n_2} x_3^{n_3}
x_4^{n_4}$ with $n_2+n_3+n_4\leq 2$ required by  physical
constraints.  Then we divide each monomial $x_2^{n_2} x_3^{n_3}
x_4^{n_4}$ by Gr\"{o}bner basis and collect all monomials in the
remainder. These monomials collected from the remainder times
${1\over D_0 D_1}$ give the integrand basis.

A technical issue of above algorithm is the ordering of ISPs in the
constructing of Gr\"{o}bner basis. Different ordering gives, in
general, different Gr\"{o}bner basis and different sets of
representative elements, although they are equivalent to each other.
Once a particular ordering is chosen, we should stick to it through
the whole calculation to avoid inconsistency. For instance, if the
ordering is chosen as $x_3>x_4>x_2$ we get 9 integrand basis as
\bea \left\{1,{x_2},{x_2}^2,{x_3},{x_2}
   {x_3},{x_3}^2,{x_4},{x_2} {x_4},{x_4}^2\right\}~.~~~\eea
This integrand basis can be used to expand bubble topology. In order
to get the coefficients of integrand basis analytically, we should
first put $\ell=x_2 e_1+x_1 e_2+ x_4 e_3+ x_3 e_4$ back into
integrand  after subtracting all box and triangle contributions.
Then we can replace $x_1$ by \eref{bubble-x1}, and get a polynomial
$A(x_2,x_3,x_4)$. The next step is to divide this polynomial by
Gr\"{o}nber basis and obtain the remainder. This algorithm,
different from previous parametrization method,  ensures that the
remainder is nothing but the linear combination of monomials in
integrand basis with coefficients we want to find.

If using the parametrization method,  we can  replace\footnote{This
parametrization works for almost every value of $x_2$ except $x_2=0$
and $x_2=\a_{11}$ where the variety is degenerate.} $x_3=
-\a_{12}{x_2\over x_4} (1-{x_2\over \a_{11}})$ in the expression
$A(x_2,x_3,x_4)$ as well as integrand basis, and get
\bean & & \frac{c(6) {x_2}^4 \a_{12}^2}{{x_4}^2 \a_{11}^2}-\frac{2 c(6)
   {x_2}^3 \a_{12}^2}{{x_4}^2 \a_{11}}+\frac{c(5) {x_2}^3
   \a_{12}}{{x_4} \a_{11}}+\frac{c(6) {x_2}^2
   \a_{12}^2}{{x_4}^2} +\frac{c(4) {x_2}^2 \a_{12}}{{x_4}
   \a_{11}}-\frac{c(5) {x_2}^2 \a_{12}}{{x_4}}\nn & &-\frac{c(4) {x_2}
   \a_{12}}{{x_4}}+c(3){x_2}^2+c(8) {x_2} {x_4}+c(2)
   {x_2}+c(9) {x_4}^2+c(7) {x_4}+c(1)~.~~~\eean
Since we have already used the equation to reduce one variable
further, remaining variables $x_2, x_4$ are totally free variables.
 What we
need to do is to compare each independent monomial $x_4^a x_2^b$
($a,b$ could be negative integers) at both sides. The
parametrization method can also be used for numerical fitting.  We
only need to write down enough linear equations to solve
coefficients by taking sufficient numerical values $(x_2,x_4)$ at
both sides.

Similar to triangle topology, the variety defined by
\eref{bubble-x234} is irreducible for general kinematic
configuration. However, when $K_1^2=0$\footnote{For one-loop theory,
bubble basis with $K_1^2=0$ vanishes after integration, but it is
necessary at the integrand level.} we have $\a_{12}=0$ by our
construction, thus equation \eref{bubble-x234} is reduced to
$x_3x_4=0$. In other words, the  variety is degenerated to two
branches: one with $x_3=0$ and $x_2, x_4$ as free parameters, and
another with $x_4=0$ and $x_2,x_3$ as free parameters. Each branch
can detect six coefficients out of nine integrand basis, while three
basis $\{1,x_2,x_2^2\}$ can be detected by both branches. Thus we
have $6+6-3=9$, and both branches are necessary to find all
coefficients of integrand basis analytically or numerically.

\subsection{One-loop tadpole topology}

In this case, we choose arbitrary two independent momenta to
construct the momentum basis. Since there is only one propagator
$D_0$, all four variables $x_i$, $i=1,2,3,4$ are ISPs and the
variety is defined by equation
\bea D_0= x_1 x_2 +x_3 x_4~.~~~\eea
Requiring the total dimension of monomials to be no larger than one,
we get following basis
\bea \{1,x_1,x_2,x_3,x_4\}~.~~~\eea
This variety is irreducible and we can parameterize it by solving
$x_1=-{x_3 x_4\over x_2}$. Thus after putting $x_1$ back to
integrand after subtracted all contributions from boxes, triangles
and bubbles, we can read out coefficients of one-loop tadpole
integrand basis by comparing monomials of $x_2^a x_3^b x_4^c$.

\section{A premiere: some trivial two-loop topologies}

Starting from this section, we will discuss the integrand basis and
variety of various two-loop topologies classified in \eref{4-n123}
using the same method presented in previous section for one-loop
topologies. Before we discuss non-trivial topologies, there are some
topologies whose integrand basis and structure of variety are quite
simple. These include two cases. The first case is all topologies of
type (A), where two one-loop sub-structures share only one single
vertex. The second case is all topologies having maximal number of
propagators, {\sl i.e.}, 8 propagators for pure 4-dimensional
two-loop diagrams.

\EPSFIGURE[ht]{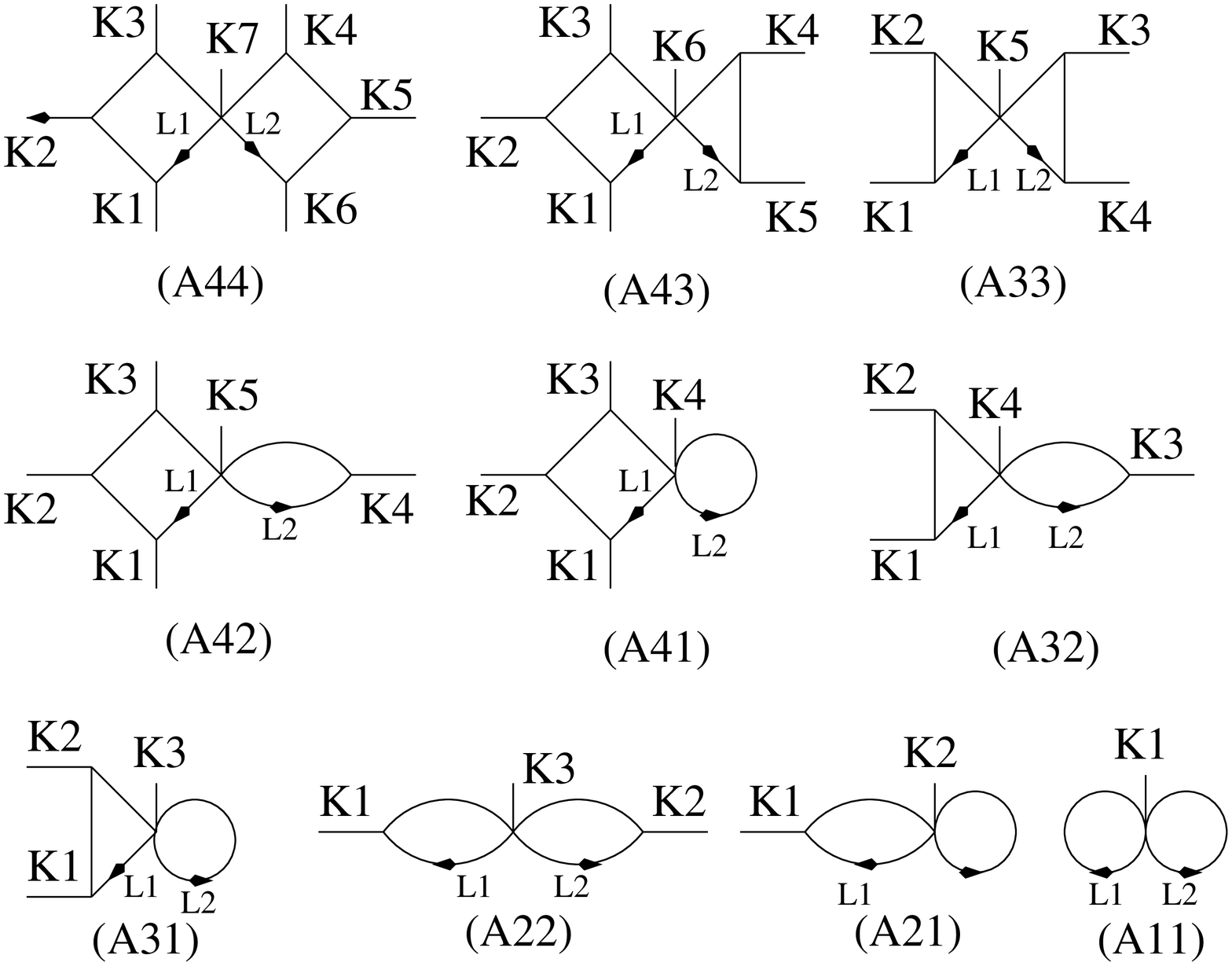,width=12.5cm} {The type (A) contains
$10$ topologies with $n_3=0$. Every topology is denoted by (Anm)
where $n,m$ are the numbers of propagators of the  left and right
one-loop sub-topologies respectively. These diagrams are drawn in
most general form, and some external momenta, for instance $K_7$ in
(A44) diagram, could be absent. All external momenta are out-going
while convention of loop momenta is labeled by  arrows in each
diagram. \label{4d-A} }
%

\subsection{Two-loop topologies of type (A)}

All two-loop topologies of type (A) can be found in Figure \eref{4d-A}.
Since there is no propagator involving both $\ell_1,\ell_2$,
integrand basis and variety defined by propagators will be double
copy of corresponding two one-loop sub-topologies with minor
modification. This modification comes from constraints of total
degree of monomials in integrand basis. Taking topology (A33) as an
example, for the left one-loop sub-topology, we can use $K_1,K_2$ to
construct momentum basis $e_1, e_2, e_3, e_4$, thus $x_3=\ell_1\cdot
e_3$, $x_4=\ell_1\cdot e_4$ will become ISPs after solving linear
equations. Similarly for the right one-loop sub-topology, we can use
$K_3, K_4$ to construct another momentum basis $\W e_1,\W e_2,\W
e_3,\W e_4$, thus $y_3=\ell_2\cdot \W e_3$, $y_4=\ell_2\cdot \W e_4$
become ISPs. The representative elements of integrand basis for
(A33) can be given by monomial $x_3^{n_3} x_4^{n_4} y_3^{m_3}
y_4^{m_4}$. From the left one-loop sub-topology we have  constraint
$n_3+n_4\leq 3$ because along the loop there are only three
vertices. Similarly we have $m_3+m_4\leq 3$ from right one-loop
sub-topology. However, since there are only five vertices along
whole two-loop topology we should have $n_3+n_4+m_3+m_4\leq 5$.
Under these conditions $x_3^3 y_3^3$ should be excluded from
integrand basis and we get $7\times 7-4=45$ basis for (A33).

The variety is also the union of varieties of corresponding two
one-loop sub-topologies, so its structure can be easily inferred. To
determine coefficients of integrand basis, similar procedures as
presented in previous section can be applied, such as  Gr\"{o}bner
basis method or parametrization method.

\subsection{Topologies with  eight propagators}

Besides topologies of type (A), there are three special topologies
in type (B) and (C) which have maximal number (eight) of
propagators. Since there are  eight components for two loop momenta
$\ell_1,\ell_2$, putting eight propagators on-shell  will completely
freeze all eight components, thus the variety will be fixed to
isolated points. These three topologies are planar penta-box (B43)
as shown in  Figure \eref{4d-B}, and non-planar crossed
penta-triangle (C42), crossed double-box (C33) as shown in  Figure
\eref{4d-C}\footnote{Topology (A44) also has eight propagators. The
variety is simply given by four isolated points and integrand basis
has exactly four terms. These four points can be used to determine
four coefficients of basis.}.

\subsubsection{The topology (B43): planar penta-box}

For (B43) topology, we take $K_1, K_5$ to construct momentum basis
$e_i, i=1,2,3,4$ and use them to expand both loop momenta
$\ell_1,\ell_2$ with coefficients $x_i=\ell_1\cdot e_i$ and
$y_i=\ell_2\cdot e_i$. Since there are four propagators containing
only $\ell_1$, just like the one-loop box case, $x_1,x_2,x_3$ can be
solved as linear functions of $x_4$. After substituting these
solutions, $D_0=\ell_1^2$ becomes quadratic function of single
variable $x_4$
\bea D_0 & = & c_2 x_4^2+ c_1 x_4+ c_0~,~~~\eea
where $c_i$ are some functions of external momenta, which may be
complicated depending on kinematic configurations, but not important
here. Similarly, there are three propagators containing only
$\ell_2$, so like the one-loop triangle case, $y_1, y_2$ are solved
as linear functions of $y_3, y_4$. After substituting these
solutions, $\W D_0= \ell_2^2$ becomes
\bea \W D_0=  \W c_{20} y_3^2+\W c_{02} y_4^2+ \W c_{11} y_3 y_4 +\W
c_{10} y_3 +\W c_{01} y_4+ \W c_{00}~.~~~ \eea
Propagator $(\ell_1+\ell_2+K_6)^2$ can also be expressed as function
of these ISPs as
\bea \WH D_0= \sum_{ij}  d_{ij} x_4^i y_3^j+ \sum_{ij} \W d_{ij}
x_4^i y_4^j~,~i,j=0,1~,~~~ \eea
where we have used the conditions $\ell_1^2=\ell_2^2=0$.

The integrand basis is constructed  by dividing monomials $x_4^{n_4}
y_3^{m_3} y_4^{m_4}$ with conditions $n_4\leq 5$, $m_3+m_4\leq 4$
and $n_4+m_3+m_4\leq 7$ over Gr\"{o}bner basis of the ideal
generated by polynomials $D_0, \W D_0, \WH D_0$. The result is
\bea {\cal B}_{B43}=\{1, x_4, y_3, y_4\}~.~~~\label{B1-basis} \eea
The variety defined by $D_0, \W D_0, \WH D_0$ has four branches and
each branch has a single solution. Thus using four branches, we can
fit coefficients of four integrand basis analytically or numerically
by the method discussed in previous section.

Above results will not change for following specific kinematic
configurations: (1) $K_6$ or $ K_7$ or both are absent; (2) some of
$K_i$, $i=1,2,3,4,5$ are massless.

\subsubsection{The topology (C42): non-planar crossed penta-triangle}

For (C42) topology, we take $K_1, K_4$ to construct momentum basis
$e_i, i=1,2,3,4$ and use them to expand both loop momenta
$\ell_1,\ell_2$  with coefficients $x_i=\ell_1\cdot e_i$ and
$y_i=\ell_2\cdot e_i$. Since there are four propagators containing
only $\ell_1$, $x_1,x_2,x_3$ can be solved as linear functions of
$x_4$, and $D_0=\ell_1^2$ can be rewritten as a quadratic polynomial
of $x_4$
\bea D_0 & = & c_2 x_4^2+ c_1 x_4+ c_0~.~~~\eea
Coefficients $c_i$ are again some functions of external momenta
whose explicit expressions are not important here. Similarly, there
are two propagators containing only $\ell_2$, and $y_2$ can be
solved as linear function of $y_1, y_3, y_4$. However, unlike the
topologies of type (B), here we have two propagators containing both
$\ell_1, \ell_2$, {\sl i.e.}, $\WH D_0= (\ell_1+\ell_2+K_6)^2$ and
$\WH D_1= (\ell_1+\ell_2+K_6+K_7)^2$. We can get one more linear
equation $\WH D_1-\WH D_0$ and solve $y_1$ as linear function of
$x_4, y_3, y_4$. Thus we have three ISPs $(x_4,y_3,y_4)$ and three
quadratic polynomials. Using ideal generated by these three
polynomials we find the integrand basis is given by
\bea {\cal B}_{C42}=\{1, x_4, y_3, y_4\}~.~~~\label{B1-basis} \eea
The variety defined by these three quadratic equations has four
branches, and each branch is given by a point. Thus using four
branches we can find coefficients of four integrand basis. Again
above discussion does not change whether  $K_6, K_7$ are absent or
not, or any of other external momenta go to massless limit.

\subsubsection{The topology (C33): non-planar crossed double-box}

For (C33) topology, we take $K_1, K_4$ to construct momentum basis
$e_i, i=1,2,3,4$, and use them expand both loop momenta
$\ell_1,\ell_2$. We can get five linear equations from  eight
on-shell equations, and solve, for instance, $x_1,x_2,x_3,y_1,y_2$
as functions of three ISPs $(x_4, y_3, y_4)$. After substituting all
RSPs in the remaining three propagators we get three quadratic
polynomials. The variety defined by these three quadratic
polynomials is given by eight points (eight branches).  By
Gr\"{o}bner basis method, the integrand basis is given by  8
elements
\bea {\cal B}_{C33}=\{1, x_4, y_3, x_4 y_3, y_3^2, y_3^3, y_4, y_3
y_4\}~.~~~\eea
As usual, each branch of variety can detect one coefficient of
integrand basis, and  using all 8 branches, we can get all
coefficients. Again above discussion does not rely on the explicit
kinematic configuration of external momenta.

\section{Example one: planar penta-triangle}

%
\EPSFIGURE[ht]{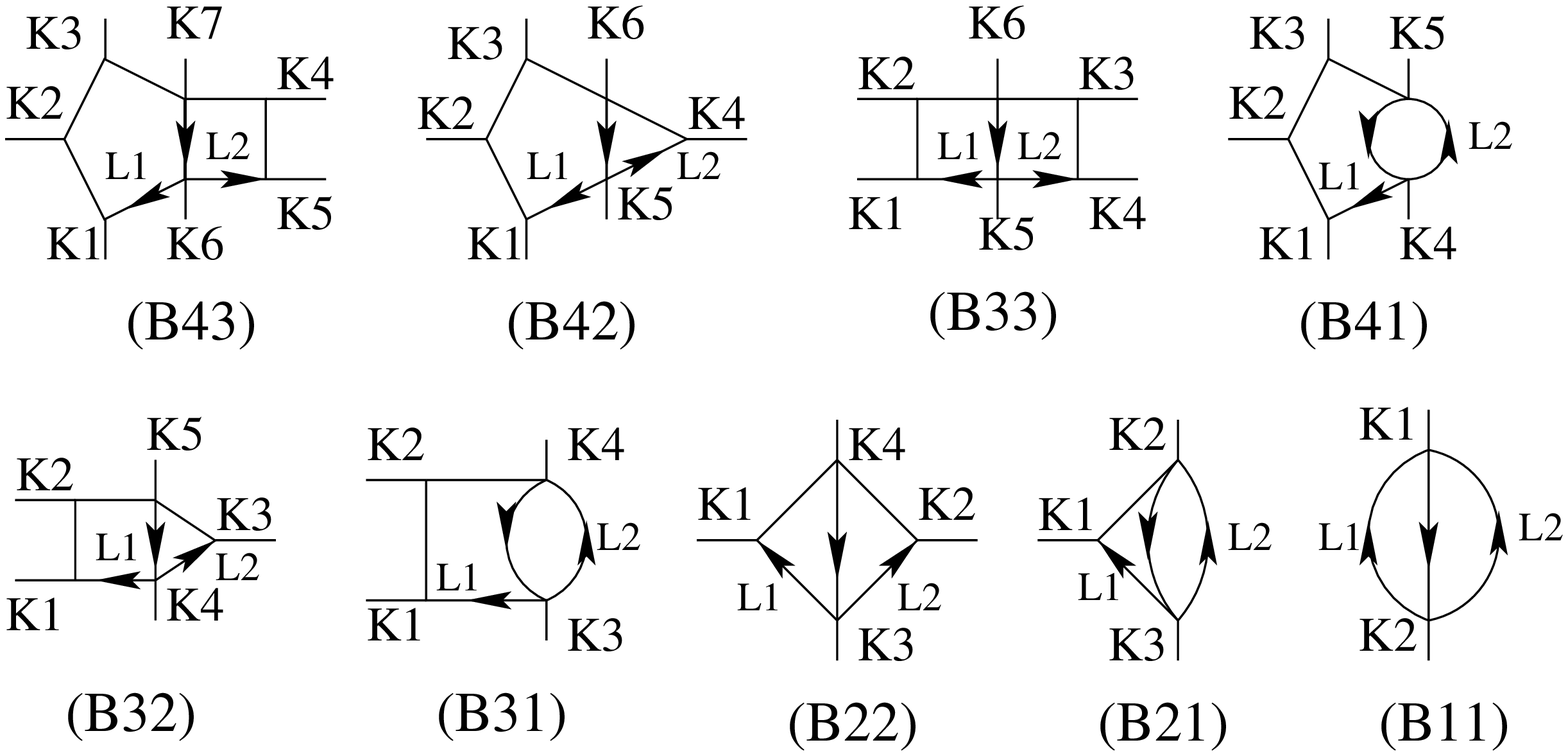,width=12.5cm} {The type (B) contains 9
togologies with $n_3=1$. Every topology is denoted by (Bnm) where
$n,m$ are the numbers of propagators containing only $\ell_1$ or
$\ell_2$. The diagrams are drawn in most general form, and some
external momenta, for instance $K_6, K_7$ in (B43), could be absent.
All external momenta are out-going while convention of loop momenta
is labeled by arrows in each diagram. \label{4d-B} }

Having understood simple topologies of planar penta-box, non-planar
crossed penta-triangle and crossed double-box, we  move to
non-trivial topologies where varieties are given by manifolds with
dimension at least one, not just isolated points. For these
topologies, analysis becomes more complicated, so we will take two
topologies as examples to illustrate various properties. The first
example we will study is planar two-loop penta-triangle topology
(B42), as shown in Figure \ref{4d-B}.

The penta-triangle topology has 7 propagators. If we choose
$K_1,K_4$ to generate momentum basis $e_i,i=1,2,3,4$, all kinematics
can be expanded as
\bea &&\ell_1=x_2 e_1+x_1 e_2+x_4 e_3 +x_3 e_4~,~~~\ell_2=y_2
e_1+y_1
e_2+y_4 e_3 +y_3 e_4~,~~~\nonumber\\
&&K_1=\alpha_{11}e_1+\alpha_{12}e_2~,~~~
K_1+K_2=\alpha_{21}e_1+\alpha_{22}e_2+\alpha_{23}e_3+\alpha_{24}e_4~,~~~\nonumber\\
&&K_1+K_2+K_3=\alpha_{31}e_1+\alpha_{32}e_2+\alpha_{33}e_3+\alpha_{34}e_4~,~~~\nonumber\\
&&K_4=\beta_{11}e_1+\beta_{12}e_2~,~~~K_5=\gamma_{11}e_1+\gamma_{12}e_2+\gamma_{13}e_3+\gamma_{14}e_4~,~~~\eea
and $K_6$ is constructed from momenta conservation. Above parameters
are general if $K_6$ is arbitrary, but when $K_6=0$ or $K_6$ is
massless, there will be relations among parameters. For example,
when $K_6=0$, we should have
\bea
\gamma_{11}=-\beta_{11}-\alpha_{31}~,~~~\gamma_{12}=-\beta_{12}-\alpha_{32}~,~~~\gamma_{13}=-\alpha_{33}~,~~~\gamma_{14}=-\alpha_{34}~,~~~\eea
and when $K_5=K_6=0$  we should  have $\gamma_{1i}=0$ and
\bea
\alpha_{31}=-\beta_{11}~,~~~\alpha_{32}=-\beta_{12}~,~~~\alpha_{33}=0~,~~~\alpha_{34}=0~.~~~\eea
These relations will be important when discussing branch structure
of variety under specific kinematic configurations.

Using above expansion, we can expand all  seven propagators
\bea &&D_0=\ell_1^2~,~~~D_1=(\ell_1-K_1)^2~,
~~~D_2=(\ell_1-K_1-K_2)^2,~~ D_3=(\ell_1-K_1-K_2-K_3)^2\nn
&&\widetilde{D}_0=\ell_2^2,~~~
\widetilde{D}_1=(\ell_2-K_4)^2,~~~~\widehat{D}_0=(\ell_1+\ell_2+K_5)^2~,~~~\eea
and use four linear equations $D_1-D_0=0, D_2-D_0=0, D_3-D_0=0$ and
$\widetilde{D}_1-\widetilde{D}_0=0$ to solve $x_1,x_2, x_3, y_2$ as
linear functions of four ISPs $(x_4,y_1,y_3,y_4)$. The results are
given by
\bea x_1&=&{\alpha_{12} (\alpha_{24} \alpha_{33}-\alpha_{23}
\alpha_{34}) \over \alpha_{12} (\alpha_{23} \alpha_{31}-\alpha_{21}
\alpha_{33})+\alpha_{11} (\alpha_{22} \alpha_{33}-\alpha_{23}
\alpha_{32})}x_4\nonumber\\
&&+{\alpha_{12} (-\alpha_{21} \alpha_{22} \alpha_{33}+\alpha_{11}
(\alpha_{22} \alpha_{33}-\alpha_{23} \alpha_{32})+\alpha_{23}
(\alpha_{31} \alpha_{32}+
\alpha_{33}\alpha_{34}-\alpha_{33}\alpha_{24}))\over \alpha_{12}
(\alpha_{23} \alpha_{31}-\alpha_{21} \alpha_{33})+\alpha_{11}
(\alpha_{22} \alpha_{33}-\alpha_{23} \alpha_{32})}~,~~~\eea
\bea x_2&=&{\alpha_{11} (\alpha_{23} \alpha_{34}-\alpha_{24}
\alpha_{33}) \over \alpha_{12} (\alpha_{23} \alpha_{31}-\alpha_{21}
\alpha_{33})+\alpha_{11} (\alpha_{22} \alpha_{33}-\alpha_{23}
\alpha_{32})}x_4\nonumber\\
&&+{\alpha_{11} (\alpha_{21} \alpha_{22} \alpha_{33}+\alpha_{12}
(\alpha_{23} \alpha_{31}-\alpha_{21} \alpha_{33})-\alpha_{23}
(\alpha_{31}\alpha_{32}+
\alpha_{33}\alpha_{34}-\alpha_{33}\alpha_{24}))\over \alpha_{12}
(\alpha_{23} \alpha_{31}-\alpha_{21} \alpha_{33})+\alpha_{11}
(\alpha_{22} \alpha_{33}-\alpha_{23} \alpha_{32})}~,~~~\eea
\bea x_3&=&{\alpha_{12} (\alpha_{21} \alpha_{34}-\alpha_{24}
\alpha_{31})+\alpha_{11} (\alpha_{24} \alpha_{32}-\alpha_{22}
\alpha_{34})\over \alpha_{12} (\alpha_{23} \alpha_{31}-\alpha_{21}
\alpha_{33})+\alpha_{11} (\alpha_{22} \alpha_{33}-\alpha_{23}
\alpha_{32})}x_4\nonumber\\
&&+{ \alpha_{11} ((-\alpha_{23} \alpha_{24}-
\alpha_{22}\alpha_{21}+\alpha_{22}\alpha_{31})
\alpha_{32}+\alpha_{12} (\alpha_{21}
\alpha_{32}-\alpha_{22}\alpha_{31})+\alpha_{22} \alpha_{33}
\alpha_{34})\over \alpha_{12} (\alpha_{23} \alpha_{31}-\alpha_{21}
\alpha_{33})+\alpha_{11} (\alpha_{22} \alpha_{33}-\alpha_{23}
\alpha_{32})}\nonumber\\
&&+{\alpha_{12} (\alpha_{23} \alpha_{24} \alpha_{31}+\alpha_{21} (
\alpha_{31}\alpha_{22}-\alpha_{31}\alpha_{32}-\alpha_{33}
\alpha_{34}))\over \alpha_{12} (\alpha_{23} \alpha_{31}-\alpha_{21}
\alpha_{33})+\alpha_{11} (\alpha_{22} \alpha_{33}-\alpha_{23}
\alpha_{32})}~,~~~\eea
and
\bea y_2=\beta_{11}(1-{y_1\over \beta_{12}})~.~~~\eea

Now we consider the remaining three equations. Firstly the equation
$D_0=0$ becomes a quadratic equation of $x_4$ and we always have 2
solutions $x_4^{\Gamma_1},x_4^{\Gamma_2}$ in $\mathbb{C}$-plane.
There is no intersection between these two solutions, so the variety
 has been split into two separate branches
parameterized by $x_4^{\Gamma}$. Remaining two equations are
\bea \W D_0=\beta_{11}(1-{y_1\over \beta_{12}})y_1+y_3y_4=0~,~~~\eea
\bea
\WH D_0=0 &=&\Big(x_2^{\Gamma}+\gamma_{11}-(x_1^{\Gamma}+\gamma_{12}){\beta_{11}\over
\beta_{12}}\Big)y_1+(x_4^{\Gamma}+\gamma_{13})y_3+(x_3^{\Gamma}+\gamma_{14})y_4\nonumber\\
&&+(x_1^{\Gamma}+\gamma_{12})\beta_{11}
+\gamma_{11}x_1^{\Gamma}+\gamma_{12}x_2^{\Gamma}+\gamma_{13}x_3^{\Gamma}+\gamma_{14}x_4^{\Gamma}+\gamma_{11}\gamma_{12}+\gamma_{13}\gamma_{14}~.~~~\eea
%


Knowing the ideal generated by these four ISPs, we can use the
Gr\"{o}bner basis method with ordering $(x_4,y_1, y_4, y_3)$ to find
integrand basis under constraints on the powers of monomial
$x_4^{d(x_4)}y_1^{d(y_1)}y_3^{d(y_3)}y_4^{d(y4)}$
\bea &&\sum_{\mbox{\tiny{all~ISPs~of~x}}}d(x_i)\leq
5~,~~~\sum_{\mbox{\tiny{all~ISPs~of~y}}}d(y_i)\leq
3~,~~~\sum_{\mbox{\tiny{all~ISPs~of~x}}}d(x_i)+\sum_{\mbox{\tiny{all~ISPs~of~y}}}d(y_i)\leq
6~~.~\eea
 Elements of integrand basis will be
different depending on actual kinematic configurations. In this
case, there are three kinds of integrand basis depending on if $K_4$
is massless or if $K_5,K_6$ are absent. For all kinematic
configurations with $K_4$ massive, the integrand basis contains 14
elements given by
\bea {\cal B}_{B42}^{I}= \{1, x_4, y_1, y_3, x_4 y_3, y_1 y_3,
y_3^2, y_3^3, y_4, y_3 y_4, y_3^2 y_4, y_4^2,
 y_3 y_4^2, y_4^3\}~.~~~\label{B2-basis}\eea
For kinematic configurations with $K_4$ massless but at most one of
$K_5,K_6$  absent, the integrand basis contains 14 elements given by
\bea {\cal B}_{B42}^{II}=\{1, x_4, y_1, y_3, x_4 y_3, y_1 y_3,
y_3^2, y_1 y_3^2, y_3^3, y_4, y_1 y_4, y_4^2,
 y_1 y_4^2, y_4^3\}~.~~~\label{B2-K4-basis}\eea
For kinematic configurations with $K_4$ massless and both
$K_5=K_6=0$, the integrand basis has 20 elements given by
\bea {\cal B}_{B42}^{III}&=&\{1, x_4, y_1, x_4 y_1, y_1^2, x_4
y_1^2, y_1^3, x_4 y_1^3, y_3, y_1 y_3,\nonumber\\
&& y_1^2 y_3, y_3^2, y_1 y_3^2, y_3^3, y_4, y_1 y_4, y_1^2 y_4,
y_4^2, y_1 y_4^2, y_4^3\}~.~~~\label{B21-K4-basis}\eea
Note that elements of integrand basis generated from different
ordering of ISPs will possibly be different, but after choosing one
ordering, there will always be three kinds of integrand basis
depending on kinematic configurations.

After given integrand basis, we need to discuss how to get their
coefficients  from integrand coming from Feynman diagrams or
unitarity cut method. As in the one-loop case, either algebraic
geometry method or parametrization method can be used.

The algebraic geometry method is illustrated as follows. Firstly we
should get integrand  ${\cal F}(\ell_1,\ell_2)$ from Feynman
diagrams or unitarity cut method after subtracting contributions
from higher topologies. After expanding $\ell_1,\ell_2$ into
momentum basis and substituting RSPs with expressions of ISPs, we
can rewrite ${\cal F}(\ell_1,\ell_2)$ as polynomials of ISPs. For
example, in this example ${\cal F}(x_4,y_1,y_3,y_4)$. Then we can
divide ${\cal F}(x_4,y_1,y_3,y_4)$ by Gr\"{o}bner basis generated
from ideal $I\equiv \Spaa{D_0, \W D_0,\WH D_0}$ with ordering
$(x_4,y_1, y_4, y_3)$. The remainder of division is linear
combinations of all terms in integrand basis with wanted
coefficients.

All coefficients can be found at the same time  using above
algebraic geometry method, but it may take long time to do so if the
number of elements is large. Instead we can use branch-by-branch
polynomial fitting method (see reference \cite{Badger:2012dv}) to
simplify problem, by finding a smaller set of coefficients at one
time. The idea can be illustrated as follows. Because
$D_0=\a(x_4-x_4^{\Gamma_1})(x_4-x_4^{\Gamma_2})$, we can divide
polynomials ${\cal F}(x_4,y_1,y_3,y_4)$ by Gr\"{o}bner basis
generated from $I_1\equiv \Spaa{(x_4-x_4^{\Gamma_1}), \W D_0,\WH
D_0}$ with ISPs ordering  $(x_4,y_1, y_4, y_3)$. After the division
${\cal F}(x_4,y_1,y_3,y_4)/I_1$, we will get remainder
\bea {\cal R}({\cal F}(x_4,y_1,y_3,y_4)/I_1) =f_1+f_2 y_3+f_3
y_3^2+f_4 y_3^3+f_5 y_4+f_6 y_3 y_4+f_7 y_3^2 y_4~,~~~\eea
with seven known coefficients $f_i$. It is easy to see that the
remainder of 14 integrand basis over $I_1$ is given by
\bea & & (1)/I_1\to 1~,~ (x_4)/I_1\to d_2~,~ (y_1)/I_1\to d_{31}y_4
+d_{32}y_3 +d_{33}~,~ (y_3)/I_1\to y_3~,~~~\nn
& &  (x_4 y_3)/I_1\to d_5 y_3~,~ (y_1 y_3)/I_1\to
 d_{61}y_3 y_4+d_{62} y_3^2+d_{63}
   y_3~,~ (y_3^2)/I_1\to y_3^2~,~ (y_3^3)/I_1\to  y_3^3~,~~~\nn
&& (y_4)/I_1\to y_4~,~ (y_3 y_4)/I_1\to  y_3 y_4~,~ (y_3^2
y_4)/I_1\to y_3^2 y_4~,~~~ \nn
& & (y_4^2)/I_1\to  d_{12,1}y_3 y_4+d_{12,2} y_4+d_{12,3}
y_3^2+d_{12,4} y_3+d_{12,5}~,~~~\nn & & (y_3 y_4^2)/I_1\to d_{13,1}
y_3^2 y_4+d_{13,2} y_3 y_4+d_{13,3} y_3^3+d_{13,4} y_3^2+d_{13,5}
y_3~,~~~\nn & & (y_4^3)/I_1\to  d_{14,1}y_3^2 y_4+d_{14,2} y_3
y_4+d_{14,3} y_4+d_{14,4} y_3^3+d_{14,5} y_3^2+d_{14,6}
y_3+d_{14,7}~,~~~ \eea
with known coefficients $d$. Thus by comparing both sides we obtain
following seven equations of 14 unknown coefficients $c_i$ from one
branch
\bea f_1 & = & c_1+ c_2 d_2+c3
d_{33}+c_{12}d_{12,5}+c_{14}d_{14,7}~,~~~\nn f_2 & = & c_3
d_{32}+c_4+c_5 d_5+c_6
d_{63}+c_{12}d_{12,4}+c_{13}d_{13,5}+c_{14}d_{14,6}~,~~~\nn f_3 & =
& c_6 d_{62}+c_7+c_{12} d_{12,3}+c_{13}
d_{13,4}+c_{14}d_{14,5}~,~~~\nn f_4 & = & c_8
+c_{13}d_{13,3}+c_{14}d_{14,4}~,~~~\nn f_5 & = &  c_3
d_{51}+c_9+c_{12}d_{12,2}+c_{14}d_{14,3}~,~~~\nn f_6 & = & c_6
d_{61}+c_{10}+c_{12} d_{12,1}+c_{13}d_{13,2}+c_{14}d_{14,2}~,~~~\nn
f_7 & = &  c_{11} +c_{13} d_{13,1}+c_{14} d_{14,1}~.~~~\eea
Similarly, we can  divide polynomials ${\cal F}(x_4,y_1,y_3,y_4)$ by
Gr\"{o}bner basis generated from another branch $I_2\equiv
\Spaa{(x_4-x_4^{\Gamma_2}), \W D_0,\WH D_0}$ with the same ISPs
ordering $(x_4,y_1, y_4, y_3)$. After that we can get another seven
equations relating $\W f_i$ to $c_i$ with other known coefficients
$\W d$. With this modified algebraic method, we can get a smaller
set of coefficients in each branch. In this example each branch can
be used to write down seven equations (we will say that this branch
can detect seven coefficients). Combining results of both branches
we get 14 independent equations, and they can be used to solve 14
coefficients of $c_i$.

Besides algebraic geometry method, it is also possible to find
coefficients by parametrization method. This method is tightly
related to the branch-by-branch fitting  method. In this example, we
can use $D_0$ to solve $x_4$ and get two solutions. Then we put one
solution $x_4^{\Gamma_i}$ to $\W D_0, \WH D_0$, and use one
variable, for example, $y_4$ to express $y_1, y_3$. Finally we put
$y_1(y_4), y_3(y_4)$ back to the identity
\bea {\cal F}(x_4^{\Gamma_i},y_1(y_4),y_3(y_4),y_4)=\sum_{k=1}^{14}
c_k {\cal B}_{B42,k}(y_4)~,~~~\eea
and find coefficients $c_i$ by comparing both sides. This method is
very useful to evaluate coefficients analytically or numerically. In
this example, we only need to take arbitrary $7$ values of $y_4$ to
produce seven equations from each branch, and solve 14 linear
equations by combining two branches to find all coefficients.

\subsection{Structure of variety under various kinematic configurations}

For some kinematic configurations, for instance, some of external
momenta being massless or absent, the variety will split into
different branches.
In this example, as we have mentioned, no matter what kinematic
configuration is, we always have  two solutions
$x_4^{\Gamma_1},x_4^{\Gamma_2}$ from equation $\ell_1^2=0$. Thus we
will focus on the two remaining  equations $\W D_0, \WH D_0$ with
$x_4$ replaced by two solutions $x_4^{\Gamma_1}, x_4^{\Gamma_2}$.
Since in general $x_4^{\Gamma_1}\neq x_4^{\Gamma_2}$, branches
parameterized by different $x_4^{\Gamma_i}$ will not intersect with
each other.

When $K_4$ is massive, $\beta_{11}\neq 0$, the on-shell equation
$\widetilde{D}_0=0$ is not degenerate. If we take $y_1=\tau$ as free
parameter, $\widetilde{D}_0=0$ becomes non-degenerate conic section
of variables $y_3, y_4$, while
 $\widehat{D}_0=0$ becomes linear
equation of variables $y_3, y_4$. Using following two equations
\bea
\widetilde{D}_0=0~:~~~y_3y_4+F(\tau)=0~,~~~\widehat{D}_0=0~:~~~ay_3+by_4+c(\tau)=0~,~~~\eea
where $a,b$ are some constants, $F(\tau)$ is second order function
of $\tau$, and $c(\tau)$ is linear function of $\tau$,  we can solve
\bea y_4={ac(\tau)\pm \sqrt{a^2[c(\tau)^2+4abF(\tau)]}\over
-2ab}~.~~~\eea
$y_4$ is a rational function of $\tau$ if $c(\tau)^2+4abF(\tau)$
inside the square root is a perfect square. Using the explicit
expressions of $F(\tau)$, $c(\tau)$ and $a,b$ we find the
discriminant of quadratic function $c(\tau)^2+4abF(\tau)$ to be
\bea
(x_1^{\Gamma}x_2^{\Gamma}+x_3^{\Gamma}x_4^{\Gamma})(\beta_{11}-{\Xi\over
\beta_{12}})+{(x_2^{\Gamma}+\gamma_{11}+\beta_{11})(x_1^{\Gamma}+\gamma_{12}+\beta_{12})+(x_4^{\Gamma}+\gamma_{13})(x_3^{\Gamma}+\gamma_{14})\over
\beta_{12}}\Xi~,~~~\eea
where
\bea
\Xi=\gamma_{11}x_1^{\Gamma}+\gamma_{12}x_2^{\Gamma}+\gamma_{13}x_3^{\Gamma}+\gamma_{14}x_4^{\Gamma}+\gamma_{11}\gamma_{12}+\gamma_{13}\gamma_{14}~,~~~\eea
and $x_i^\Gamma$ denotes the solution of $x_i$ with
$x_4=x_4^\Gamma$. The first term in above result vanishes because
$D_0=x_1^{\Gamma}x_2^{\Gamma}+x_3^{\Gamma}x_4^{\Gamma}=0$. Generally
the second term will not be zero, but if $K_5=0$, {\sl i.e.},
$\gamma_{1i}=0$, we have $\Xi=0$. Similarly, if $K_5\neq 0$ but
$K_6=0$, using
 $\gamma_{11}=-\beta_{11}-\alpha_{31}$,
$\gamma_{12}=-\beta_{12}-\alpha_{32}$, $\gamma_{13}=-\alpha_{33}$
and $\gamma_{14}=-\alpha_{34}$, the second term becomes
\bea
{-x_1^{\Gamma}\alpha_{31}-x_2^{\Gamma}\alpha_{32}-x_3^{\Gamma}\alpha_{33}-x_4^{\Gamma}\alpha_{34}+\alpha_{31}\alpha_{32}+\alpha_{33}\alpha_{34}\over
\beta_{12}}\Xi={D_3-D_0\over 2\beta_{12}}\Xi~,~~~\eea
which vanishes by on-shell equation $D_3=D_0=0$. In these cases,
$c(\tau)^2+4abF(\tau)$ is a perfect square, and we can get two
solutions which are rational functions of free parameter $y_1$ for
each solution $x_4^\Gamma$. In other words, each original
irreducible branch will split into two branches in these specific
kinematic configurations.  In total we get four branches denoted by
${V}^{\Gamma_1,\Pi_1}, {V}^{\Gamma_1,\Pi_2}$ and
${V}^{\Gamma_2,\Pi_1},{V}^{\Gamma_2,\Pi_2}$. Each branch can detect
4 coefficients. Two branches ${V}^{\Gamma_1,\Pi_1},
{V}^{\Gamma_1,\Pi_2}$ intersect at a single point. Similarly, the
two branches ${V}^{\Gamma_2,\Pi_1},{V}^{\Gamma_2,\Pi_2}$ intersect
at another point. There is no intersection among other combination
of branches. This matches the number of integrand basis since
$4\times 4-2=14$.

If $K_4$ is massless, {\sl i.e.}, $\beta_{11}=0$, but at most one of
$K_5, K_6$ is absent, then $\widetilde{D}_0=0=y_3 y_4$. There are
two branches parameterized by $y_3=0$ with $y_4$ free parameter or
$y_4=0$ with $y_3$ free parameter. Considering the remaining linear
equation $\widehat{D}_0=0$ of $(y_1,y_3,y_4)$, it is easy to see
there are also four branches $V^{\Gamma_1,\Pi_1},
V^{\Gamma_1,\Pi_2}$ and $V^{\Gamma_2,\Pi_1},V^{\Gamma_2,\Pi_2}$. The
intersection pattern of these four branches is the same as in
previous paragraph\footnote{Besides branch structure of variety, the
integrand basis \eref{B2-basis} need to be modified too. The reason
is that in the case $K_4^2=0$, we have $\W D_0=y_3 y_4$, thus
elements such as $y_3 y_4, y_3^2 y_4, y_3 y_4^2$ could be divided by
$\W D_0$. They should be excluded from integrand basis. The modified
integrand basis is given by \eref{B2-K4-basis}.}.

For specific kinematic configuration where $K_4$ is massless and
both $K_5,K_6$ are absent,  {\sl the dimension of variety will
increase from one to two, and the integrand basis is given by 20
elements as shown in \eref{B21-K4-basis} instead of 14 elements}.
This can be explained by noticing that $y_1$ disappears from the
three equations
\bea D_0 & = & -{\a_{24}\over
\a_{23}}x_4^2+{\a_{23}\a_{24}-\a_{12}\a_{21}+\a_{21}\a_{22}\over
\a_{23}}x_4~, ~ \W D_0= y_3 y_4 ~,~~~\nn \WH D_0 & = & x_4
y_3-{\a_{24}\over
\a_{23}}x_4y_4+{\a_{21}\a_{22}-\a_{21}\a_{12}+\a_{23}\a_{24}\over
\a_{23}}y_4~~~~\eea
in this specific kinematic configuration. The variety is given by
two branches. One branch is parameterized by $x_4=0,y_4=0$ with
$y_1, y_3$ free parameters, and the other branch,  by $x_4=( \a_{21}
\a_{22}-\a_{12} \a_{21} + \a_{23} \a_{24})/\a_{24}$, $y_3=0$ with
$y_1, y_4$ free parameters.  Each branch can detect 10 coefficients
and there is no intersection between them, so adding them up we can
detect all 20 coefficients.

\section{Example two: non-planar crossed double-triangle}

Our second example will be non-planar crossed double-triangle
topology (C22) as shown in Figure \ref{4d-C}. Different from planar
penta-triangle topology (B42), the variety of (C22) is
two-dimensional, so the intersection between different branches
could be one-dimensional variety instead of single points. Topology
(C22) also has symmetry of relabeling $(K_1, K_2, K_3)$ as well as
symmetry of relabeling $(K_4,K_5)$. Discussion of different
kinematic configurations can be simplified by using these
symmetries.
\EPSFIGURE[ht]{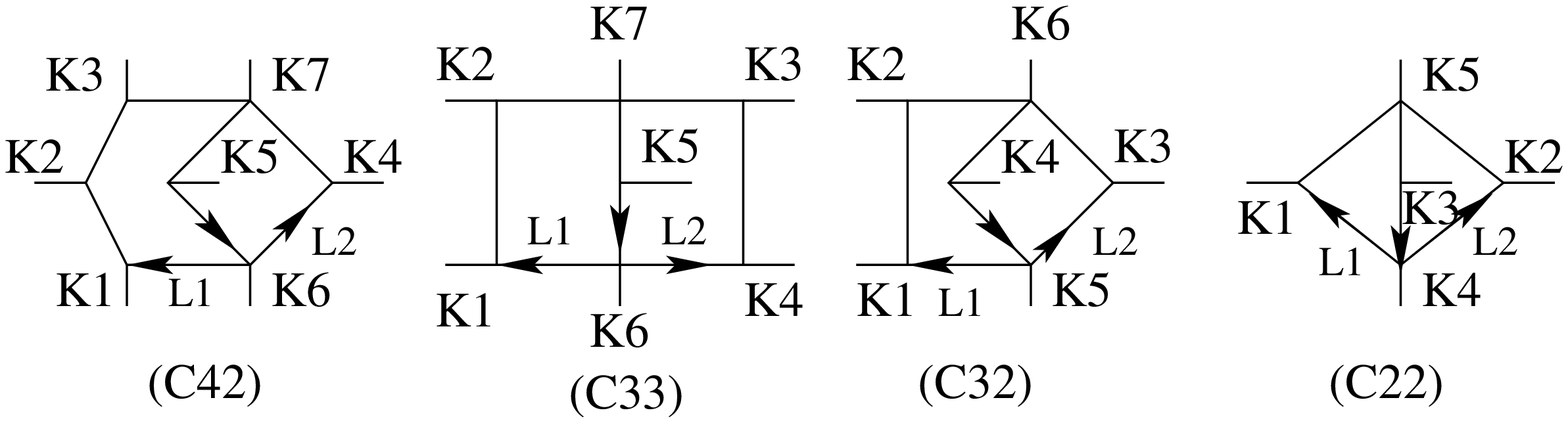,width=12.5cm} {The type (C) contains
all 4 topologies with $n_3=2$. Every topology is denoted by (Cnm)
where $n,m$ are the numbers of propagators containing only $\ell_1$
or $\ell_2$ respectively. The diagrams are drawn in most general
form, thus some external momenta, for instance $K_6, K_7$ in (C42),
could be absent. All external momenta are out-going while convention
of loop momenta is labeled by arrows in each diagram. \label{4d-C} }

For this topology, we use $K_1, K_2$ to construct momentum basis
$e_i$ and use them to expand external and loop momenta as
\bea K_1& = & \a_{11} e_1+ \a_{12} e_2~,~K_2=\b_{11} e_1+ \b_{12}
e_2~,~K_4=\sum_{i=1}^{4} \c_{1i} e_i~,~K_3+K_4=\sum_{i=1}^{4}
\c_{2i} e_i~,~~~\nn
\ell_1 & = & x_2 e_1+ x_1 e_2+ x_4 e_3 + x_3 e_4~,~\ell_2= y_2
e_1+y_1 e_2+ y_4 e_3+ y_3 e_4~.~~~\eea
With this expansion, 6 propagators can be rewritten as functions of
8 variables. The two propagators  containing only $x_i$ variables
are given by
\bea D_0 & = & \ell_1^2=2(x_1 x_2+x_3 x_4)~,~D_1= (\ell_1-K_1)^2
=D_0+ 2\a_{11}\a_{12}-2(\a_{11}x_1+\a_{12} x_2)~.~~~\eea
The two propagators  containing only $y_i$ variables are given by
\bea \W D_0 & = & \ell_2^2= 2(y_1 y_2+ y_3 y_4)~,~\W
D_1=(\ell_2-K_2)^2 =\W D_0 +2\b_{11}\b_{12}-2(\b_{11} y_1+\b_{12}
y_2)~.~~~\eea
The remaining two propagators contain both $x_i,y_i$ variables
\bea \WH D_0 & = & (\ell_1+\ell_2+K_4)^2= D_0+ \W D_0+2 (\c_{11}
\c_{12}+\c_{13}\c_{14})+2 (x_1 y_2+x_2 y_1+x_3 y_4+x_4 y_3) \nn & &
~~~+2(\c_{11}(x_1+y_1)+\c_{12}(x_2+y_2)+\c_{13}(x_3+y_3)
+\c_{14}(x_4 +y_4))~,~~~\nn
\WH D_1 & = & (\ell_1+\ell_2+K_4+K_3)^2= D_0+ \W D_0+2 (\c_{21}
\c_{22}+\c_{23}\c_{24})+2 (x_1 y_2+x_2 y_1+x_3 y_4+x_4 y_3) \nn & &
~~~+2(\c_{21}(x_1+y_1)+\c_{22}(x_2+y_2)+\c_{23}(x_3+y_3)
+\c_{24}(x_4 +y_4))~.~~~\eea
From three linear equations $D_1-D_0=0, \W D_1-\W D_0=0$ and $\WH
D_1-\WH D_0=0$  we can solve $x_1, y_2, x_2$
%
%
%
%
%
%
%
%
as functions of five ISPs $(x_3, x_4, y_1, y_3, y_4)$. Substituting
these solutions back into $D_0, \W D_0, \WH D_0$ we get three
polynomial equations, which define the variety of this topology.

We will consider various kinematic configurations where $K_4,K_5$
could be absent, or some of $K_1,K_2,K_3$ are massless. In order to
make the kinematic configuration clear, we use the notation
$\mbox{C22}_{(U,P)}^{(L,N,R)}$, where each $L,N,R$ could be either
$M$ or $m$ representing massive or massless limit of $K_1,K_3,K_2$
respectively. $U,P$ could be either $K_4,K_5$ if they are non-zero
or $\oslash$ if  they are absent. In this notation, for example,
$\mbox{C22}_{(K_4,\oslash)}^{(M,M,m)}$ represents kinematic
configuration with $K_1,K_3$ massive, $K_2$ massless, $K_4$ non-zero
and $K_5$ absent.

\subsection{The integrand basis}

To determine the integrand basis, we take all possible monomials
$x_3^{d(x_3)} x_4^{d(x_4)} y_1^{d(y_1)} y_3^{d(y_3)} y_4^{d(y_4)}$
under conditions
\bea &&\sum_{\mbox{\tiny{all~ISPs~of~x}}}d(x_i)\leq
4~,~~~\sum_{\mbox{\tiny{all~ISPs~of~y}}}d(y_i)\leq
4~,~~~\sum_{\mbox{\tiny{all~ISPs~of~x}}}d(x_i)+\sum_{\mbox{\tiny{all~ISPs~of~y}}}d(y_i)\leq
5~,~~~\eea
and divide them by Gr\"{o}bner basis generated from $D_0, \W D_0,
\WH D_0$ with ISPs' ordering $ x_3> y_3> x_4>y_4> y_1$ (we will use
the same ordering through this example).
For different kinematic configurations, the number and elements of
integrand basis can be different as demonstrated in previous
example.

After checking all 24 different kinematic configurations, we find
that there are in total 6 different kinds of integrand basis. For
kinematic configurations where at least one of $K_4,K_5$ are
non-zero and $K_1,K_2$ are massive, the integrand basis contains 100
elements given by
\bea &&{\cal B}_{C22}^{I} = \nn && \{1, x_3, x_4, x_3 x_4, x_4^2,
x_3 x_4^2, x_4^3, x_3 x_4^3, x_4^4, y_1, x_3 y_1,
 x_4 y_1, x_3 x_4 y_1, x_4^2 y_1, x_3 x_4^2 y_1, x_4^3 y_1, x_3 x_4^3 y_1,
 x_4^4 y_1, y_1^2,\nn & &  x_3 y_1^2, x_4 y_1^2, x_4^2 y_1^2, x_4^3 y_1^2, y_1^3, x_3 y_1^3,
  x_4 y_1^3, x_4^2 y_1^3, y_1^4, x_3 y_1^4, x_4 y_1^4, y_1^5, y_3, x_3 y_3, x_4 y_3,
 x_4^2 y_3, x_4^3 y_3, x_4^4 y_3, y_1 y_3,\nn & &  x_4 y_1 y_3, x_4^2 y_1 y_3, x_4^3 y_1 y_3,
 y_1^2 y_3, x_4 y_1^2 y_3, y_1^3 y_3, x_4 y_1^3 y_3, y_1^4 y_3, y_3^2, x_3 y_3^2,
 x_4 y_3^2, y_1 y_3^2, x_4 y_1 y_3^2, y_1^2 y_3^2, x_4 y_1^2 y_3^2, \nn & &
 y_1^3 y_3^2, y_3^3, x_3 y_3^3, x_4 y_3^3, y_1 y_3^3, x_4 y_1 y_3^3,
 y_1^2 y_3^3, y_3^4, x_3 y_3^4, x_4 y_3^4, y_1 y_3^4, y_3^5, y_4, x_4 y_4, x_4^2 y_4,
  x_4^3 y_4, x_4^4 y_4, y_1 y_4, \nn & & x_4 y_1 y_4, x_4^2 y_1 y_4, x_4^3 y_1 y_4, y_1^2 y_4,
 x_4 y_1^2 y_4, x_4^2 y_1^2 y_4, y_1^3 y_4, x_4 y_1^3 y_4, y_1^4 y_4, y_4^2,
 x_4 y_4^2, x_4^2 y_4^2, x_4^3 y_4^2, y_1 y_4^2, x_4 y_1 y_4^2, \nn & & x_4^2 y_1 y_4^2,
 y_1^2 y_4^2, x_4 y_1^2 y_4^2, y_1^3 y_4^2, y_4^3, x_4 y_4^3, x_4^2 y_4^3,
 y_1 y_4^3, x_4 y_1 y_4^3, y_1^2 y_4^3, y_4^4, x_4 y_4^4, y_1 y_4^4, y_4^5\}~.~~~\label{C43-Basis}\eea
For kinematic configurations where at least one of $K_4,K_5$ are
non-zero and $K_1$ is massive, $K_2$ is massless, the integrand
basis still contains 100 elements, and is given by replacing one
element from \eref{C43-Basis}
\bea {\cal B}_{C22}^{II}= {\cal B}_{C22}^{I}-\{x_4^2 y_4^3\}+\{x_4^2
y_1^2 y_3\}~.~~~\label{C43K2-Basis}\eea
For kinematic configurations where at least one of $K_4,K_5$ are
non-zero and $K_1$ is massless, the integrand basis contains 98
elements, and is given by removing 17 elements from \eref{C43-Basis}
while adding another 15 elements:
\bea {\cal B}_{C22}^{III}& = & {\cal B}_{C22}^{I}-\{x_3 x_4, x_3
x_4^2, x_3 x_4^3, x_3 x_4 y_1, x_3 x_4^2 y_1, x_3 x_4^3 y_1, x_4^2
y_3,
 x_4^3 y_3, x_4^4 y_3, x_4^2 y_1 y_3, x_4^3 y_1 y_3, y_3^5,\nn & &  x_4^2 y_1 y_4,
 x_4^3 y_1 y_4, x_4^2 y_1^2 y_4, x_4^2 y_1 y_4^2, y_4^5\}+
\{x_3^2, x_3^3, x_3^4, x_3^2 y_1, x_3^3 y_1, x_3^4 y_1, x_3^2 y_1^2, x_3^3 y_1^2,
 x_3^2 y_1^3, x_3^2 y_3, \nn & & x_3^3 y_3, x_3^4 y_3, x_3^2 y_3^2, x_3^3 y_3^2,
 x_3^2 y_3^3\}~.~~~\label{C43K1-Basis}\eea
If both $K_4,K_5$ are absent  and $K_2$ is massive, the integrand
basis contains 96 elements, and is given by removing 22 elements
from \eref{C43-Basis} while adding another 18 elements
\bea {\cal B}_{C22}^{IV} &= & {\cal B}_{C22}^{I}-\{x_3 x_4, x_3
x_4^2, x_3 x_4^3, x_3 x_4 y_1, x_3 x_4^2 y_1, x_3 x_4^3 y_1, x_3
y_1^2,
 x_4^2 y_1^2, x_4^3 y_1^2, x_3 y_1^3, x_4^2 y_1^3, x_3 y_1^4,\nn & & x_4^2 y_3, x_4^3 y_3,
 x_4^4 y_3, x_4^2 y_1 y_3, x_4^3 y_1 y_3, y_1 y_3^4, y_3^5, x_4^2 y_1^2 y_4,
 y_1 y_4^4, y_4^5\}+ \{x_3^2, x_3^3, x_3^4, x_3^2 y_1, x_3^3 y_1,\nn & &  x_3^4 y_1, x_3^2 y_3, x_3^3 y_3,
 x_3^4 y_3, x_3 y_1 y_3, x_3^2 y_1 y_3, x_3^3 y_1 y_3, x_3^2 y_3^2, x_3^3 y_3^2,
 x_3 y_1 y_3^2, x_3^2 y_1 y_3^2, x_3^2 y_3^3, x_3 y_1 y_3^3\}~.~~~\label{C41-Basis}\eea
If both $K_4,K_5$ are absent, $K_2$ is massless, and at least one of
$K_1,K_3$ are massive, the integrand basis contains 96 elements, and
is given by replacing 9 elements from \eref{C41-Basis}
\bea {\cal B}_{C22}^{V}& = & {\cal B}_{C22}^{IV}-\{x_4 y_3^2, x_4
y_1 y_3^2, x_4 y_1^2 y_3^2, x_4 y_3^3, x_4 y_1 y_3^3, x_4 y_3^4,
 x_4 y_1^2 y_4, x_4 y_1^3 y_4, x_4 y_1^2 y_4^2\}\nn & & +\{x_3 y_1^2,
 x_3^2 y_1^2, x_3^3 y_1^2, x_4^2 y_1^2, x_4^3 y_1^2, x_3 y_1^3,
 x_3^2 y_1^3, x_4^2 y_1^3, x_3 y_1^4\}~.~~~\label{C41K2-Basis}\eea
Finally if both $K_4,K_5$ are absent, and all $K_1,K_2,K_3$ are
massless,  the integrand basis contains 144 elements, which is given
by
\bea&& {\cal B}_{C22}^{VI}=  \nn &&\{1, x_1, x_1^2, x_1^3, x_1^4,
x_2, x_1 x_2, x_1^2 x_2, x_1^3 x_2, x_2^2, x_1 x_2^2,
 x_1^2 x_2^2, x_2^3, x_1 x_2^3, x_2^4, y_1, x_1 y_1, x_1^2 y_1, x_1^3 y_1, x_1^4 y_1,\nn & &
  x_2 y_1, x_1 x_2 y_1, x_1^2 x_2 y_1, x_1^3 x_2 y_1, x_2^2 y_1, x_1 x_2^2 y_1,
 x_1^2 x_2^2 y_1, x_2^3 y_1, x_1 x_2^3 y_1, x_2^4 y_1, y_1^2, x_1 y_1^2, x_1^2 y_1^2,
  x_1^3 y_1^2,\nn & &  x_2 y_1^2, x_1 x_2 y_1^2, x_1^2 x_2 y_1^2, x_2^2 y_1^2,
 x_1 x_2^2 y_1^2, x_2^3 y_1^2, y_1^3, x_1 y_1^3, x_1^2 y_1^3, x_2 y_1^3,
 x_1 x_2 y_1^3, x_2^2 y_1^3, y_1^4, x_1 y_1^4, x_2 y_1^4,\nn & &  y_2, x_1 y_2, x_1^2 y_2,
 x_1^3 y_2, x_1^4 y_2, x_2 y_2, x_1 x_2 y_2, x_1^2 x_2 y_2, x_1^3 x_2 y_2, x_2^2 y_2,
 x_1 x_2^2 y_2, x_1^2 x_2^2 y_2, x_2^3 y_2, x_1 x_2^3 y_2,\nn & &  x_2^4 y_2, y_1 y_2,
 x_1 y_1 y_2, x_1^2 y_1 y_2, x_1^3 y_1 y_2, x_2 y_1 y_2, x_1 x_2 y_1 y_2,
 x_1^2 x_2 y_1 y_2, x_2^2 y_1 y_2, x_1 x_2^2 y_1 y_2, x_2^3 y_1 y_2, \nn & & y_1^2 y_2,
 x_1 y_1^2 y_2, x_1^2 y_1^2 y_2, x_2 y_1^2 y_2, x_1 x_2 y_1^2 y_2, x_2^2 y_1^2 y_2,
 y_1^3 y_2, x_1 y_1^3 y_2, x_2 y_1^3 y_2, y_2^2, x_1 y_2^2, x_1^2 y_2^2, x_1^3 y_2^2,\nn & &
  x_2 y_2^2, x_1 x_2 y_2^2, x_1^2 x_2 y_2^2, x_2^2 y_2^2, x_1 x_2^2 y_2^2,
 x_2^3 y_2^2, y_1 y_2^2, x_1 y_1 y_2^2, x_1^2 y_1 y_2^2, x_2 y_1 y_2^2,
 x_1 x_2 y_1 y_2^2, x_2^2 y_1 y_2^2,\nn & &  y_1^2 y_2^2, x_1 y_1^2 y_2^2,
 x_2 y_1^2 y_2^2, y_2^3, x_1 y_2^3, x_1^2 y_2^3, x_2 y_2^3, x_1 x_2 y_2^3,
 x_2^2 y_2^3, y_1 y_2^3, x_1 y_1 y_2^3, x_2 y_1 y_2^3, y_2^4, x_1 y_2^4,
 x_2 y_2^4, \nn & & y_3, x_1 y_3, x_1^2 y_3, x_1^3 y_3, x_1^4 y_3, y_1 y_3, x_1 y_1 y_3,
 x_1^2 y_1 y_3, x_1^3 y_1 y_3, y_1^2 y_3, x_1 y_1^2 y_3, x_1^2 y_1^2 y_3, y_1^3 y_3,
 x_1 y_1^3 y_3,  y_3^2,\nn & & x_1 y_3^2, x_1^2 y_3^2, x_1^3 y_3^2, y_1 y_3^2, x_1 y_1 y_3^2,
  x_1^2 y_1 y_3^2, y_1^2 y_3^2, x_1 y_1^2 y_3^2, y_3^3, x_1 y_3^3, x_1^2 y_3^3,
 y_1 y_3^3, x_1 y_1 y_3^3, y_3^4, x_1 y_3^4\}~.~~~\label{C41K1K2K3-Basis}\eea
%

\subsection{Structure of variety under various kinematic
configurations}

Having given the integrand basis we move to the discussion of
variety determined by six propagators under various kinematic
configurations.

\subsubsection{Kinematic configurations with  $K_4,K_5$ non-zero}

Given the integrand basis, the focus  becomes finding their
coefficients. As mentioned above, the computation can be simplified
 using branch-by-branch method, thus it is important to study the
structure of variety in various kinematic configurations. For
general case where both $K_4,K_5$ are non-zero  and $K_1,K_2,K_3$
are massive, the variety defined by six on-shell  equations is
irreducible, {\sl i.e.}, there is only one branch with dimension
two. All 100 coefficients of integrand basis \eref{C43-Basis} should
be determined at the same time using this irreducible branch.

The variety will split into two branches when one of $K_1,K_2,K_3$
is massless, this corresponds to kinematic configurations
$\mbox{C22}_{(K_4,K_5)}^{(M,M,m)}$,
$\mbox{C22}_{(K_4,K_5)}^{(M,m,M)}$ and
$\mbox{C22}_{(K_4,K_5)}^{(m,M,M)}$. It is easy to see that when
$K_2^2=0$, we have $\b_{11}=0$, thus $\W D_0=y_3 y_4$. Similarly
when $K_1^2=0$, we have $\a_{12}=0$, thus $D_0=x_3 x_4$. For
$K_3^2=0$, we could use the massless condition $ (\gamma_{21} -
\gamma_{11}) (\gamma_{22} - \gamma_{12}) + (\gamma_{23} -
    \gamma_{13}) (\gamma_{24} - \gamma_{14})=0$
to solve $\c_{24}$, and substitute it back to $\W D_0, \WH D_0$ to
solve $y_3, x_4$. After putting solutions of $y_3, x_3$ back to
$D_0$, the numerator of $D_0$ is factorized into two factors, i.e.,
there are two branches.

Above procedure, although  straightforward, could be  complicated
and probably miss some branches in certain kinematic configurations.
An alternative and better way of finding branches of variety is to
use Macaulay2\cite{Macaulay2}.

Let us take kinematic configuration
$\mbox{C22}_{(K_4,K_5)}^{(M,M,m)}$ as an example to illustrate the
structure of these two branches. In this example, one branch is
characterized by $y_3=0$ and the other branch by $y_4=0$. For the
first branch, only 65 elements are left after putting $y_3=0$ to
integrand basis \eref{C43K2-Basis}. Dividing these 65 monomials over
Gr\"{o}bner basis generated from  equations defining this branch, we
find that only 59 of them are independent. So we can only find 59
coefficients of integrand basis \eref{C43K2-Basis}. Similarly, for
the second branch, 66 elements are left after putting $y_4=0$, and
only 59 are independent after dividing them by Gr\"{o}bner basis
generated from definition equations of this branch. Both branches
are varieties of dimension two, and their intersection is an
irreducible variety of dimension one. The one-dimensional
intersection  can detect 18 coefficients, thus we can find all
$59+59-18=100$ coefficients  using both branches.

If two of $K_1,K_2,K_3$ are massless, {\sl i.e.}, kinematic
configurations $\mbox{C22}_{(K_4,K_5)}^{(M,m,m)}$,
$\mbox{C22}_{(K_4,K_5)}^{(m,m,M)}$ and
$\mbox{C22}_{(K_4,K_5)}^{(m,M,m)}$, the variety will further split
into 4 branches. Take kinematic configuration
$\mbox{C22}_{(K_4,K_5)}^{(m,M,m)}$ as an example, massless
conditions of $K_1,K_2$ will reduce $D_0= x_3 x_4$ and $\W D_0 = y_3
y_4$. It is easy to see that there are 4 branches characterized by
$V_1:(x_3=0,y_3=0)$, $V_2:(x_3=0,y_4=0)$, $V_3:(x_4=0,y_3=0)$ and
$V_4:(x_4=0,y_4=0)$. Using algebraic or other methods, one can find
that each branch can detect $34$ coefficients. A naive summation of
these 4 branches gives $34\times 4=136$ coefficients, which is
larger than the number of integrand basis. This means that there are
intersections among 4 branches. By analyzing intersections among all
possible combinations of branches, we find that intersections for
pairs $(V_1, V_2)$, $(V_1, V_3)$, $(V_4,V_3)$, $(V_4, V_2)$ are
irreducible one-dimensional varieties\footnote{Each one-dimensional
intersection can detect $10$ coefficients for this example. With
information of other intersections, we can make following counting.
Since intersection of three or four branches detects $2$
coefficients, each intersection of two branches will detect $10-2=8$
independent coefficients, thus each branch will independently detect
$34-8-8-2=16$ coefficients that can not be detected by other
branches. Adding all together we have $16\times 4+8\times 4+2=98$
coefficients as it should be.}, and intersections for  pairs $(V_1,
V_4)$ and $(V_2, V_3)$ are isolated points. Intersections of three
or four branches are again above two isolated points.

If we assume kinematic configuration to be
$\mbox{C22}_{(K_4,K_5)}^{(m,m,m)}$ where all $K_1,K_2,K_3$ are
massless, the variety is given by eight branches, {\sl i.e.}, each
branch of previous paragraph has further split into two branches.
The first two branches $V_1,V_2$ characterized by $x_3=y_3=0$ (or
the seventh and eighth branches $V_7,V_8$ characterized by
$x_4=y_4=0$) can detect 19 and 21 coefficients respectively, and 34
coefficients can be detected by using two branches. This can be
checked by noticing that the intersection of these two branches can
detect 6 coefficients, so $19+21-6=34$. Similarly, each of the third
and fourth branches $V_3,V_4$ characterized by $x_3=y_4=0$ (or the
fifth and sixth branches $V_5,V_6$ characterized by $x_4=y_3=0$) can
detect $20$ coefficients, and 34 coefficients can be detected by
using two branches. This can also be checked by noticing that their
intersection can detect $6$ coefficients, so $20+20-6=34$. We also
need to clarify the intersection pattern among eight branches. There
are no intersections shared by five or more branches. The
intersections of following six pairs $(V_1, V_2,V_3, V_4)$,
$(V_5,V_6,V_7,V_8)$, $(V_1, V_2,V_5, V_6)$, $(V_3,V_4,V_7,V_8)$,
$(V_1, V_3,V_6, V_8)$, $(V_2,V_4,V_5,V_7)$ are single points.
Intersections of every three branches are also single points, which
are inherited from corresponding intersection of every four branches
(for example, intersection point of $(V_1, V_2, V_3)$ coming from
intersection point of $(V_1, V_2, V_3, V_4)$). No new intersecting
points besides the ones of every four branches are found for
intersections of every three branches. The intersections of every
two branches are possibly one-dimensional varieties or single
points. In order to express the intersection pattern, we will use
following notation $V_1\cap V_2= (d|m)$ where $d$ is the dimension
of variety (so $d=1$ for one-dimension and $d=0$ for points) and $m$
is the number of coefficients detected by the intersection. Thus all
possible intersections between pairs $(V_i,V_j)$ are given by
\bean (1|6) & = & V_1\cap V_2 =V_2\cap V_4= V_2\cap V_6=V_3\cap
V_4=V_3\cap V_8=V_5\cap V_6=V_6\cap V_8=V_7\cap V_8~,~~~\nn
(1|5) & = & V_1\cap V_3 =V_1\cap V_6=V_4\cap V_7=V_5\cap V_7~,~~~\nn
(0|1) & = & V_1\cap V_4=V_1\cap V_5=V_1\cap V_8=V_2\cap V_3=V_2\cap
V_7= V_3\cap V_6=V_3\cap V_7=V_4\cap V_5\nn & =&V_4\cap V_8=V_5\cap
V_8=V_6\cap V_7~.~~~ \eean
%

\subsubsection{Kinematic configurations with one of $K_4,K_5$ absent}

For $\mbox{C22}_{K_4,\oslash}$ or $\mbox{C22}_{K_5,\oslash}$, {\sl
i.e.}, one of $K_4, K_5$ absent,  the variety is given by two
branches\footnote{This can be seen by solving $y_3,x_3$ using $\W
D_0=0$ and  $\WH D_0=0$ equations and putting solutions back to
$D_0$, which is factorized to two pieces. One can also use Macaulay2
to find branches. From now on, we will not discuss how to get
branches.} even without imposing massless conditions of $K_i,
i=1,2,3$. Each branch can detect 64 coefficients of integrand basis
and their one-dimension intersection can detect 28 coefficients. By
using two branches all $64+64-28=100$ coefficients can be detected.

For kinematic configurations $\mbox{C22}_{K_4/K_5,\oslash}^{m,M,M}$,
$\mbox{C22}_{K_4/K_5,\oslash}^{M,m,M}$ and
$\mbox{C22}_{K_4/K_5,\oslash}^{M,M,m}$, the variety is given by four
branches. To illustrate the structure of  branches, let us take
$\mbox{C22}_{K_4/K_5,\oslash}^{M,M,m}$ as an example. Each branch is
2-dimensional variety and can detect 21 coefficients. Let us use
$V_1, V_2$ to denote two branches characterized by $y_3=0$, and
$V_3, V_4$ to denote two branches characterized by $y_4=0$. We find
that these four branches will intersect at a single point. Among
intersections of every three branches, non-trivial two intersecting
points exist for pair $(V_1, V_2, V_4)$ and $(V_2, V_3, V_4)$. Pair
$(V_1, V_3)$ intersects at a point, while intersections of all other
five pairs of every two branches are one-dimension varieties. Among
them $V_1\cap V_2, V_3\cap V_4$ can detect 11 coefficients while
$V_1\cap V_4, V_2\cap V_3$ can detect 6 coefficients and $V_2\cap
V_4$, $10$ coefficients. It is also worth to mention that though
having same four branches, the intersection pattern of
$\mbox{C22}_{K_4/K_5,\oslash}^{m,M,M}$,
$\mbox{C22}_{K_4/K_5,\oslash}^{M,m,M}$ and
$\mbox{C22}_{K_4/K_5,\oslash}^{M,M,m}$ are different from these of
$\mbox{C22}_{K_4,K_5}^{m,m,M}$,$\mbox{C22}_{K_4,K_5}^{m,M,m}$, and
$\mbox{C22}_{K_4,K_5}^{M,m,m}$.

Next let us discuss kinematic configurations
$\mbox{C22}_{K_4/K_5,\oslash}^{m,m,M}$,
$\mbox{C22}_{K_4/K_5,\oslash}^{m,M,m}$ and
$\mbox{C22}_{K_4/K_5,\oslash}^{M,m,m}$. For these cases, the variety
is given by six branches. Taking
$\mbox{C22}_{K_4/K_5,\oslash}^{m,M,m}$ as an example, the first two
branches $V_1, V_2$ characterized by $x_3=y_3=0$ can detect 19 and
21 coefficients respectively, and the intersection of these two
branches can detect 6 coefficients, thus we have 34 coefficients by
using both branches. The third branch $V_3$ characterized by
$x_3=y_4=0$ can detect 34 coefficients. Similarly, the fourth branch
$V_4$ characterized by $x_4=y_3=0$ can also detect 34 coefficients.
The last two branches $V_5, V_6$ characterized by $x_4=y_4=0$ can
detect 19 and 21 coefficients respectively, and by using both
branches one can detect 34 coefficients. It is interesting to notice
that these six branches are split from corresponding 4 branches of
$\mbox{C22}_{K_4,K_5}^{m,M,m}$. We will again clarify the
intersection pattern of these six branches. No intersections exist
for every five or six branches. For intersections of every four
branches, pair $(V_1, V_3, V_4, V_5)$ and $(V_2,V_3, V_4, V_6)$
intersect at single points. Apart from the inherit intersecting
points of four branches, there are also pairs of every three
branches $(V_1, V_2, V_3)$, $(V_4, V_5, V_6)$, $(V_1, V_2, V_4)$,
$(V_3, V_5, V_6)$ that intersect at different single points. For
intersections of every two branches, $(V_1, V_5)$, $(V_2, V_6)$
intersect at one single points,  $(V_3, V_4)$ intersects at two
points, and $(V_{3,4}, V_{2,6})$, $(V_1, V_2)$, $(V_5,V_6)$
intersect at one-dimensional variety which can detect 6
coefficients, while $(V_{3,4}, V_{1,5})$ also intersect at
one-dimensional variety which can detect 5 coefficients.

For kinematic configuration $\mbox{C22}_{K_4/K_5,\oslash}^{m,m,m}$
where all $K_1,K_2,K_3$ are massless, the variety  splits to eight
branches. The branch structure is the same as
$\mbox{C22}_{K_4,K_5}^{m,m,m}$. Two branches $V_1,V_2$ characterized
by $x_3=y_3=0$ as well as two branches $V_7,V_8$ characterized by
$x_4=y_4=0$ can detect 19 and 21 coefficients respectively, while
two branches $V_3,V_4$ characterized by $x_3=y_4=0$ and two branches
$V_5,V_6$ characterized by $x_4=y_3=0$ can detect 20 coefficients
respectively. These eight branches intersect at a single point,
while all intersections among every seven, six, five, four or three
branches are also located at the same point. There are 28 possible
intersecting pairs of  two branches, among them 12 are
one-dimensional varieties, and intersections of the remaining 16
pairs are the same single point as the intersection of eight
branches. For the $12$ one-dimensional variety, 8 of them coming
from $(V_{2}, V_{1,4,5})$, $(V_8, V_{3,6,7})$, $(V_3, V_4)$ and
$(V_5, V_6)$ can detect 6 coefficients individually, while the other
four  coming from $(V_1, V_{3,6})$ and $(V_7, V_{4,5})$ can detect 5
coefficients.

\subsubsection{Kinematic configurations with both $K_4,K_5$ absent}

For kinematic configuration $\mbox{C22}_{\oslash,\oslash}$, since
$K_4=K_5=0$, momentum conservation ensures $K_3=-K_1-K_2$. $K_1,
K_2$ are still independent, so we can use them to construct momentum
basis $e_i$. For this simple case, we can write down analytic
expressions and make discussion more transparent.

Using parametrization $K_1=\a_{11} e_1+\a_{12}e_2$ and
$K_2=\b_{11}e_1+\b_{12} e_2$, the three non-linear cut equations can
be given by
\bea D_0 & = & x_3 x_4 + {\a_{11}\a_{12}\over
\beta_{12}}(1-{y_1\over \beta_{12}})y_1~,~~~ \nn
\W D_0 & = & y_3 y_4 + \b_{11} (1 - {y_1\over \b_{12}})y_1 ~,~~~ \nn
\WH D_0 & = & x_4 y_3 + x_3 y_4 +{\a_{12}\b_{11}+\a_{11}\b_{12}\over
\b_{12}}(1 - {y_1\over \b_{12}})y_1~~~\label{C41-CutEqn} \eea
after eliminating all RSPs. If $K_1,K_2,K_3$ are massive, the
variety is given by following six branches defined by ideals:
\bea V^{C22^{(M,M,M)}_{(\oslash,\oslash)}}_1&  = & \{y_3, x_3,
y_1\}~,~V^{C22^{(M,M,M)}_{(\oslash,\oslash)}}_2 = \{y_3 ,
  x_3, y_1 - \b_{12}  \}~,~~~\nn
V^{C22^{(M,M,M)}_{(\oslash,\oslash)}}_3 &=&  \{y_4, y_1,
x_4\}~,~V^{C22^{(M,M,M)}_{(\oslash,\oslash)}}_4  = \{y_4,
  y_1 - \b_{12},
  x_4\}~,~~~\nn
V^{C22^{(M,M,M)}_{(\oslash,\oslash)}}_5& =&  \{y_3 y_4 +
    \b_{11} (1 - y_1/\b_{12})y_1, y_3 \a_{12} -
    x_3 \b_{12},  y_4 \a_{11} - x_4 \b_{11}\}~,~~~\nn
V^{C22^{(M,M,M)}_{(\oslash,\oslash)}}_6 & = & \{y_3 y_4 +
    \b_{11} (1 - y_1/\b_{12})y_1, -y_3 \a_{11} +
    x_3 \b_{11}, y_4 \a_{12} - x_4 \b_{12}\}~.~~~\label{C41-Branch}\eea
Among these six branches, four of them $V_i,i=1,2,3,4$ will
 detect 19 coefficients individually and two of them $V_i,i=5,6$, 36
coefficients. The physical picture is following. Each branch of
$\mbox{C22}^{(M,M,M)}_{(K_4/K_5,\oslash)}$ will split into three
branches with two branches detecting 19 coefficients and one branch
detecting 36 coefficients. The intersection pattern of six branches
is following. No intersections exist for six or every five branches.
Each combination of $(V_2, V_4, V_5, V_6)$ and $(V_1, V_3, V_5,
V_6)$ intersects at a single point. No new intersection points exist
for intersections of three branches. For intersections of $15$ pairs
$(V_i,V_j)$, there are no intersections among $4$ pairs $(V_{1,3},
V_{2,4})$, while $(V_1, V_3)$ intersects at one point, and $(V_5,
V_6)$ intersects at two points. Intersections of remaining $9$ pairs
are one-dimensional variety.

If one or two momenta of $K_1,K_2,K_3$ are massless, {\sl i.e.},
kinematic configurations $\mbox{C22}^{(m,M,M)}_{(\oslash,\oslash)}$,
$\mbox{C22}^{(M,m,M)}_{(\oslash,\oslash)}$ ,
$\mbox{C22}^{(M,M,m)}_{(\oslash,\oslash)}$,
$\mbox{C22}^{(m,m,M)}_{(\oslash,\oslash)}$,
$\mbox{C22}^{(m,M,m)}_{(\oslash,\oslash)}$ and
$\mbox{C22}^{(M,m,n)}_{(\oslash,\oslash)}$, the variety still has
six branches. Definition of branches are still the same as
\eref{C41-Branch} for $\mbox{C22}^{(m,M,M)}_{(\oslash,\oslash)}$,
$\mbox{C22}^{(M,m,M)}_{(\oslash,\oslash)}$,
$\mbox{C22}^{(m,m,M)}_{(\oslash,\oslash)}$, but will be different
for $\mbox{C22}^{(M,M,m)}_{(\oslash,\oslash)}$,
$\mbox{C22}^{(m,M,m)}_{(\oslash,\oslash)}$,
$\mbox{C22}^{(M,m,m)}_{(\oslash,\oslash)}$ where the first four
branches are the same as  \eref{C41-Branch}, and the last two
branches change to
\bea V_5^{\mbox{C22}^{(M,M,m),(m,M,m),(M,m,m)}_{(\oslash,\oslash)}}
& = & \{y_3 , -y_4 \a_{12} + x_4 \b_{12},
 x_3 y_4 +  \a_{11} (1-y_1/\b_{12})y_1\}~,~~~\nn
 V_6^{\mbox{C22}^{(M,M,m),(m,M,m),(M,m,m)}_{(\oslash,\oslash)}} & = &
 \{y_4, x_4 y_3 + \a_{11} (1-y_1/\b_{12})y_1 , -y_3 \a_{12} + x_3 \b_{12}\}~.~~~\label{C41-K2-Branch}\eea

For the last kinematic configuration
$\mbox{C22}^{(m,m,m)}_{(\oslash,\oslash)}$, the external momenta are
extremely degenerated since we must either have $\la_1\sim \la_2\sim
\la_3$ or $\W \la_1\sim \W\la_2\sim \W\la_3$. In other words, we can
not use $K_1, K_2$ to construct momentum basis $e_i$. One possible
choice of momentum basis is the massless momenta $K_1, K_2$, $e_3,
e_4$ satisfying\footnote{We can always have this choice. For
example, if $K_1=\la_1\W\la_1$, $K_2=\la_1\W\la_2$ we can take
$e_3=c_3\la_2\W\la_1$ and $e_4=c_4\la_2\W\la_2$. Similarly if
$K_1=\la_1\W\la_1$, $K_2=\la_2 \W\la_1$, we can take
$e_3=c_3\la_1\W\la_2$, $e_4=c_4\la_2\W\la_2$.}
\bea K_1\cdot K_2=K_1\cdot e_3=K_2\cdot e_4 =e_3\cdot e_4=0~,~
K_1\cdot e_4  =  K_2\cdot e_3=1~.~~~ \eea
With this momentum basis we can expand loop momentum $\ell_1$ as
\bea \ell_1 & = & (\ell_1\cdot e_4)  K_1+ (\ell_1\cdot e_3)K_2+
(\ell_1\cdot K_2) e_3 + (\ell_1\cdot K_1)e_4\equiv x_1 K_1+x_2
K_2+x_3 e_3+ x_4 e_4~,~~~\eea
and similarly for $\ell_2$. Then the six propagators are given by
\bea D_0 & = & \ell_1^2= 2 (x_1 x_4+x_2 x_3)~,~D_1=(\ell_1-K_1)^2
=D_0- 2 x_4~,~~~\nn
\W D_0 & = & \ell_2^2=2(y_1 y_4+y_2 y_3)~,~\W D_1=(\ell_2-K_2)^2= \W
D_0-2 y_4~,~~~\nn
\WH D_0 & = & (\ell_1+\ell_2)^2=
2(x_1+y_1)(x_4+y_4)+2(x_2+y_2)(x_3+y_3)~,~~~\nn \WH D_1& = &
(\ell_1+\ell_2+K_3)^2= \WH D_0-2 (x_4+y_4)-2(x_3+y_3)~.~~~\eea
Solving these equations we find
\bea x_4=0~,~y_4=0~,~x_3=-y_3~,~~~\eea
and there are only two non-linear equations left
\bea D_0= x_2 y_3~,~\W D_0=y_2 y_3~.~~~\eea
Now we have five ISPs $(x_1, x_2, y_1, y_2, y_3)$ and two non-linear
equations. The integrand basis is  given by $144$ monomials of ISPs
 under degree conditions as already shown.
The variety is given by two branches. The first one characterized by
$y_3=0$ is dimension four variety, and the second one characterized
by $x_2=y_2=0$ is dimension three. The first branch can detect 114
coefficients while the second branch can detect 49 coefficients.
Their intersection is two-dimensional variety, which can detect 19
coefficients.

The splitting of branches of different kinematic configurations is
summarized in Figure \ref{C4-branch}.
\EPSFIGURE[ht]{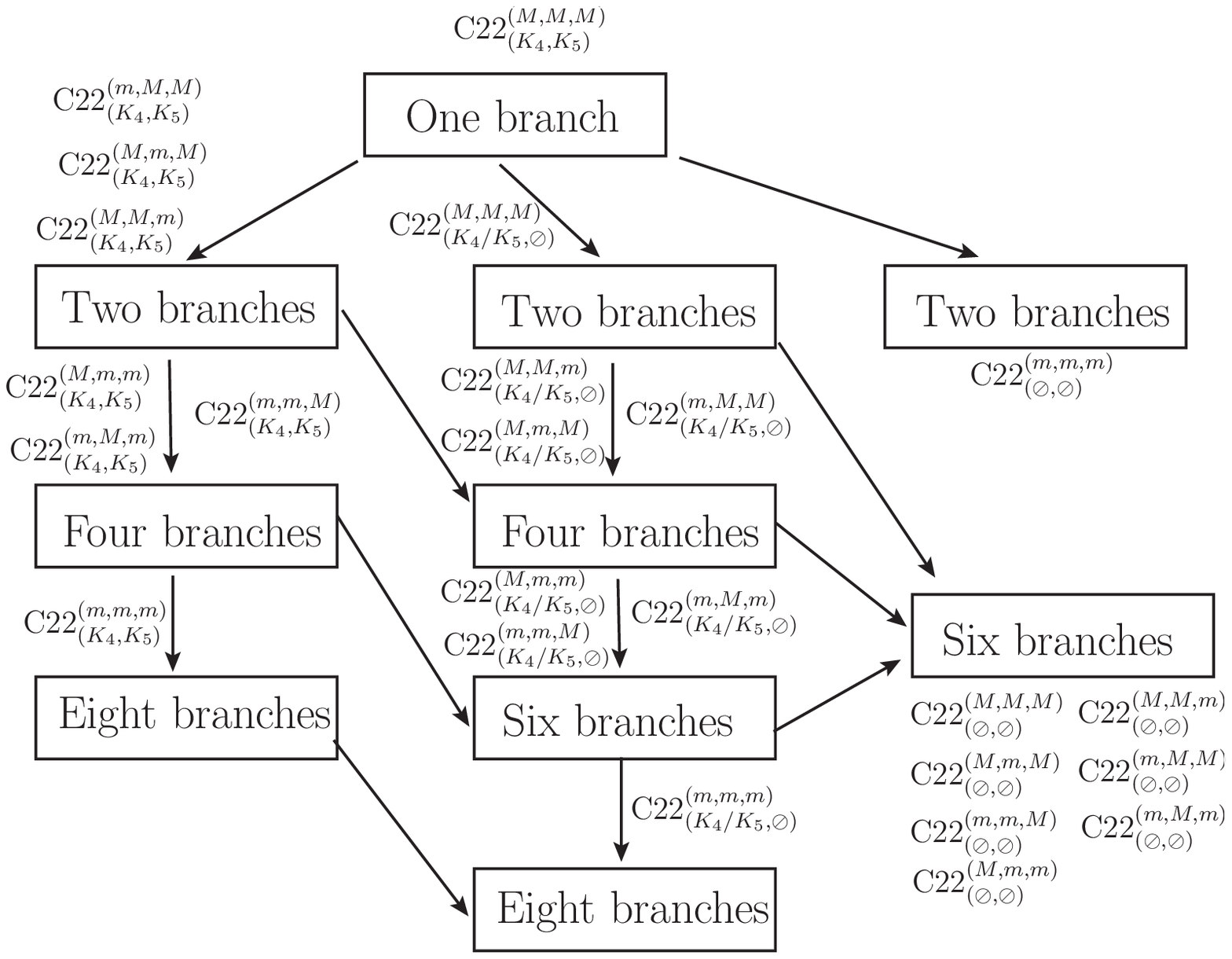,width=16cm} {The splitting of variety
into branches under different kinematic configurations. All branches
are dimension two varieties, except the most degenerated case
$\mbox{C22}^{(m,m,m)}_{(\oslash,\oslash)}$ where one branch is
dimension four and the other, dimension three. The arrows indicate
how  branches split when one more specific kinematic condition is
imposed.  \label{C4-branch} }
%

\section{Remaining two-loop topologies}

After demonstrating methods and various properties with above planar
penta-triangle and non-planar crossed double-triangle examples, we
will present results for the remaining two-loop topologies in this
section. We will omit many details but  show only main results.

\subsection{The topology (C32): non-planar crossed box-triangle}

There is only one topology left for type (C), i.e., the crossed
box-triangle topology (C32). We use $K_1, K_3$ to construct momentum
basis $e_i$. From seven on-shell equations, we can solve, for
instance, $x_1,x_2,y_1,y_2$ as linear functions of four ISPs $(x_3,
x_4, y_3, y_4)$. The remaining three propagators are  quadratic
functions of ISPs. For general kinematic configuration, the
expression and solution of cut equations are tedious, so we will not
explicitly write them down here.

{\bf Integrand basis:} In general, the variety defined by these
three remaining quadratic cut equations is irreducible and dimension
one.  Using Gr\"{o}bner basis method under ISPs ordering
$(y_4,y_3,x_4,x_3)$ and the renormalization conditions
\bea &&\sum_{\mbox{\tiny{all~ISPs~of~x}}}d(x_i)\leq
5~,~~~\sum_{\mbox{\tiny{all~ISPs~of~y}}}d(y_i)\leq
4~,~~~\sum_{\mbox{\tiny{all~ISPs~of~x}}}d(x_i)+\sum_{\mbox{\tiny{all~ISPs~of~y}}}d(y_i)\leq
6~,~~~\eea
we can get integrand basis for various kinematic configurations.
There are all together four kinds of integrand basis depending on
the massless limits of $K_1,K_3$ since we have chosen $K_1,K_3$ to
generate momentum basis. For all kinematic configurations with
$K_1,K_3$ massive, the integrand basis contains 38 elements given by
\bea   {\cal B}_{C32}^{I} &=&  \{1, x_3, x_3^2, x_3^3, x_3^4, x_3^5,
x_3^6, x_4, x_3 x_4, x_3^2 x_4, x_3^3 x_4,
 x_3^4 x_4, x_3^5 x_4, y_3, x_3 y_3, x_3^2 y_3, x_3^3 y_3,\nn & & x_3^4 y_3, x_3^5 y_3,
 x_4 y_3,  x_3 x_4 y_3, x_3^2 x_4 y_3, x_3^3 x_4 y_3, x_3^4 x_4 y_3, y_3^2, x_3 y_3^2,
 x_4 y_3^2, y_3^3, x_3 y_3^3, x_4 y_3^3, \nn && y_3^4, x_3 y_3^4, x_4 y_3^4, y_4, x_3 y_4,
 y_3 y_4, y_3^2 y_4, y_3^3 y_4\}~.~~~\label{C33-basis}\eea
For all kinematic configurations with $K_1$ massless while $K_3$
massive, the integrand basis still contains 38 elements, and is
given by replacing 9 elements in ${\cal B}_{C32}^{I}$
\bea {\cal B}_{C32}^{II}&= &{\cal B}_{C32}^{I}-\{x_3 x_4, x_3^2 x_4,
x_3^3 x_4, x_3^4 x_4, x_3^5 x_4, x_3 x_4 y_3, x_3^2 x_4 y_3
 x_3^3 x_4 y_3, x_3^4 x_4 y_3\} \nn & & + \{x_4^2, x_4^3, x_4^4, x_4^5, x_4^6, x_4^2 y_3, x_4^3 y_3, x_4^4 y_3, x_4^5 y_3\}~.~~~\label{C3K1-basis}
\eea
For all kinematic configurations with $K_1$ massive while $K_3$
massless, the integrand basis contains 38 elements, and is given by
replacing 6 elements in ${\cal B}_{C32}^{I}$
\bea {\cal B}_{C32}^{III}&= &{\cal B}_{C32}^{I}-\{x_4 y_3^2, x_4
y_3^3, x_4 y_3^4, y_3 y_4, y_3^2 y_4, y_3^3 y_4\}  + \{y_4^2, x_3
y_4^2, y_4^3, x_3 y_4^3, y_4^4, x_3 y_4^4\}~.~~~\label{C3K3-basis}
\eea
Finally for all kinematic configurations with both $K_1,K_3$
massless, the 38 elements of integrand basis are given by replacing
fifteen elements in ${\cal B}_{C32}^{I}$
\bea {\cal B}_{C32}^{IV}&= &{\cal B}_{C32}^{I}-\{x_3 x_4, x_3^2 x_4,
x_3^3 x_4, x_3^4 x_4, x_3^5 x_4, x_3 x_4 y_3, x_3^2 x_4 y_3, x_3^3
x_4 y_3, x_3^4 x_4 y_3,\nn & & x_4 y_3^2, x_4 y_3^3, x_4 y_3^4, y_3
y_4, y_3^2 y_4, y_3^3 y_4\}  + \{x_4^2, x_4^3, x_4^4, x_4^5, x_4^6,
x_4^2 y_3, x_4^3 y_3, x_4^4 y_3,\nn && x_4^5 y_3,  y_4^2, x_3 y_4^2,
y_4^3, x_3 y_4^3, y_4^4, x_3 y_4^4\}~.~~~\label{C3K1K3-basis} \eea

To discuss the structure of variety, we again use the notation
$\mbox{C32}_{(U,P)}^{(L,N,R)}$ where now $U,P$ could either be
$K_5,K_6$ or $\oslash$ representing corresponding $K_5,K_6$ absent.
$L$ will be $m$ if at least one momentum of $K_1,K_2$ is massless,
and $R$ will be $m$ if $K_3$ is massless, while $N$ will be $m$ if
$K_4$ is massless. Otherwise they will be  $M$.

The number of branches under various kinematic configurations is
summarized in table \ref{C3-n-table}.
\begin{table}[h]
  \centering
\begin{tabular}{|c|c|c|c|}
  \hline
    \backslashbox{$(L,N,R)$}{$(U,P)$} & $(K_5,K_6)$ & $(K_5/K_6,\oslash)$ & $(\oslash,\oslash)$  \\ \hline
     $(M,M,M)$ &  1  & 2  & 4 \\ \hline
  $(m,M,M), (M,m,M), (M,M,m)$ & 2 & 4 & 6 \\ \hline
$(m,m,M), (m,M,m), (M,m,m)$ & 4 &  6 & 6 for $(M,m,m)$/8  \\  \hline
 $(m,m,m)$ & 8 & 8 & 8 \\
  \hline
\end{tabular}
  \caption{Number of branches of various kinematic
  configurations for non-planar crossed box-triangle topology. The kinematic configurations are denoted
  by $\mbox{C32}_{(U,P)}^{(L,N,R)}$.}\label{C3-n-table}
\end{table}
For each kinematic configuration, one should use all branches to
find all 38 coefficients of integrand basis. We can also use
branch-by-branch polynomial fitting method to simplify calculations.

\EPSFIGURE[ht]{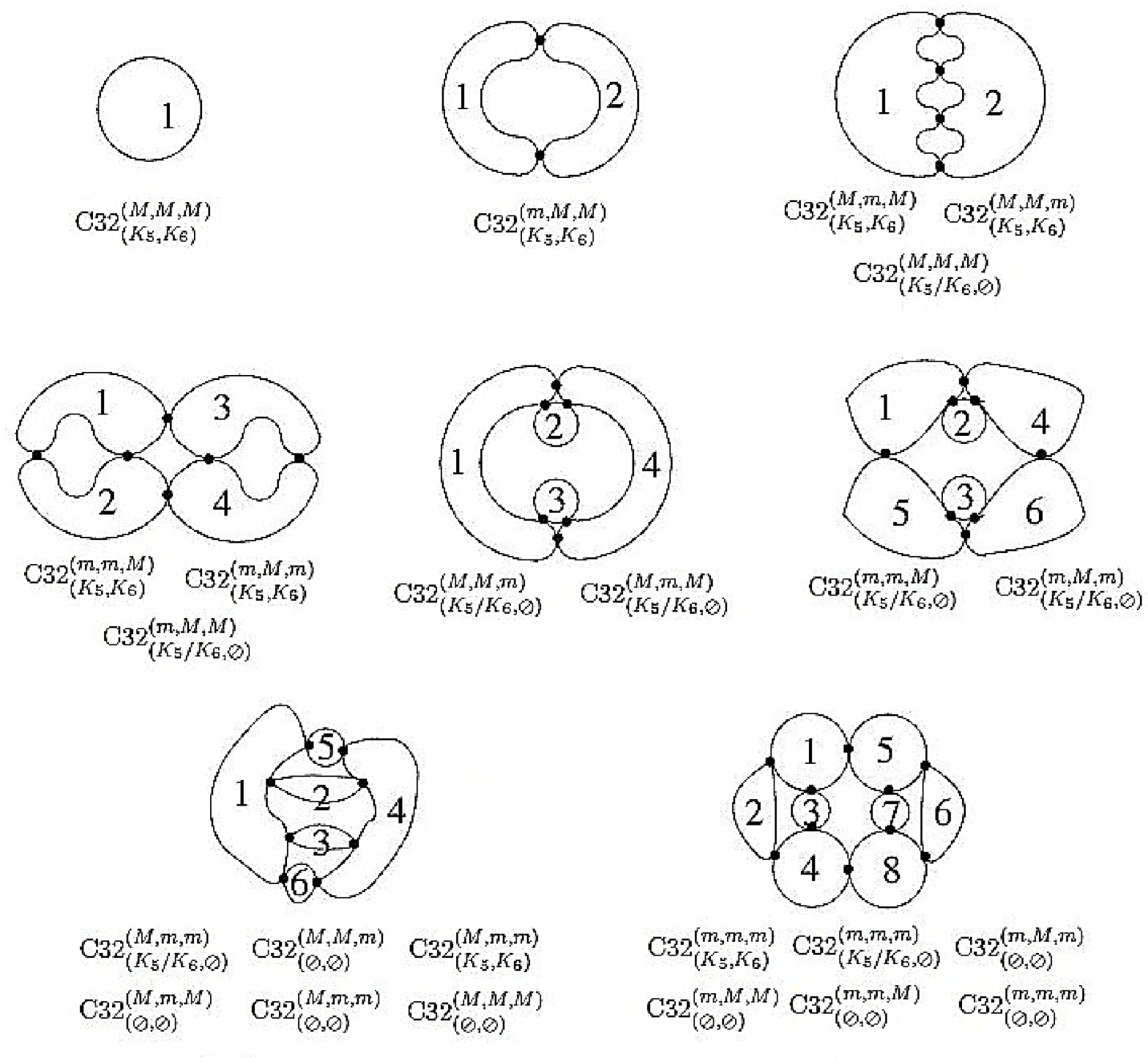,width=16cm} {Intersections of branches
for various kinematic configurations of non-planar crossed
box-triangle topology (C32). Each branch $V_i$ is represented by a
closed loop and denoted by $i$, while black dot is the intersecting
point. Kinematic configurations for each pattern are listed below
each diagram. \label{C3-inte} }

{\bf Variety with one branch: } For the most general kinematics
$\mbox{C32}_{(K_5,K_6)}^{(M,M,M)}$, the variety is irreducible with
dimension one. All 38 coefficients should be found  using this
branch.

{\bf Variety with two branches: } For kinematic configurations
\bea
\mbox{C32}_{(K_5,K_6)}^{(M,M,m)}~,~\mbox{C32}_{(K_5,K_6)}^{(M,m,M)}~,~
\mbox{C32}_{(K_5,K_6)}^{(m,M,M)}~,~\mbox{C32}_{(K_5/K_6,\oslash)}^{(M,M,M)}~,~~~\eea
the variety is given by two branches with dimension one. These
branches will intersect at points. More explicitly, for
$\mbox{C32}_{(K_5,K_6)}^{(m,M,M)}$, two branches intersect at two
isolated points, while for $\mbox{C32}_{(K_5,K_6)}^{(M,M,m)}$,
$\mbox{C32}_{(K_5,K_6)}^{(M,m,M)}$ and
$\mbox{C32}_{(K_5/K_6,\oslash)}^{(M,M,M)}$, two branches intersect
at four points.

{\bf Variety with four branches: } For kinematic configurations
\bea &&~\mbox{C32}_{(K_5,K_6)}^{(m,m,M)}~,~
\mbox{C32}_{(K_5,K_6)}^{(m,M,m)}~,~\mbox{C32}_{(K_5/K_6,\oslash)}^{(M,M,m)}~,~
\mbox{C32}_{(K_5/K_6,\oslash)}^{(M,m,M)}~,~\mbox{C32}_{(K_5/K_6,\oslash)}^{(m,M,M)}~,~
~~~\eea
the variety is given by four branches with dimension one. The
intersection pattern among four branches $V_1,V_2,V_3,V_4$ can be
shown as follows. For $\mbox{C32}_{(K_5,K_6)}^{(m,m,M)}$,
$\mbox{C32}_{(K_5,K_6)}^{(m,M,m)}$ and
$\mbox{C32}_{(K_5/K_6,\oslash)}^{(m,M,M)}$, the only non-zero
intersections are given by $V_1\cap V_2=V_3\cap V_4=(0|2)$, $V_1\cap
V_3=V_2\cap V_4=(0|1)$. For
$\mbox{C32}_{(K_5/K_6,\oslash)}^{(M,M,m)}$ and
$\mbox{C32}_{(K_5/K_6,\oslash)}^{(M,m,M)}$, non-zero intersections
are given by $V_1\cap V_4=(0|2)$, $V_1\cap V_{2}=V_2\cap V_4=(0|1)$,
$V_1\cap V_3=V_3\cap V_4=(0|1)$.

{\bf Variety with six branches: } For kinematic configurations
\bea
&&\mbox{C32}_{(K_5,K_6)}^{(M,m,m)}~,~\mbox{C32}_{(\oslash,\oslash)}^{(M,M,M)}~,~
\mbox{C32}_{(K_5/K_6,\oslash)}^{(m,M,m)}~,~\mbox{C32}_{(K_5/K_6,\oslash)}^{(m,m,M)}~,~
\nn & &
\mbox{C32}_{(K_5/K_6,\oslash)}^{(M,m,m)}~,~\mbox{C32}_{(\oslash,\oslash)}^{(M,M,m)}~,
~\mbox{C32}_{(\oslash,\oslash)}^{(M,m,M)}~,~
\mbox{C32}_{(\oslash,\oslash)}^{(M,m,m)}~,~~~\eea
the variety is given by six branches with dimension one. These
branches again intersect at points. For
$\mbox{C32}_{(K_5/K_6,\oslash)}^{(m,M,m)}$ and
$\mbox{C32}_{(K_5/K_6,\oslash)}^{(m,m,M)}$, each pair of $(V_1,
V_2)$, $(V_2, V_4)$, $(V_4, V_1)$, $(V_5, V_6)$, $(V_6, V_3)$,
$(V_3, V_5)$, $(V_1, V_5)$, $(V_4, V_6)$ intersects at one single
point. For $\mbox{C32}_{(K_5,K_6)}^{(M,m,m)}$,
$\mbox{C32}_{(\oslash,\oslash)}^{(M,M,M)}$,
$\mbox{C32}_{(K_5/K_6,\oslash)}^{(M,m,m)}$,
$\mbox{C32}_{(\oslash,\oslash)}^{(M,M,m)}$,
$\mbox{C32}_{(\oslash,\oslash)}^{(M,m,M)}$ and
$\mbox{C32}_{(\oslash,\oslash)}^{(M,m,m)}$, each pair of $(V_1,
V_i)$ and $(V_4, V_i)$ for $i=2,3,5,6$ intersects at one single
point.

{\bf Variety with eight branches: } For kinematic configurations
\bea
\mbox{C32}_{(K_5,K_6)}^{(m,m,m)}~,~\mbox{C32}_{(K_5/K_6,\oslash)}^{(m,m,m)}~,~
\mbox{C32}_{(\oslash,\oslash)}^{(m,M,m)}~,~\mbox{C32}_{(\oslash,\oslash)}^{(m,m,M)}~,~
\mbox{C32}_{(\oslash,\oslash)}^{(m,m,m)}~,~\mbox{C32}_{(\oslash,\oslash)}^{(m,M,M)}~,~~~~\eea
the variety is given by eight branches with dimension one. There
will be single intersecting point for each pair of following ten
combinations: $(V_1, V_2)$, $(V_1, V_3)$, $(V_1, V_5)$, $(V_2,
V_4)$, $(V_3, V_4)$, $(V_4, V_8)$, $(V_5, V_6)$, $(V_5, V_7)$,
$(V_6, V_8)$ and $(V_7, V_8)$.

The intersection pattern of branches for each kinematic
configurations is shown in Figure \eref{C3-inte}.


\subsection{The topology (B41): planar penta-bubble}

 From on-shell  equations of six propagators we can
get three linear equations for pure $\ell_1$, and reduce four RSPs
$(x_1,x_2,x_3,x_4)$ to one. Exception happens when $K_4=K_5=0$,
$D_4=(\ell_1-K_1-K_2-K_3)^2=\ell_1^2$ from momentum conservation,
and the independent linear equations containing pure $\ell_1$ reduce
to two. In this case we get two ISPs from $x_i$. There is no linear
equation for pure $\ell_2$, so all four $y_i$ are ISPs. Adding them
together there will be 5 ISPs (or 6 ISPs for the case $K_4=K_5=0$).

We use $K_1,K_3$ to construct momentum basis $e_i$. After solving
linear equations we can express  remaining three quadratic equations
with ISPs. Using Gr\"{o}bner basis method with ISPs' ordering
$(x_4,y_1,y_2,y_3,y_4)$ for  kinematic configurations
$\mbox{B41}_{(K_4,K_5)}$, $\mbox{B41}_{(K_4/K_5,\oslash)}$ under
renormalization conditions
\bea &&\sum_{\mbox{\tiny{all~ISPs~of~x}}}d(x_i)\leq
5~,~~~\sum_{\mbox{\tiny{all~ISPs~of~y}}}d(y_i)\leq
2~,~~~\sum_{\mbox{\tiny{all~ISPs~of~x}}}d(x_i)+\sum_{\mbox{\tiny{all~ISPs~of~y}}}d(y_i)\leq
5~.~~\eea
we can get integrand basis with $18$ elements. We have three kinds
of integrand basis for kinematic configurations
$\mbox{B41}_{(K_4,K_5)}$, $\mbox{B41}_{(K_4/K_5,\oslash)}$ according
to kinematics of $K_1,K_3$ since we have chose $K_1, K_3$ to
generate momentum basis.  The first kind is suitable for all
kinematic configurations of $\mbox{B41}_{(K_4,K_5)}$, or $K_3$
massive while others arbitrary for $\mbox{B41}_{(K_4,\oslash)}$, or
$K_1$ massive while others arbitrary for
$\mbox{B41}_{(K_5,\oslash)}$. It is given by  18 elements
\bea \mathcal{B}_{B41}^{I}=\{1, x_4, y_1, y_2, x_4y_2, y_2^2, y_3,
y_1y_3, y_2y_3, y_3^2, y_4, x_4y_4, y_1y_4, y_2y_4, x_4y_2y_4,
y_3y_4, y_4^2, x_4y_4^2\}~.~~~\eea
The second kind is suitable for kinematic configurations with $K_1$
massless while others arbitrary for $\mbox{B41}_{(K_5,\oslash)}$.
The 18 elements of integrand basis are given by replacing one
element in $\mathcal{B}_{B41}^{I}$
\bea
\mathcal{B}_{B41}^{II}=\mathcal{B}_{B4}^{I}-\{x_4y_2y_4\}+\{x_4y_2^2\}~.~~~\eea
The third kind is suitable for kinematic configurations with $K_3$
massless while others arbitrary for $\mbox{B41}_{(K_4,\oslash)}$.
The 18 elements of integrand basis  are given by replacing three
elements in $\mathcal{B}_{B41}^{I}$
\bea
\mathcal{B}_{B41}^{III}=\mathcal{B}_{B41}^{I}-\{x_4y_2,y_2^2,x_4y_2y_4\}+\{x_4y_1,y_1^2,x_4y_1^2\}~.~~~\eea
For all kinematic configurations of $\mbox{B41}_{(\oslash,\oslash)}$
where ISPs are given by six variables with ordering
$(x_3,x_4,y_1,y_2,y_3,y_4)$, we  get 83 elements for integrand basis
\bea \mathcal{B}_{B41}^{IV}&=&\{1, x_3, x_3^2, x_3^3, x_3^4, x_3^5,
x_4, x_4^2, x_4^3, x_4^4, x_4^5, y_1,x_3y_1, x_3^2y_1, x_3^3y_1,
x_3^4y_1, x_4y_1, x_4^2y_1, x_4^3y_1, x_4^4y_1,\nonumber\\
&& y_1^2, x_3y_1^2, x_3^2y_1^2, x_3^3y_1^2, x_4y_1^2, x_4^2y_1^2,
x_4^3y_1^2, y_2, x_3y_2, x_3^2y_2, x_3^3y_2, x_3^4y_2, x_4y_2,
x_4^2y_2, x_4^3y_2, x_4^4y_2,\nonumber\\
&& y_2^2, x_3y_2^2, x_3^2y_2^2, x_3^3y_2^2, x_4y_2^2, x_4^2y_2^2,
x_4^3y_2^2, y_3, x_3y_3, x_3^2y_3, x_3^3y_3, x_3^4y_3, x_4y_3,
y_1y_3, x_3y_1y_3,\nonumber\\
&& x_3^2y_1y_3, x_3^3y_1y_3, x_4y_1y_3, y_2y_3, x_3y_2y_3,
x_3^2y_2y_3, x_3^3y_2y_3, x_4y_2y_3, y_3^2, x_3y_3^2, x_3^2y_3^2,
x_3^3y_3^2, x_4y_3^2,\nonumber\\
&& y_4, x_4y_4, x_4^2y_4, x_4^3y_4, x_4^4y_4, y_1y_4, x_4y_1y_4,
x_4^2y_1y_4, x_4^3y_1y_4, y_2y_4, x_4y_2y_4, x_4^2y_2y_4,
x_4^3y_2y_4, \nonumber\\
&&y_3y_4, x_4y_3y_4, y_4^2, x_4y_4^2, x_4^2y_4^2,
x_4^3y_4^2\}~.~~~\eea
The reason we have 83 elements instead of 18 is that, for
$\mbox{B41}_{(K_4,K_5)}$ and $\mbox{B41}_{(K_4/K_5,\oslash)}$, $x_4$
is determined by quadratic equation, {\sl i.e.}, the maximal power
of $x_4$ is two, while for $\mbox{B41}_{(\oslash,\oslash)}$ we can
have $d(x_3)+d(x_4)\leq 5$.

In order to simplify the calculations of coefficients, we need to
discuss the branch structure of variety. For
$\mbox{B41}_{(K_4,K_5)}$ and $\mbox{B41}_{(K_4/K_5,\oslash)}$, there
is a quadratic equation of single variable $x_4$, and we can always
get two solutions of $x_4$ in $\mathbb{C}$-plane no matter what the
momentum configuration of $K_1,K_2,K_3$ is. Thus there will always
be two separate branches characterized by two solutions
$x_4^{\Gamma_1}, x_4^{\Gamma_2}$. For $\mbox{B41}_{(K_4,K_5)}$,
these two branches with dimension two will not split further. Using
each branch we can detect 9 coefficients of integrand basis, and
since there is no intersection between two branches, we can detect
all $9+9=18$ coefficients  using both branches. For
$\mbox{B41}_{(K_4/K_5,\oslash)}$, each branch will split further
into two branches, so there will be in total four branches:
$V_1,V_2$ characterized by $x_4^{\Gamma_1}$ and $V_3,V_4$
characterized by $x_4^{\Gamma_2}$. Each branch can detect 6
coefficients. The two branches characterized by $x_4^{\Gamma_i}$
will intersect at one-dimensional variety with intersection pattern
$(1|3)=V_1\cap V_2$ and $(1|3)=V_3\cap V_4$. So  using two branches
of each $x_4^{\Gamma_i}$ we can detect $6+6-3=9$ coefficients, and
in total $9+9=18$ coefficients  using all 4 branches.

For kinematic configurations $\mbox{B41}_{(\oslash,\oslash)}$, $x_3,
x_4$ are both ISPs, so the quadratic equation of $(x_3,x_4)$ could
not be factorized  into two separate pieces in general. If all
$K_1,K_2,K_3$ are massive, the variety is given by two branches.
Each branch is 3-dimensional, and the intersection of these two
branches is 2-dimensional. Each branch can detect 58 coefficients,
while the intersection of them is $(2|33)=V_1\cap V_2$. If at least
one momentum of $K_1,K_3$ is massive, the variety will split into
four 3-dimensional branches $V_1,V_2$ and $V_3, V_4$. Using $V_1$ or
$V_3$ we can detect 28 coefficients of integrand basis, while using
$V_2$ or $V_4$ we can detect 36 coefficients. The intersection of
these four branches is 1-dimensional, and it can detect 3
coefficients. Intersections of every three branches are also the
same 1-dimensional variety as the one given by intersection of four
branches. For intersections of every two branches, $(V_1,V_3)$ and
$(V_2,V_4)$ are inherited from the intersection of four branches,
which is 1-dimensional variety. The intersections of $(V_1,V_4)$ and
$(V_2,V_3)$ are 2-dimensional. Their intersection pattern is
$(2|6)=V_1\cap V_4$, $(2|6)=V_2\cap V_3$. The intersections of
$(V_1,V_2)$ and $(V_3, V_4)$ are also 2-dimensional, from which  18
coefficients can be detected  using each intersection.

For the special kinematic configuration of
$\mbox{B41}_{(\oslash,\oslash)}$ with both $K_1,K_3$  massless, the
three quadratic on-shell  equations reduce to
\bea x_3x_4=0~,~~~y_1y_2+y_3y_4=0~,~~~x_3y_4+x_4y_3=0~.~~~\eea
Besides the ordinary four branches
\bea &&V_1:~~~ x_3=0~,~y_2=0~,~y_3=0~,~x_4~,~y_1~,~y_4~,~\mbox{free
parameters}~,~~~\nonumber\\
&&V_2:~~~ x_3=0~,~y_1=0~,~y_3=0~,~x_4~,~y_2~,~y_4~,~\mbox{free
parameters}~,~~~\nonumber\\
&&V_3:~~~ x_4=0~,~y_2=0~,~y_4=0~,~x_3~,~y_1~,~y_3~,~\mbox{free
parameters}~,~~~\nonumber\\
&&V_4:~~~ x_4=0~,~y_1=0~,~y_4=0~,~x_3~,~y_2~,~y_3~,~\mbox{free
parameters}~,~~~\eea
there is also another embedded branch given by the ideal
\bea V_5:~~~\{x_4y_3+x_3y_4, y_1y_2+y_3y_4, x_4^2, x_3x_4,
x_3^2\}~.~~~\eea
Each of the ordinary branches can detect 28 coefficients, while
$V_5$ can detect 37 coefficients. These five branches intersect at
one single point, and intersections of every four branches are also
the same point. For intersections of every three branches, we have
$(1|6)=(V_1,V_2,V_5)$, $(1|6)=(V_3,V_4,V_5)$, $(1|3)=(V_1,V_3,V_5)$,
$(1|3)=(V_2,V_4,V_5)$, and they are all  different 1-dimensional
varieties. The other intersections of every three branches are
inherited from the same point of intersection of five branches. For
intersections of every two branches, $(V_1,V_4)$, $(V_2,V_3)$ are
 at the same point of intersection of five branches, while
$(V_1,V_3)$, $(V_2,V_4)$ are the same 1-dimensional varieties of
intersections of $(V_1,V_3,V_5)$ and $(V_2,V_4,V_5)$ respectively.
Intersections of other combinations of pairs are 2-dimensional and
we have $(2|12)=V_1\cap V_5$, $(2|12)=V_2\cap V_5$, $(2|12)=V_3\cap
V_5$, $(2|12)=V_4\cap V_5$, $(2|15)=V_1\cap V_2$, $(2|15)=V_3\cap
V_4$. They are all  different 2-dimensional varieties.

\subsection{The topology (B33): planar double-box}

This topology has been discussed in details in many other
papers\cite{Gluza:2010ws, Kosower:2011ty,Larsen:2012sx,
CaronHuot:2012ab }, here we will briefly summarize some results. We
use $K_1, K_4$ to construct momentum basis $e_i, i=1,2,3,4$ and all
kinematics can be expanded by this basis. The seven on-shell
equations can be reduced to three quadratic equations with four
variables after solving four linear equations. Since there are two
linear equations for $x_i$ variables and two for $y_i$, by solving
them we can get 4 ISPs $(x_2,x_4,y_1,y_4)$ for instance. Then
$D_0=0$ becomes a conic section of $(x_2,x_4)$, $\widetilde{D}_0=0$
becomes a conic section of $(y_1,y_4)$ and $\widehat{D}_0=0$ is a
quadratic equation of $(x_2,x_4,y_1,y_4)$. Variety defined by these
three quadratic equations will be reducible if any of
$K_1,K_2,K_3,K_4$ is massless, or any of $K_5,K_6$ is zero.

The renormalization conditions
\bea &&\sum_{\mbox{\tiny{all~ISPs~of~x}}}d(x_i)\leq
4~,~\sum_{\mbox{\tiny{all~ISPs~of~y}}}d(y_i)\leq
4~,~\sum_{\mbox{\tiny{all~ISPs~of~x}}}d(x_i)+\sum_{\mbox{\tiny{all~ISPs~of~y}}}d(y_i)\leq
6~~~~\eea
constrain all possible monomials
$x_2^{d(x_2)}x_4^{d(x_4)}y_1^{d(y_1)}y_4^{d(y_4)}$. We can get 32
elements for integrand basis after dividing them by Gr\"{o}bner
basis generated from three quadratic equations with ISPs' ordering
$(x_2,x_4,y_1,y_4)$. Integrand basis for different kinematic
configurations can be arranged to four kinds according to the
kinematics of $K_1,K_4$, which we have chosen to generate momentum
basis $e_i$. If $K_1,K_4$ are massive, integrand basis is given by
following $32$ elements
\bea \mathcal{B}_{B33}^{I}&=&\{1, x_2, x_4, x_2x_4, x_4^2, x_2x_4^2,
x_4^3, x_2x_4^3, x_4^4, y_1, x_4y_1, x_4^2y_1, x_4^3y_1, x_4^4y_1,
y_4, x_2y_4, x_4y_4, x_4^2y_4,\nonumber\\
&& x_4^3y_4, x_4^4y_4, y_1y_4, x_4y_1y_4, y_4^2, x_4y_4^2, y_1y_4^2,
x_4y_1y_4^2, y_4^3, x_4y_4^3, y_1y_4^3, x_4y_1y_4^3, y_4^4,
x_4y_4^4\}~.~~~\eea
If $K_1$ is massless while $K_4$ is massive, integrand basis is
given by replacing 6 elements in $\mathcal{B}_{B33}^{I}$ as follows
\bea \mathcal{B}_{B33}^{II}=\mathcal{B}_{B33}^{I}-\{x_2x_4,
x_2x_4^2, x_2x_4^3, x_4^2y_1, x_4^3y_1, x_4^4y_1\}+\{x_2^2, x_2^3,
x_2^4, x_2^2y_4, x_2^3y_4, x_2^4y_4\}~.~~~\eea
If $K_4$ is massless while $K_1$ is massive, integrand basis is
given by replacing 6 elements in $\mathcal{B}_{B33}^{I}$ as follows
\bea \mathcal{B}_{B33}^{III}=\mathcal{B}_{B33}^{I}-\{y_1y_4,
x_4y_1y_4, y_1y_4^2, x_4y_1y_4^2, y_1y_4^3, x_4y_1y_4^3\}+\{y_1^2,
x_4y_1^2, y_1^3, x_4y_1^3, y_1^4, x_4y_1^4\}~.~~~\eea
Finally if both $K_1,K_4$ are massless, integrand basis is given by
replacing 12 elements in $\mathcal{B}_{B33}^{I}$, which are exactly
6 elements from the second kind of integrand basis plus the other 6
elements from the third kind of integrand basis
\bea \mathcal{B}_{B33}^{IV}&=&\{1, x_2, x_2^2, x_2^3, x_2^4, x_4,
x_4^2, x_4^3, x_4^4, y_1, x_4y_1, y_1^2, x_4y_1^2, y_1^3, x_4y_1^3,
y_1^4, x_4y_1^4, y_4,\nonumber\\
&& x_2y_4, x_2^2y_4, x_2^3y_4, x_2^4y_4, x_4y_4, x_4^2y_4, x_4^3y_4,
x_4^4y_4, y_4^2, x_4y_4^2, y_4^3, x_4y_4^3, y_4^4,
x_4y_4^4\}~.~~~\eea

We use notation $\mbox{B33}_{(U,P)}^{(L,R)}$ to denote different
kinematic configurations, where again $U,P$ could be either
$K_5,K_6$ or $\oslash$, and $L$ is denoted by $m$ if at least one
momentum of $K_1,K_2$ (or $K_3,K_4$ for $R$) is massless, otherwise
it will be denoted by $M$.

For general kinematic configuration
$\mbox{B33}_{(K_5,K_6)}^{(M,M)}$, the variety is irreducible with
dimension one. For kinematic configurations
$\mbox{B33}_{(K_5,K_6)}^{(m,M)}$, $\mbox{B33}_{(K_5,K_6)}^{(M,m)}$,
$\mbox{B33}_{(K_5/K_6,\oslash)}^{(M,M)}$ and
$\mbox{B33}_{(\oslash,\oslash)}^{(M,M)}$, the variety splits into
two branches. Each branch can detect 17 coefficients of integrand
basis, and two branches intersect at two points, which exactly gives
$17+17-2=32$ coefficients when using both two branches.

For $\mbox{B33}_{(K_5,K_6)}^{(m,m)}$, the variety is given by four
branches. Each branch can detect 9 coefficients. There is no
intersection for four branches or every three branches, while each
of following pairs $(V_1, V_2)$, $(V_2, V_3)$, $(V_3, V_4)$ and
$(V_4,V_1)$ intersects at a single point. Thus  when combining them
together we can find $9\times 4-4=32$ coefficients. For kinematic
configurations $\mbox{B33}_{(K_5/K_6,\oslash)}^{(m,M)}$ ,
$\mbox{B33}_{(K_5/K_6,\oslash)}^{(M,m)}$,
$\mbox{B33}_{(\oslash,\oslash)}^{(m,M)}$ and
$\mbox{B33}_{(\oslash,\oslash)}^{(M,m)}$, the variety is also given
by four branches. Branches $V_1,V_3$ can detect 5 coefficients
individually while branches $V_2,V_4$ can  detect 13 coefficients
individually. The non-zero intersections among branches are still
single points between following pairs $(V_1, V_2)$, $(V_2, V_3)$,
$(V_3, V_4)$, $(V_4, V_1)$.

For kinematic configurations
$\mbox{B33}_{(K_5/K_6,\oslash)}^{(m,m)}$ and
$\mbox{B33}_{(\oslash,\oslash)}^{(m,m)}$, the variety is given by
six branches. Among these six branches, $V_1, V_4$ can detect $9$
coefficients while $V_2, V_3, V_5, V_6$ can detect 5 coefficients.
Non-zero intersections exist only for following pairs  $(V_1, V_2)$,
$(V_2, V_3)$, $(V_3, V_4)$, $(V_4, V_5)$, $(V_5, V_6)$ and $(V_6,
V_1)$, and each intersection is a single point. So using all six
branches we can detect $2\times 9+4\times 5-6=32$ coefficients.

Results presented here are consistent with those found in
\cite{Gluza:2010ws, Kosower:2011ty,Larsen:2012sx, CaronHuot:2012ab
}. In our discussion, the variety will be reducible for kinematic
configurations that any of $K_1,K_2,K_3$ $K_4$ is massless, or any
of $K_5,K_6$ is zero. These configurations correspond to the
existence of three-vertex $\oplus$ or $\ominus$. The distribution of
$\oplus$ and $\ominus$ will generate different kinematical solutions
to the heptacut constraints, which, in our language, are irreducible
branches of the variety after primary decomposition. Each
irreducible branch can be seen as a Riemann sphere, and the
intersecting points between two branches is precisely the poles of
heptacut Jacobian. According to these mapping, we can reconstruct
the global structures of double-box topology shown in the references
from irreducible branches and their intersections. The 32 elements
of integrand basis are sufficient to expand double-box amplitude at
the integrand-level, yet they are still redundant after loop
integration. Only after eliminating the redundancy using IBP method
for instance can we get integral basis shown in \cite{Gluza:2010ws}.

\subsection{The topology (B32): planar box-triangle}

For this topology, we can get two linear equations for
$(x_1,x_2,x_3,x_4)$ and one linear equation for $(y_1,y_2,y_3,y_4)$,
and reduce 8 RSPs to 5 ISPs $(x_3,x_4,y_1,y_3,y_4)$. We use
$K_1,K_3$ to construct momentum basis $e_i$. Under the
renormalization conditions
\bea &&\sum_{\mbox{\tiny{all~ISPs~of~x}}}d(x_i)\leq
4~,~~~\sum_{\mbox{\tiny{all~ISPs~of~y}}}d(y_i)\leq
3~,~~~\sum_{\mbox{\tiny{all~ISPs~of~x}}}d(x_i)+\sum_{\mbox{\tiny{all~ISPs~of~y}}}d(y_i)\leq
5~~~\eea
we can get integrand basis using the Gr\"{o}bner basis with ordering
$(x_3,x_4,y_1,y_3,y_4)$. For $\mbox{B32}_{(K_4,K_5)}$ and
$\mbox{B32}_{(K_4/K_5,\oslash)}$, we  get 69 elements for integrand
basis, but these elements may be different. The difference can be
classified by the kinematics of $K_1,K_3$, and there are in total 4
kinds of integrand basis. The first kind is for all configurations
with $K_1,K_3$ massive, and the 69 elements are given by
\bea \mathcal{B}_{B32}^{I}&=&\{1, x_3, x_4, x_3x_4, x_4^2, x_3x_4^2,
x_4^3, x_3x_4^3, x_4^4, y_1, x_4y_1, x_4^2y_1, x_4^3y_1, x_4^4y_1,
y_3, x_3y_3, x_4y_3, x_4^2y_3,\nonumber\\\
&& x_4^3y_3, x_4^4y_3, y_1y_3, x_4y_1y_3, y_3^2, x_3y_3^2, x_4y_3^2,
x_4^2y_3^2, x_4^3y_3^2, y_1y_3^2, x_4y_1y_3^2, y_3^3, x_3y_3^3,
x_4y_3^3, x_4^2y_3^3, \nonumber\\
&&y_4, x_3y_4, x_4y_4, x_3x_4y_4, x_4^2y_4, x_3x_4^2y_4, x_4^3y_4,
x_3x_4^3y_4, x_4^4y_4, y_1y_4, x_4y_1y_4, x_4^2y_1y_4,
x_4^3y_1y_4,\nonumber\\
&& y_3y_4, x_4y_3y_4, x_4^2y_3y_4, x_4^3y_3y_4, y_1y_3y_4,
x_4y_1y_3y_4, y_3^2y_4, x_4y_3^2y_4, x_4^2y_3^2y_4, y_4^2, x_3y_4^2,
x_4y_4^2,\nonumber\\
&& x_4^2y_4^2, x_4^3y_4^2, y_1y_4^2, x_4y_1y_4^2, y_3y_4^2,
x_4y_3y_4^2, x_4^2y_3y_4^2, y_4^3,
 x_3y_4^3, x_4y_4^3, x_4^2y_4^3\}~.~~~\eea
The second kind is for configurations with $K_1$ massless while
$K_3$ massive, and the 69 elements are given by replacing 15
elements in $\mathcal{B}_{B32}^{I}$
\bea \mathcal{B}_{B32}^{II}&=&\mathcal{B}_{B32}^{I}-\{x_3x_4,
x_3x_4^2, x_3x_4^3, x_4^2y_1, x_4^3y_1, x_4^4y_1, x_4^2y_3^2,
x_4^3y_3^2,
x_4^2y_3^3, x_3x_4y_4, x_3x_4^2y_4,\nonumber\\
&& x_3x_4^3y_4,
 x_4^2y_1y_4, x_4^3y_1y_4, x_4^2y_3^2y_4\}+\{x_3^2, x_3^3, x_3^4, x_3^2y_3,
x_3^3y_3, x_3^4y_3, x_3^2y_3^2,
x_3^3y_3^2, \nonumber\\
&&x_3^2y_3^3, x_3^2y_4, x_3^3y_4, x_3^4y_4, x_3^2y_4^2,
 x_3^3y_4^2, x_3^2y_4^3\}~.~~~\eea
The third kind is for configurations with $K_3$ massless while $K_1$
massive, and the 69 elements are given by replacing 12 elements in
$\mathcal{B}_{B32}^{I}$
\bea \mathcal{B}_{B32}^{III}&=&\mathcal{B}_{B32}^{I}-\{y_3y_4,
x_4y_3y_4, x_4^2y_3y_4, x_4^3y_3y_4, y_1y_3y_4, x_4y_1y_3y_4,
y_3^2y_4,
x_4y_3^2y_4, x_4^2y_3^2y_4,\nonumber\\
&& y_3y_4^2, x_4y_3y_4^2,
 x_4^2y_3y_4^2\}+\{y_1^2, x_4y_1^2, y_1^3, x_4y_1^3, x_4^2y_1y_3, x_4^3y_1y_3,
y_1^2y_3, x_4y_1^2y_3, \nonumber\\
&&x_4^2y_1y_3^2, y_1^2y_4, x_4y_1^2y_4,
 x_4^2y_1y_4^2\}~.~~~\eea
The last kind of integrand basis is for configurations with
$K_1,K_3$ massless, and the 69 elements are given by replacing 23
elements in $\mathcal{B}_{B32}^{I}$
\bea \mathcal{B}_{B32}^{IV}&=&\mathcal{B}_{B32}^{I}-\{x_3x_4,
x_3x_4^2, x_3x_4^3, x_4^2y_1, x_4^3y_1, x_4^4y_1, x_3x_4y_4,
x_3x_4^2y_4,
x_3x_4^3y_4, x_4^2y_1y_4, x_4^3y_1y_4, \nonumber\\
&&y_3y_4, x_4y_3y_4, x_4^2y_3y_4, x_4^3y_3y_4, y_1y_3y_4,
x_4y_1y_3y_4, y_3^2y_4, x_4y_3^2y_4, x_4^2y_3^2y_4, y_3y_4^2,
x_4y_3y_4^2,\nonumber\\
&& x_4^2y_3y_4^2\}+ \{x_3^2, x_3^3, x_3^4, y_1^2, x_4y_1^2, y_1^3,
 x_4y_1^3, x_3^2y_3, x_3^3y_3, x_3^4y_3, y_1^2y_3, x_4y_1^2y_3, x_3^2y_3^2,
 x_3^3y_3^2,\nonumber\\
&& x_3^2y_3^3, x_3^2y_4, x_3^3y_4, x_3^4y_4, y_1^2y_4, x_4y_1^2y_4,
x_3^2y_4^2, x_3^3y_4^2, x_3^2y_4^3\}~.~~~\eea

The integrand basis for $\mbox{B32}_{(\oslash,\oslash)}$ can be
distinguished by kinematics of $K_3$. If $K_3$ is massive, we still
get 69 elements, while if $K_3$ is massless, we can get 77 elements
instead of 69. The number of elements changes because in the
specific momentum configuration, the sub-triangle-loop is
0m-triangle, so variety is 3-dimensional, while in other momentum
configurations the variety is 2-dimensional. The 69 elements for
$K_3$ massive case are given by
\bea \mathcal{B}_{B32}^{V}&=&\{1, x_3, x_3^2, x_3^3, x_3^4, x_4,
x_4^2, x_4^3, x_4^4, y_1, x_3y_1, x_3^2y_1, x_3^3y_1, x_3^4y_1,
x_4y_1,
x_4^2y_1, x_4^3y_1, x_4^4y_1, y_3, \nonumber\\
&&x_3y_3, x_3^2y_3, x_3^3y_3, x_3^4y_3, x_4y_3, y_1y_3, x_3y_1y_3,
x_3^2y_1y_3, x_3^3y_1y_3, x_4y_1y_3, y_3^2, x_3y_3^2, x_3^2y_3^2,
x_3^3y_3^2,\nonumber\\
&& x_4y_3^2, y_1y_3^2, x_3y_1y_3^2, x_3^2y_1y_3^2, x_4y_1y_3^2,
y_3^3, x_3y_3^3, x_3^2y_3^3, x_4y_3^3, y_4, x_4y_4, x_4^2y_4,
x_4^3y_4, x_4^4y_4,\nonumber\\
&& y_1y_4, x_4y_1y_4, x_4^2y_1y_4, x_4^3y_1y_4, y_3y_4, x_4y_3y_4,
y_1y_3y_4, x_4y_1y_3y_4, y_3^2y_4, x_4y_3^2y_4, y_4^2, x_4y_4^2,\nonumber\\
&& x_4^2y_4^2, x_4^3y_4^2, y_1y_4^2, x_4y_1y_4^2,x_4^2y_1y_4^2,
y_3y_4^2, x_4y_3y_4^2, y_4^3, x_4y_4^3, x_4^2y_4^3\}~,~~~\eea
which is different from the previous four kinds of
$\mbox{B32}_{(K_4,K_5)}$ and $\mbox{B32}_{(K_4/K_5,\oslash)}$. For
configurations with $K_3$ massless, the 77 elements are given by
\bea \mathcal{B}_{B32}^{VI}&=&\{1, x_3, x_3^2, x_3^3, x_3^4, x_4,
x_4^2, x_4^3, x_4^4, y_1, x_3y_1, x_3^2y_1, x_3^3y_1, x_3^4y_1,
x_4y_1, x_4^2y_1, x_4^3y_1,\nonumber\\
&& x_4^4y_1, y_1^2, x_3y_1^2, x_3^2y_1^2, x_3^3y_1^2, x_4y_1^2,
x_4^2y_1^2, x_4^3y_1^2, y_1^3, x_3y_1^3, x_3^2y_1^3, x_4y_1^3,
x_4^2y_1^3, y_3,\nonumber\\
&& x_3y_3, x_3^2y_3, x_3^3y_3, x_3^4y_3, x_4y_3, y_1y_3, x_3y_1y_3,
x_3^2y_1y_3, x_3^3y_1y_3, x_4y_1y_3, y_1^2y_3,
x_3y_1^2y_3,\nonumber\\
&& x_3^2y_1^2y_3, x_4y_1^2y_3, y_3^2, x_3y_3^2, x_3^2y_3^2,
x_3^3y_3^2, y_1y_3^2, x_3y_1y_3^2, x_3^2y_1y_3^2, y_3^3, x_3y_3^3,
x_3^2y_3^3, y_4, \nonumber\\
&&x_4y_4, x_4^2y_4, x_4^3y_4, x_4^4y_4, y_1y_4, x_4y_1y_4,
x_4^2y_1y_4, x_4^3y_1y_4, y_1^2y_4, x_4y_1^2y_4, x_4^2y_1^2y_4,
y_4^2, \nonumber\\
&&x_4y_4^2, x_4^2y_4^2, x_4^3y_4^2, y_1y_4^2, x_4y_1y_4^2,
x_4^2y_1y_4^2, y_4^3, x_4y_4^3, x_4^2y_4^3\}~.~~~\eea

After obtained integrand basis, we move to the discussions of branch
structure of variety. For the most general momentum configuration
$\mbox{B32}_{(K_4,K_5)}$ with all external momenta massive, the
variety has only one irreducible branch, but for some kinematic
configurations it will split into many branches. We will use the
notation $\mbox{B32}_{(U,P)}^{(L,R)}$, where as usual $U,P$ could be
$K_4,K_5$ or $\oslash$, while $L$ is denoted by $m$ if at least one
momentum of $K_1,K_2$ (or $K_3$ for $R$) is massless, otherwise it
is denoted by $M$. The number of branches  of variety can be
summarized in table (\ref{B5branch}).
\begin{table}[h]
  \centering
  \begin{tabular}{|c|c|c|c|}
    \hline
    \backslashbox{$(U,P)$}{$(L,R)$} & $(K_4,K_5)$ & $(K_4/K_5,\oslash)$ & $(\oslash,\oslash)$ \\
     \hline
    $(M,M)$ & 1 & 2 & 2 \\
    \hline
    $(m,M),(M,m)$ & 2 & 4 & 4 for $(m,M)$; 2+1 for $(M,m)$ \\
    \hline
    $(m,m)$ & 4 & 6 & 2+1 \\
    \hline
  \end{tabular}
  \caption{Number of branches of some kinematic configurations for planar box-triangle topology
(B32). Generally each branch is 2-dimensional variety, but for
momentum configurations $\mbox{B32}_{(\oslash,\oslash)}^{(M,m)}$ and
$\mbox{B32}_{(\oslash,\oslash)}^{(m,m)}$, there is an extra branch
of dimension one. We write it as $2+1$ to emphasize the difference.
}\label{B5branch}
\end{table}

{\bf Variety with one branch: } For kinematic configuration
$\mbox{B32}_{(K_4,K_5)}^{(M,M)}$, the variety is irreducible with
dimension two. All 69 coefficients should be calculated using this
branch.

{\bf Variety with two branches: } For kinematic configurations
\bea
\mbox{B32}_{(K_4,K_5)}^{(m,M)}~,~\mbox{B32}_{(K_4,K_5)}^{(M,m)}~,~\mbox{B32}_{(K_4/K_5,\oslash)}^{(M,M)}
~,~\mbox{B32}_{(\oslash,\oslash)}^{(M,M)}~,~~~\nonumber\eea
the variety has two branches. For $\mbox{B32}_{(K_4,K_5)}^{(m,M)}$,
each branch can detect 38 coefficients, and intersection of these
two branches is 1-dimensional with intersection pattern
$(1|7)=V_1\cap V_2$. For $\mbox{B32}_{(K_4,K_5)}^{(M,m)}$,
$\mbox{B32}_{(K_4/K_5,\oslash)}^{(M,M)}$ and
$\mbox{B32}_{(\oslash,\oslash)}^{(M,M)}$, each branch can detect 42
coefficients, and intersection pattern of these two branches is
$(1|15)=V_1\cap V_2$.

{\bf Variety with four branches: } For kinematic configurations
\bea
\mbox{B32}_{(K_4,K_5)}^{(m,m)}~,~\mbox{B32}_{(K_4/K_5,\oslash)}^{(m,M)}~,~
\mbox{B32}_{(K_4/K_5,\oslash)}^{(M,m)}~,~\mbox{B32}_{(\oslash,\oslash)}^{(m,M)}~,~~\nonumber\eea
the variety has four branches. For $\mbox{B32}_{(K_4,K_5)}^{(m,m)}$,
each branch can  detect 23 coefficients individually, while for
$\mbox{B32}_{(K_4/K_5,\oslash)}^{(m,M)}$ and
$\mbox{B32}_{(\oslash,\oslash)}^{(m,M)}$, each of $V_1$ and $V_4$
can detect 17 coefficients, and each of $V_2$ and $V_3$ can detect
29 coefficients. The intersections of branches for these kinematic
configurations are following. These four branches will intersect at
a single point, while intersections of every three branches are also
the same point. For intersections of every two branches, $(V_1,V_4)$
and $(V_2,V_3)$ intersect at the same single point, and
intersections for other pairs are $(1|4)=V_1\cap V_3$,
$(1|4)=V_2\cap V_4$, $(1|8)=V_1\cap V_2$, $(1|8)=V_3\cap V_4$. They
are four different 1-dimensional varieties. For
$\mbox{B32}_{(K_4/K_5,\oslash)}^{(M,m)}$, each of $V_1$ and $V_3$
can detect 10 coefficients, while each of $V_2$ and $V_4$ can detect
36 coefficients. There is no intersection for four branches, while
$(V_1,V_2,V_4)$ intersects at a single point, and $(V_2,V_3,V_4)$
intersects at another single point. There is no intersection between
$(V_1,V_3)$, while the intersection of $(V_2,V_4)$ is 1-dimensional
$(1|9)=V_2\cap V_4$. For other intersections of every two branches,
we have $(1|4)=V_1\cap V_2$, $(1|4)=V_1\cap V_4$, $(1|4)=V_2\cap
V_3$ and $(1|4)=V_3\cap V_4$. They are different 1-dimensional
varieties.

{\bf Variety with six branches: } For kinematic configurations
\bea \mbox{B32}_{(K_4/K_5,\oslash)}^{(m,m)}~,~~~\nonumber\eea
the variety has six branches $V_1,V_2,V_3,V_4,V_5,V_6$. Each of
$V_1$ and $V_4$ can detect 10 coefficients, and  each of $V_2$ and
$V_5$ can detect 17 coefficients, while each of $V_3$ and $V_6$ can
detect 23 coefficients. There are no intersections among six
branches and every five branches. The only non-zero intersection of
every four branches is $(V_2,V_3,V_5,V_6)$, and they intersect at a
single point. For intersections of every three branches,
$(V_2,V_3,V_5)$, $(V_2,V_3,V_6)$, $(V_2,V_5,V_6)$ and
$(V_3,V_5,V_6)$ will intersect at the same point as intersection of
$(V_2,V_3,V_5,V_6)$. $(V_1,V_2,V_3)$ will intersect at different
single point, and $(V_4,V_5,V_6)$ will intersect at another
different single point. For intersections of every two branches,
$(V_2,V_5)$ and $(V_3,V_6)$ will intersect at the same point as
intersection of $(V_2,V_3,V_5,V_6)$, while intersection pattern of
other pairs are $(1|4)=V_1\cap V_2$, $(1|4)=V_1\cap V_3$,
$(1|4)=V_2\cap V_6$, $(1|4)=V_3\cap V_5$, $(1|4)=V_4\cap V_5$,
$(1|4)=V_4\cap V_6$ and $(1|5)=V_2\cap V_3$, $(1|5)=V_5\cap V_6$.
They are all different 1-dimensional varieties.

{\bf Variety with 2+1 branches: } For kinematic configurations
\bea
\mbox{B32}_{(\oslash,\oslash)}^{(m,m)}~,~\mbox{B32}_{(\oslash,\oslash)}^{(M,m)}~,~~~\eea
the integrand basis contains 77 elements, and the three quadratic
equations reduce to
\bea x_3x_4=0~,~~y_3y_4=0~,~~~x_4y_3+x_3y_4=0~.~~~\eea
There will be three branches. Two branches $V_1,V_2$ are given by
$x_3=0, y_3=0$ with $y_1,x_4,y_4$ as free parameters and
$x_4=0,y_4=0$ with $y_1,x_3,y_3$ as free parameters. These two
branches are 3-dimensional. The third branch $V_3$ is embedded in
these two branches, and it is given by the ideal
\bea V_3~:~~~\{x_3^2,x_3x_4,x_4^2,y_3^2,
y_3y_4,y_4^2,x_3y_4+y_3x_4\}~.~~~\eea
Geometrically it is just the 1-dimensional variety
$x_3=x_4=y_3=y_4=0$ with $y_1$ as free parameter. Although the third
branch is the intersection of $V_1,V_2$ geometrically, from the
point of algebraic geometry, it is an independent branch. Each $V_1$
or $V_2$ can detect 39 coefficients, while $V_3$ can detect 27
coefficients. Since geometrically $V_3$ is the intersection of
$V_1,V_2$, it is clear that intersections of these three branches or
every two branches are the same 1-dimensional variety, thus we have
$(1|4)=V_1\cap V_2$, $(1|14)=V_1\cap V_3=V_2\cap V_3$,
$(1|4)=V_1\cap V_2\cap V_3$ .

\subsection{The topology (B31): planar box-bubble}

This topology contains a sub-loop of bubble structure. When
$K_3=K_4=0$, there is no difference between propagators
$D_0=\ell_0^2$ and $D_2=(\ell_0-K_1-K_2)^2$ because of momentum
conservation. This will effectively eliminate one on-shell equation.
For $\mbox{B31}_{(K_3,K_4)}$ and $\mbox{B31}_{(K_3/K_4,\oslash)}$
there are five independent on-shell equations, and from which we can
get two linear equations for $(x_1,x_2,x_3,x_4)$. By solving these
linear equations we can reduce 8 variables to 6 ISPs. For
$\mbox{B31}_{(\oslash,\oslash)}$, we have four independent on-shell
 equations, thus we can only construct one linear equation for
$(x_1,x_2,x_3,x_4)$. In this case we get 7 ISPs.

For $\mbox{B31}_{(K_3,K_4)}$ and $\mbox{B31}_{(K_3/K_4,\oslash)}$ we
can use $K_1,K_2$ to construct momentum basis $e_i$, while for
$\mbox{B31}_{(\oslash,\oslash)}$ there are only two external legs,
we should choose another auxiliary momentum together with one of
$K_1,K_2$ to construct momentum basis $e_i$. By expand all momenta
with this basis, we  get, for instance, 6 ISPs
$(x_3,x_4,y_1,y_2,y_3,y_4)$ for $\mbox{B31}_{(K_3,K_4)}$,
$\mbox{B31}_{(K_3/K_4,\oslash)}$, and 7 ISPs
$(x_2,x_3,x_4,y_1,y_2,y_3,y_4)$ for
$\mbox{B31}_{(\oslash,\oslash)}$. Under the  renormalization
conditions
\bea &&\sum_{\mbox{\tiny{all~ISPs~of~x}}}d(x_i)\leq
4~,~~~\sum_{\mbox{\tiny{all~ISPs~of~y}}}d(y_i)\leq
2~,~~~\sum_{\mbox{\tiny{all~ISPs~of~x}}}d(x_i)+\sum_{\mbox{\tiny{all~ISPs~of~y}}}d(y_i)\leq
4~~~\eea
 we can get integrand basis using
Gr\"{o}bner basis method with ordering $(x_3,x_4,y_1,y_2,y_3,y_4)$
for $\mbox{B31}_{(K_3,K_4)}$, $\mbox{B31}_{(K_3/K_4,\oslash)}$ and
$(x_2,x_3,x_4,y_1,y_2,y_3,y_4)$ for
$\mbox{B31}_{(\oslash,\oslash)}$. For all possible momentum
configurations of $\mbox{B31}_{(K_3,K_4)}$ and
$\mbox{B31}_{(K_3/K_4,\oslash)}$, the integrand basis contains 65
elements given by
\bea \mathcal{B}_{B31}^{I}&=&\{1, x_3, x_3^2, x_3^3, x_3^4, x_4,
x_4^2, x_4^3, x_4^4, y_1, x_3y_1, x_3^2y_1, x_3^3y_1, x_4y_1,
x_4^2y_1, x_4^3y_1, y_1^2, x_3y_1^2,\nonumber\\
&& x_3^2y_1^2, x_4y_1^2, x_4^2y_1^2, y_2, x_3y_2, x_3^2y_2,
x_3^3y_2, x_4y_2, x_4^2y_2, x_4^3y_2, y_2^2, x_3y_2^2, x_3^2y_2^2,
x_4y_2^2, x_4^2y_2^2,\nonumber\\
&& y_3, x_3y_3, x_3^2y_3, x_3^3y_3, x_4y_3, y_1y_3, x_3y_1y_3,
x_3^2y_1y_3, x_4y_1y_3, y_2y_3, x_3y_2y_3, x_3^2y_2y_3,\nonumber\\
&& x_4y_2y_3, y_3^2, x_3y_3^2, x_3^2y_3^2, x_4y_3^2, y_4, x_4y_4,
x_4^2y_4, x_4^3y_4, y_1y_4, x_4y_1y_4, x_4^2y_1y_4,
y_2y_4,\nonumber\\
&& x_4y_2y_4, x_4^2y_2y_4, y_3y_4, x_4y_3y_4, y_4^2, x_4y_4^2,
x_4^2y_4^2\}~.~~~\eea
For all possible momentum configurations of
$\mbox{B31}_{(\oslash,\oslash)}$, the integrand basis contains 145
elements given by
\bea \mathcal{B}_{B31}^{II}&=&\{1, x_2, x_3, x_2x_3, x_3^2,
x_2x_3^2, x_3^3, x_2x_3^3, x_3^4, x_4, x_2x_4, x_3x_4, x_2x_3x_4,
x_3^2x_4,
x_2x_3^2x_4, x_3^3x_4,\nonumber\\
&& x_4^2, x_2x_4^2, x_3x_4^2, x_2x_3x_4^2, x_3^2x_4^2, x_4^3,
x_2x_4^3, x_3x_4^3, x_4^4, y_1, x_3y_1, x_3^2y_1, x_3^3y_1, x_4y_1,
x_3x_4y_1,\nonumber\\
&& x_3^2x_4y_1, x_4^2y_1, x_3x_4^2y_1, x_4^3y_1, y_1^2, x_3y_1^2,
x_3^2y_1^2, x_4y_1^2, x_3x_4y_1^2, x_4^2y_1^2, y_2, x_2y_2, x_3y_2,
x_2x_3y_2,\nonumber\\
&& x_3^2y_2, x_2x_3^2y_2, x_3^3y_2, x_4y_2, x_2x_4y_2, x_3x_4y_2,
x_2x_3x_4y_2, x_3^2x_4y_2, x_4^2y_2, x_2x_4^2y_2,
x_3x_4^2y_2,\nonumber\\
&& x_4^3y_2, y_2^2, x_3y_2^2, x_3^2y_2^2, x_4y_2^2, x_3x_4y_2^2,
x_4^2y_2^2, y_3, x_2y_3, x_3y_3, x_2x_3y_3, x_3^2y_3, x_2x_3^2y_3,
x_3^3y_3, \nonumber\\
&&x_4y_3, x_2x_4y_3, x_3x_4y_3, x_2x_3x_4y_3, x_3^2x_4y_3, x_4^2y_3,
x_2x_4^2y_3, x_3x_4^2y_3, x_4^3y_3, y_1y_3, x_3y_1y_3,
\nonumber\\
&&x_3^2y_1y_3, x_4y_1y_3, x_3x_4y_1y_3, x_4^2y_1y_3, y_2y_3,
x_2y_2y_3, x_3y_2y_3, x_2x_3y_2y_3, x_3^2y_2y_3,
x_4y_2y_3,\nonumber\\
&& x_2x_4y_2y_3, x_3x_4y_2y_3, x_4^2y_2y_3, y_3^2, x_2y_3^2,
x_3y_3^2, x_2x_3y_3^2, x_3^2y_3^2, x_4y_3^2, x_2x_4y_3^2,
x_3x_4y_3^2, x_4^2y_3^2,\nonumber\\
&& y_4, x_2y_4, x_3y_4, x_3^2y_4, x_3^3y_4, x_4y_4, x_2x_4y_4,
x_3x_4y_4, x_3^2x_4y_4, x_4^2y_4, x_2x_4^2y_4, x_3x_4^2y_4,
x_4^3y_4, \nonumber\\
&&y_1y_4, x_3y_1y_4, x_3^2y_1y_4, x_4y_1y_4, x_3x_4y_1y_4,
x_4^2y_1y_4, y_2y_4, x_2y_2y_4, x_3y_2y_4, x_3^2y_2y_4,
x_4y_2y_4,\nonumber\\
&& x_2x_4y_2y_4, x_3x_4y_2y_4, x_4^2y_2y_4, y_3y_4, x_2y_3y_4,
x_3y_3y_4, x_3^2y_3y_4, x_4y_3y_4, x_2x_4y_3y_4,
x_3x_4y_3y_4,\nonumber\\
&& x_4^2y_3y_4, y_4^2, x_2y_4^2, x_3y_4^2, x_4y_4^2, x_2x_4y_4^2,
x_3x_4y_4^2, x_4^2y_4^2\}~.~~~\eea

After obtained the integrand basis, we  analyze branch structure of
variety.

{\bf Branches of $\mbox{B31}_{(K_3,K_4)}$:} The  variety will split
into two branches $V_1,V_2$ when at least one momentum of $K_1,K_2$
is massless. These two branches are 3-dimensional, and their
intersection is 2-dimensional. Each branch can detect 37
coefficients, while 9 coefficients can be detected by their
intersection. So using both branches we can detect $37+37-9=65$
coefficients.

{\bf Branches of $\mbox{B31}_{(K_3/K_4,\oslash)}$:} There are two
3-dimensional branches if both $K_1,K_2$ are massive. Each branch
can detect 46 coefficients, and 27 coefficients can be detected by
their 2-dimensional intersection. When at least one momentum of
$K_1,K_2$ is massless, generally the variety will split into four
branches $V_1,V_2,V_3,V_4$. Each of $V_1$ and $V_3$ can detect 22
coefficients, while each of $V_2$ and $V_4$ can detect 30
coefficients. Intersection of all four branches is 1-dimensional,
and we have $(1|3)=V_1\cap V_2\cap V_3\cap V_4$. Intersection of
every three branches is the same 1-dimensional variety as
intersection of four branches. For intersections of every two
branches, $(V_1,V_3)$ and $(V_2,V_4)$ will intersect at the same
1-dimensional variety as intersection of four branches, and all
other intersections are 2-dimensional. The intersection pattern is
$(2|6)=V_1\cap V_4$, $(2|6)=V_2\cap V_3$ and $(2|15)=V_1\cap V_2$,
$(2|15)=V_3\cap V_4$. They are different 2-dimensional varieties.
However, for the specific momentum configurations with
 $\mbox{B31}_{(K_3,\oslash)}$ where both $K_2,K_3$ are massless, or
$\mbox{B31}_{(K_4,\oslash)}$ where both $K_1,K_4$ are massless, the
three quadratic equations reduce to
\bea x_3x_4=0~,~~y_1y_2+y_3y_4=0~,~~x_4y_3+x_3y_4=0~.~~~\eea
There are in total five branches. Four ordinary branches are given
by
\bea &&V_1:~~y_3=0~,~y_1=0~,~x_3=0~,~x_4~,~y_2~,~y_4~,~\mbox{free
parameters}~,~~~\nonumber\\
&&V_2:~~y_3=0~,~y_2=0~,~x_3=0~,~x_4~,~y_1~,~y_4~,~\mbox{free
parameters}~,~~~\nonumber\\
&&V_3:~~y_4=0~,~y_1=0~,~x_4=0~,~x_3~,~y_2~,~y_3~,~\mbox{free
parameters}~,~~~\nonumber\\
&&V_4:~~y_4=0~,~y_2=0~,~x_4=0~,~x_3~,~y_1~,~y_3~,~\mbox{free
parameters}~.~~~\eea
The fifth branch is given by the ideal
\bea V_5:~~\{x_4y_3+x_3y_4, y_1y_2+y_3y_4, x_4^2, x_3x_4,
x_3^2\}~.~~~\eea
All these branches are 3-dimensional, and each $V_1,V_2,V_3,V_4$ can
detect 22 coefficients while $V_5$ can detect 37 coefficients. All
five branches intersect at a single point. Intersection of every
four branches is also the same single point. For intersections of
every three branches, it is $(1|6)=V_1\cap V_2\cap V_5$,
$(1|6)=V_3\cap V_4\cap V_5$, $(1|3)=V_1\cap V_3\cap V_5$ and
$(1|3)=V_2\cap V_4\cap V_5$, and they are different 1-dimensional
variety. For intersections of every two branches, $(V_1,V_4)$ and
$(V_2,V_3)$ are still the same single point, $(V_1,V_3)$ is the same
1-dimensional variety as intersection of $(V_1,V_3,V_5)$, and
$(V_2,V_4)$ is the same 1-dimensional variety as intersection of
$(V_2,V_4,V_5)$. The intersections of all other pairs are
2-dimensional, and we have $(2|12)=V_1\cap V_2$, $(2|12)=V_3\cap
V_4$,  $(2|12)=V_1\cap V_5$, $(2|12)=V_2\cap V_5$, $(2|12)=V_3\cap
V_5$, $(2|12)=V_4\cap V_5$. They are all different 2-dimensional
varieties. Using these five branches, we can detect 65 coefficients
of integrand basis.

{\bf Branches of $\mbox{B31}_{(\oslash,\oslash)}$:} There are only
two external legs, and none of them can be massless, so we have only
one momentum configuration with both $K_1,K_2$ massive. The
integrand basis contains 145 elements. There are two branches of
dimension four, and 110 coefficients can be detected by each of
them. Intersection of these 2 branches is 3-dimensional, and using
it we can detect 75 coefficients. So all 145 coefficients can be
detected  using these two branches.

\subsection{The topology (B22): planar double-triangle}

For the double-triangle topology (B22), we can use $K_1, K_2$ to
construct momentum basis. When $K_3= K_4=0$, $K_1$ and $K_2$ are not
independent, and we use $K_1$ and another auxiliary momentum to
construct momentum basis. There are five propagators and  using two
linear equations $D_0-D_1=0$, $\widetilde{D}_0-\widetilde{D}_1=0$,
we can solve $x_1, y_2$. So there are six ISPs
$(x_2,x_3,x_4,y_1,y_3,y_4)$ and three quadratic equations left.

This topology has $Z_2$ symmetry between $K_1, K_2$ and $Z_2$
symmetry between $K_3, K_4$, we will take the notation
$\mbox{B22}_{(U,P)}^{(L,R)}$ where $U,P$ could be $K_3,K_4$ or
$\oslash$, and $L$ is denoted by $m$ if $K_1$ (or $K_2$ for $R$) is
massless, otherwise it is denoted by $M$. It is worth to notice when
$K_3=K_4=0$, we have $K_1=-K_2$, thus to get non-zero contribution,
$K_1, K_2$ should be massive. In other words, we do not need to
consider kinematic configurations
$\mbox{B22}_{\oslash,\oslash}^{m,M}$,
$\mbox{B22}_{\oslash,\oslash}^{M,m}$ and
$\mbox{B22}_{\oslash,\oslash}^{m,m}$.

Using Gr\"{o}bner basis defined from three quadratic equations with
ordering $(x_2,x_3,x_4,y_1,y_3,y_4)$, under the  renormalization
conditions of monomials
\bea &&\sum_{\mbox{\tiny{all~ISPs~of~x}}}d(x_i)\leq
3~,~\sum_{\mbox{\tiny{all~ISPs~of~y}}}d(y_i)\leq
3~,~\sum_{\mbox{\tiny{all~ISPs~of~x}}}d(x_i)+\sum_{\mbox{\tiny{all~ISPs~of~y}}}d(y_i)\leq
4~,~~~\eea
we can get 111 elements for integrand basis. The explicit form of
these elements depends on the kinematics of $K_1,K_2$, which we have
chosen to generate momentum basis $e_i$. There are in total four
kinds of integrand basis. For momentum configurations with $K_1,K_2$
massive, the 111 elements are given by
\bea \mathcal{B}_{B22}^{I}=&&\{1, x_2, x_3, x_2x_3, x_3^2, x_2x_3^2,
x_3^3, x_4, x_2x_4, x_3x_4, x_2x_3x_4, x_3^2x_4, x_4^2, x_2x_4^2,
x_3x_4^2, x_4^3, y_1,\nonumber\\
&& x_3y_1,
 x_3^2y_1, x_3^3y_1, x_4y_1, x_3x_4y_1, x_3^2x_4y_1, x_4^2y_1, x_3x_4^2y_1, x_4^3y_1,
y_3, x_2y_3, x_3y_3, x_2x_3y_3,\nonumber\\
&& x_3^2y_3,
 x_2x_3^2y_3, x_3^3y_3, x_4y_3, x_2x_4y_3, x_3x_4y_3, x_2x_3x_4y_3, x_3^2x_4y_3,
x_4^2y_3, x_2x_4^2y_3, x_3x_4^2y_3,\nonumber\\
&& x_4^3y_3, y_1y_3,
 x_3y_1y_3, x_3^2y_1y_3, x_4y_1y_3, x_3x_4y_1y_3, x_4^2y_1y_3, y_3^2, x_2y_3^2,
x_3y_3^2, x_2x_3y_3^2, x_3^2y_3^2, \nonumber\\
&&x_4y_3^2,
 x_3x_4y_3^2, x_4^2y_3^2, y_1y_3^2, x_3y_1y_3^2, x_4y_1y_3^2, y_3^3, x_2y_3^3,
x_3y_3^3, x_4y_3^3, y_4, x_2y_4, x_3y_4, x_3^2y_4,\nonumber\\
&&
 x_3^3y_4, x_4y_4, x_2x_4y_4, x_3x_4y_4, x_3^2x_4y_4, x_4^2y_4, x_2x_4^2y_4,
x_3x_4^2y_4, x_4^3y_4, y_1y_4, x_3y_1y_4, x_3^2y_1y_4,\nonumber\\
&&
 x_4y_1y_4, x_3x_4y_1y_4, x_4^2y_1y_4, y_3y_4, x_3y_3y_4, x_3^2y_3y_4, x_4y_3y_4,
x_3x_4y_3y_4, x_4^2y_3y_4, y_1y_3y_4, \nonumber\\
&&x_3y_1y_3y_4,
 x_4y_1y_3y_4, y_3^2y_4, x_3y_3^2y_4, x_4y_3^2y_4, y_4^2, x_2y_4^2, x_3y_4^2,
x_4y_4^2, x_2x_4y_4^2, x_3x_4y_4^2, x_4^2y_4^2,\nonumber\\
&&
 y_1y_4^2, x_3y_1y_4^2, x_4y_1y_4^2, y_3y_4^2, x_3y_3y_4^2, x_4y_3y_4^2, y_4^3,
x_2y_4^3, x_3y_4^3,
 x_4y_4^3\}~.~~~\eea
The second kind of integrand basis is for configurations with $K_1$
massless and $K_2$ massive, and the 111 elements are given by
replacing 19 elements in $\mathcal{B}_{B22}^{I}$
\bea \mathcal{B}_{B22}^{II}&=&\mathcal{B}_{B22}^{I}-\{x_3x_4,
x_2x_3x_4, x_3^2x_4, x_3x_4^2, x_3x_4y_1, x_3^2x_4y_1, x_3x_4^2y_1,
x_3x_4y_3,
x_2x_3x_4y_3,\nonumber\\
&& x_3^2x_4y_3, x_3x_4^2y_3, x_3x_4y_1y_3, x_3x_4y_3^2, x_3x_4y_4,
x_3^2x_4y_4, x_3x_4^2y_4, x_3x_4y_1y_4, x_3x_4y_3y_4,\nonumber\\
&& x_3x_4y_4^2\}+ \{x_2^2, x_2^3, x_2^2x_3, x_2^2x_4, x_2^2y_3,
x_2^3y_3, x_2^2x_3y_3, x_2^2x_4y_3, x_2^2y_3^2, x_2x_4y_3^2,
x_2^2y_4,\nonumber\\
&& x_2^3y_4, x_2x_3y_4, x_2^2x_3y_4, x_2x_3^2y_4, x_2^2x_4y_4,
x_2^2y_4^2, x_2x_3y_4^2, x_3^2y_4^2\}~.~~~\eea
The third kind of integrand basis is for configurations with $K_2$
massless and $K_1$ massive, and 111 elements are given by replacing
15 elements in $\mathcal{B}_{B22}^{I}$
\bea \mathcal{B}_{B22}^{III}&=&\mathcal{B}_{B22}^{I}-\{y_3y_4,
x_3y_3y_4, x_3^2y_3y_4, x_4y_3y_4, x_3x_4y_3y_4, x_4^2y_3y_4,
y_1y_3y_4,
x_3y_1y_3y_4, x_4y_1y_3y_4,\nonumber\\
&& y_3^2y_4,x_3y_3^2y_4,
 x_4y_3^2y_4, y_3y_4^2, x_3y_3y_4^2, x_4y_3y_4^2\}+\{y_1^2, x_3y_1^2, x_3^2y_1^2,
x_4y_1^2, x_3x_4y_1^2,
 x_4^2y_1^2,\nonumber\\
 &&y_1^3, x_3y_1^3, x_4y_1^3, y_1^2y_3, x_3y_1^2y_3, x_4y_1^2y_3,
 y_1^2y_4, x_3y_1^2y_4, x_4y_1^2y_4\}~.~~~\eea
Finally the fourth kind of integrand basis is for configurations
with $K_1,K_2$ massless, and 111 elements are given by replacing 33
elements in $\mathcal{B}_{B22}^{I}$
\bea \mathcal{B}_{B22}^{IV}&=&\mathcal{B}_{B22}^{I}-\{x_3x_4,
x_2x_3x_4, x_3^2x_4, x_3x_4^2, x_3x_4y_1, x_3^2x_4y_1, x_3x_4^2y_1,
x_3x_4y_3,
x_2x_3x_4y_3, x_3^2x_4y_3, \nonumber\\
&&x_3x_4^2y_3, x_3x_4y_1y_3, x_3x_4y_3^2, x_3x_4y_4, x_3^2x_4y_4,
x_3x_4^2y_4, x_3x_4y_1y_4, y_3y_4, x_3y_3y_4,
x_3^2y_3y_4,\nonumber\\
&& x_4y_3y_4, x_3x_4y_3y_4, x_4^2y_3y_4, y_1y_3y_4, x_3y_1y_3y_4,
x_4y_1y_3y_4, y_3^2y_4, x_3y_3^2y_4, x_4y_3^2y_4, x_3x_4y_4^2,
\nonumber\\
&&y_3y_4^2, x_3y_3y_4^2,x_4y_3y_4^2\}+\{x_2^2, x_2^3, x_2^2x_3,
x_2^2x_4, y_1^2, x_3y_1^2, x_3^2y_1^2, x_4y_1^2, x_4^2y_1^2, y_1^3,
x_3y_1^3, x_4y_1^3, \nonumber\\
&&x_2^2y_3, x_2^3y_3, x_2^2x_3y_3, x_2^2x_4y_3, y_1^2y_3,
x_3y_1^2y_3, x_4y_1^2y_3, x_2^2y_3^2, x_2x_4y_3^2, x_2^2y_4,
x_2^3y_4, x_2x_3y_4,\nonumber\\
&& x_2^2x_3y_4, x_2x_3^2y_4, x_2^2x_4y_4, y_1^2y_4, x_3y_1^2y_4,
x_4y_1^2y_4, x_2^2y_4^2, x_2x_3y_4^2, x_3^2y_4^2\}~.~~~\eea

After obtained integrand basis, we discuss the branch structure of
variety. The number of branches for different kinematic
configurations is summarized in table \ref{B7branch}.
\begin{table}[h]
  \centering
  \begin{tabular}{|c|c|c|c|}
    \hline
    \backslashbox{$(U,P)$}{$(L,R)$} & $(K_3,K_4)$ & $(K_3/K_4,\oslash)$ & $(\oslash,\oslash)$ \\
     \hline
    $(M,M)$ & 1 & 2 & 2 \\
    \hline
    $(m,M),(M,m)$ & 2 & 4 &  \\
    \hline
    $(m,m)$ & 4 & 6 & \\
    \hline
  \end{tabular}
  \caption{Number of branches for some kinematic configurations of planar double-triangle topology
(B22). Each branch is 3-dimensional variety.}\label{B7branch}
\end{table}

{\bf Variety with one branch: } For general kinematic configuration
$\mbox{B22}_{(K_3,K_4)}^{(M,M)}$, the variety is irreducible with
dimension three. All 111 coefficients of integrand basis can be
detected by this branch.

{\bf Variety with two branches: } For kinematic configurations
\bea
\mbox{B22}_{(K_3,K_4)}^{(m,M)}~,~\mbox{B22}_{(K_3,K_4)}^{(M,m)}~,~
\mbox{B22}_{(K_3/K_4,\oslash)}^{(M,M)}~,~\mbox{B22}_{(\oslash,\oslash)}^{(M,M)}~,~~~\eea
the variety is given by two branches $V_1,V_2$. For
$\mbox{B22}_{(K_3,K_4)}^{(m,M)}$ and
$\mbox{B22}_{(K_3,K_4)}^{(M,m)}$, each branch can detect 71
coefficients, and their intersection is 2-dimensional variety which
can detect 31 coefficients. For
$\mbox{B22}_{(K_4/K_5,\oslash)}^{(M,M)}$ and
$\mbox{B22}_{(\oslash,\oslash)}^{(M,M)}$, each branch can detect 77
coefficients, and their two-dimensional intersection can detect $43$
coefficients.

{\bf Variety with four branches: } For kinematic configurations
\bea
\mbox{B22}_{(K_3,K_4)}^{(m,m)}~,~\mbox{B22}_{(K_3/K_4,\oslash)}^{(m,M)}~,~\mbox{B22}_{(K_3/K_4,\oslash)}^{(M,m)}~,~~~\eea
the variety is given by four branches. Intersections of branches is
expressed by Figure \ref{B72-branch}.

\EPSFIGURE{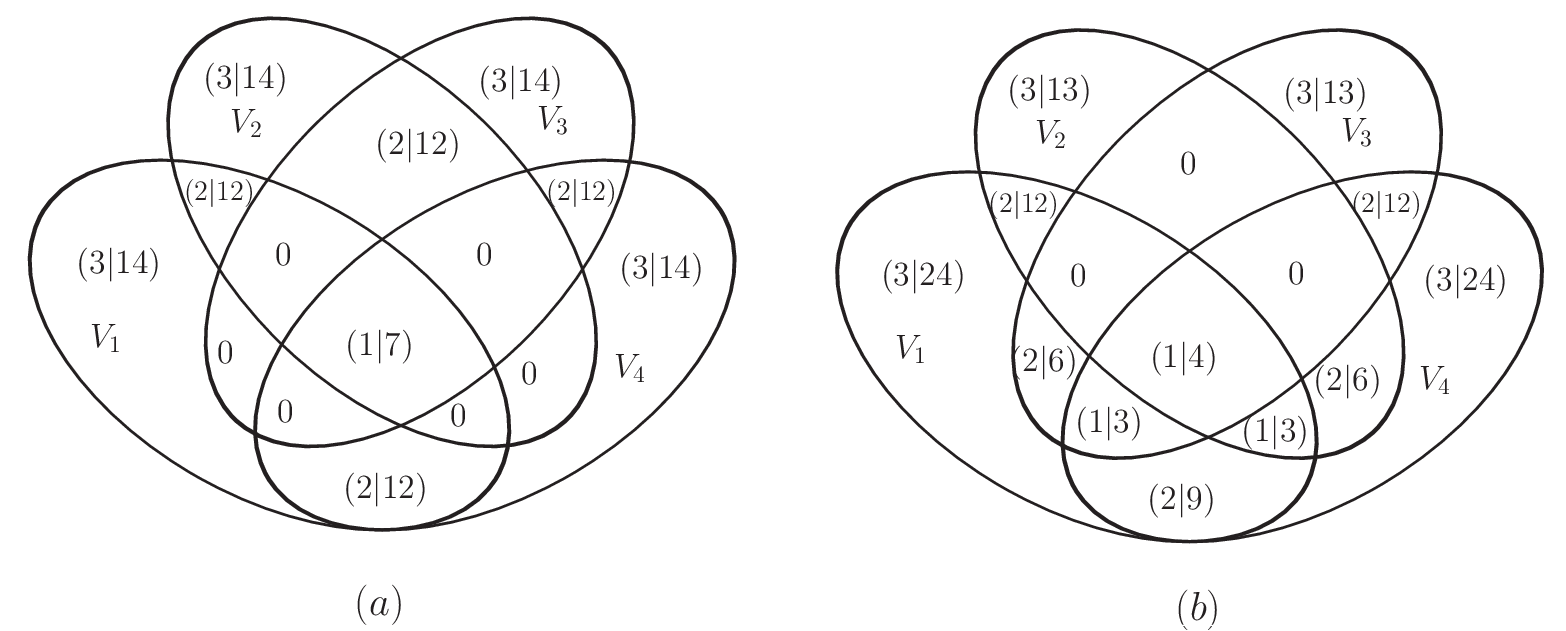,width=6.5in}{Venn diagrams for intersections
of four branches, where each ellipse represents one branch. Venn
diagram (a) is for $\mbox{B22}_{(K_3,K_4)}^{(m,m)}$, and Venn
diagram (b) is for $\mbox{B22}_{(K_3/K_4,\oslash)}^{(m,M)}$,
$\mbox{B22}_{(K_3/K_4,\oslash)}^{(M,m)}$. Number 0 means that there
is no intersection and $(d|n)$ represents that the dimension of that
intersection is $d$ and the number of coefficients detected in that
intersection is $n$. \label{B72-branch}}

{\bf Variety with six branches: } For kinematic configuration
\bea \mbox{B22}_{(K_3/K_4,\oslash)}^{(m,m)}~,~~~\eea
the variety is given by six branches $V_1,V_2,V_3,V_4,V_5,V_6$.
Among these six branches, four $V_i, i=1,2,3,4$ can detect 29
coefficients and the other two $V_5,V_6$ can detect 45 coefficients.
All six branches intersect at a single point, and intersection of
every five branches is also the same single point. For intersections
of every four branches, most of them are the same single point
inherit from intersections of every five branches except for the
following two combinations of branches $(V_1, V_3, V_5, V_6)$ and
$(V_2, V_4, V_5, V_6)$, which intersect at 1-dimensional variety
detecting $4$ coefficients. For intersections of every three
branches, besides the ones that are inherited from intersections of
four branches, there are also four pairs $(V_1, V_2, V_5)$,
$(V_1,V_2,V_6)$ and $(V_3, V_4, V_5)$, $(V_3,V_4,V_6)$ that
intersect at one-dimensional variety  detecting 4 coefficients.
Intersections of every two branches could be 2-dimensional,
1-dimensional, or single point, and they are summarized as
\bea (2|13) & = &  V_1\cap V_2= V_1\cap V_5= V_2\cap V_6=V_3\cap V_4=
 V_3\cap V_6=V_4\cap V_5~,~~~ \nn
(2|10) & = & V_2\cap V_5= V_1\cap V_6= V_4\cap V_6= V_3\cap V_5~,~~~
\nn
(1|4) & = & V_1\cap V_3=V_2\cap V_4~~,~~(0|1)= V_1\cap V_4=V_2\cap
V_3~~,~~ (1_2|7)  =  V_5\cap V_6~.~~~\eea
One interesting point is that intersection of $(V_5, V_6)$ is two
1-dimensional varieties, and to emphasize this subtlety we have used
$(1_2)$ notation.

\subsection{The topology (B21): planar triangle-bubble}

For this topology we could have two cases $\mbox{B21}_{(K_2,K_3)}$
and $\mbox{B21}_{(K_2/K_3,\oslash)}$. For $\mbox{B21}_{(K_2,K_3)}$,
since there are three external momenta, we can use $K_1, K_3$ to
construct momentum basis. For $\mbox{B21}_{(K_2/K_3,\oslash)}$,
there are only two external legs, and only one of them is
independent. So we need another auxiliary momentum together with
$K_1$ to construct momentum basis. We do not consider
$\mbox{B21}_{(\oslash,\oslash)}$ since it requires $K_1=0$, which is
tadpole-like structure. For $\mbox{B21}_{(K_2/K_3,\oslash)}$ we also
assume that both  external momenta are massive for non-vanishing
result and do not consider the kinematic configurations where any
momentum is massless.

From four propagators we can  reduce 8 variables to 7 ISPs, for
example,   $(x_2, x_3, x_4, y_1, y_2,y_3,y_4)$.
%
%
%
Using Gr\"{o}bner basis that generated from the three quadratic
equations with ordering $(y_1, x_2, y_2, x_3, y_3, x_4, y_4)$, under
the  renormalization conditions for monomials
\bea &&\sum_{\mbox{\tiny{all~ISPs~of~x}}}d(x_i)\leq
3~,~\sum_{\mbox{\tiny{all~ISPs~of~y}}}d(y_i)\leq
2~,~\sum_{\mbox{\tiny{all~ISPs~of~x}}}d(x_i)+\sum_{\mbox{\tiny{all~ISPs~of~y}}}d(y_i)\leq
3~,~~~\eea
we can get integrand basis for different kinematic configurations.
The elements of integrand basis depends on the kinematics of $K_1$.
For $\mbox{B21}_{(K_2,K_3)}$ and $\mbox{B21}_{(K_2/K_3,\oslash)}$,
if $K_1$ is massive, the integrand basis contains 80 elements given
by
\bea {\cal B}_{B21}^{I} & = & \{1, x_2, x_3, x_2 x_3, x_3^2, x_2
x_3^2, x_3^3, x_4, x_2 x_4, x_3 x_4, x_2 x_3 x_4, x_3^2 x_4, x_4^2,
x_2 x_4^2, x_3 x_4^2, x_4^3, y_1,\nn & & x_3 y_1, x_3^2 y_1, x_4
y_1, x_4^2 y_1, y_1^2, x_3 y_1^2, x_4 y_1^2, y_2, x_2 y_2, x_3 y_2,
x_2 x_3 y_2, x_3^2 y_2, x_4 y_2, x_2 x_4 y_2,\nn & &  x_3 x_4 y_2,
x_4^2 y_2, y_2^2, x_3 y_2^2, x_4 y_2^2,  y_3, x_2 y_3, x_3 y_3, x_2
x_3 y_3, x_3^2 y_3, x_4 y_3, x_2 x_4 y_3, x_3 x_4 y_3,\nn && x_4^2
y_3, y_1 y_3, x_3 y_1 y_3, x_4 y_1 y_3, y_2 y_3, x_2 y_2 y_3, x_3
y_2 y_3, x_4 y_2 y_3, y_3^2, x_2 y_3^2, x_3 y_3^2, x_4 y_3^2,
y_4,\nn & & x_2 y_4, x_3 y_4, x_2 x_3 y_4, x_3^2 y_4, x_4 y_4, x_2
x_4 y_4, x_3 x_4 y_4, x_4^2 y_4, y_1 y_4, x_3 y_1 y_4,  x_4 y_1 y_4,
y_2 y_4,\nn & & x_2 y_2 y_4, x_3 y_2 y_4, x_4 y_2 y_4, y_3 y_4, x_2
y_3 y_4, x_3 y_3 y_4, x_4 y_3 y_4, y_4^2, x_2 y_4^2, x_3 y_4^2, x_4
y_4^2\}~.~~~\label{B83-basis} \eea
If $K_1$ is massless for $\mbox{B21}_{(K_2,K_3)}$, integrand basis
is given by replacing 8 elements from ${\cal B}_{B21}^{I}$
\bea {\cal B}_{B21}^{II} &= & {\cal B}_{B21}^{I}-\{x_3 x_4, x_2 x_3
x_4, x_3^2 x_4, x_3 x_4^2, x_3 x_4 y_2, x_3 x_4 y_3, x_3 x_4 y_4,
x_2 \ y_3 y_4\}\nn & & +\{x_2^2, x_2^3, x_2^2 x_3, x_2^2 x_4, x_2^2
y_2, x_2 y_2^2, x_2^2 y_3, x_2^2 y_4\}~.~~~\eea

The variety defined by the three quadratic equations is irreducible
with dimension four for $\mbox{B21}_{(K_2,K_3)}$ with $K_1$ massive.
If $K_1$ is massless, the variety will split into two branches, and
each branch can detect 54 coefficients. Intersection of these two
branches is an irreducible 3-dimensional variety, and it can detect
28 coefficients. So  using both  branches, we can detect
$54+54-28=80$ coefficients. For $\mbox{B21}_{(K_2/K_3,\oslash)}$,
$K_1$ should be massive, and the variety has two branches. Each
branch can detect 64 coefficients, and intersection of these two
branches is an irreducible 3-dimensional variety, which can detect
48 coefficients. Using these two branches we can detect
$64+64-48=80$ coefficients of integrand basis.

\subsection{The topology (B11): planar sun-set}

For this topology, since $K_1=-K_2$, we use $K_1$ and another
auxiliary momentum to construct momentum basis. The only possible
kinematic configuration is both $K_1,K_2$ massive. There are only
three propagators and we can not construct linear equation from
on-shell  equations, thus there are 8 ISPs. The three quadratic
equations can be expressed as
\bea  D_0 & =& x_1 x_2 + x_3 x_4~,~\W D_0= y_1 y_2 + y_3 y_4~,~~~
\nn
\WH D_0 & = & x_2 y_1 + x_1 y_2 + x_4 y_3 +
 x_3 y_4 + (x_1 + y_1) \a_{11} + (x_2 + y_2) \a_{12} + \a_{11} \a_{12}~.~~~\label{B9-cut}\eea
Using Gr\"{o}bner basis with ordering $(y_4, y_3, x_4, x_3, y_2,
y_1, x_2, x_1)$, under the renormalization conditions for monomials
\bea &&\sum_{\mbox{\tiny{all~ISPs~of~x}}}d(x_i)\leq
2~,~\sum_{\mbox{\tiny{all~ISPs~of~y}}}d(y_i)\leq
2~,~\sum_{\mbox{\tiny{all~ISPs~of~x}}}d(x_i)+\sum_{\mbox{\tiny{all~ISPs~of~y}}}d(y_i)\leq
2~,~~~\eea
we can get 42 elements for integrand basis as
\bea {\cal B}_{B11}& = & \{1, x_1, x_1^2, x_2, x_1 x_2, x_2^2, x_3,
x_1 x_3, x_2 x_3, x_3^2, x_4, x_1 x_4,
 x_2 x_4, x_4^2, y_1, x_1 y_1, x_2 y_1,\nn & & x_3 y_1, x_4 y_1,  y_1^2, y_2, x_1 y_2, x_2 y_2,
 x_3 y_2, x_4 y_2, y_1 y_2, y_2^2, y_3, x_1 y_3, x_2 y_3, x_3 y_3, x_4 y_3,\nn && y_1 y_3,
 y_2 y_3, y_3^2, y_4, x_1 y_4,  x_2 y_4, x_4 y_4, y_1 y_4, y_2 y_4, y_4^2\}~.~~~ \eea
The variety defined by the three quadratic equations is irreducible
with dimension five.

\section{Conclusion}

In this paper, we use the new technique developed in
\cite{Zhang:2012ce, Mastrolia:2012an, Badger:2012dv} to classify the
two-loop integrand basis in pure four dimension space-time. Although
there are only small number of topologies for planar and non-planar
two-loop diagrams, the diverse external momentum configurations
greatly increase the number of integrand basis that we need to
discuss. Because the integrand basis and branch structure of variety
will depend on the topology as well as external kinematics, it is
necessary to classify possible sets of integrand basis and study the
evolution of variety under various kinematic limits.

The algebraic geometry methods, such as Gr\"{o}bner basis method and
multivariate polynomial division, play a crucial role in our
discussion. Using these methods, we are able to present explicit
form of integrand basis as well as detailed study of varieties, such
as branch structures and their intersections. The same methods also
allow us to determine coefficients of integrand basis.

We must emphasize that our result is only a small step towards the
practical evaluation of general two-loop amplitudes. As we have
mentioned in the introduction, the number of two-loop integrand
basis is much more than the number of two-loop integral basis and it
is highly desirable to reduce integrand basis further. One way to do
so is to use the IBP-method \cite{IBP}. However, with the time
consuming, it is not feasible at this moment.

Although our results are for  two-loop diagrams in pure
four-dimension space-time, the same analysis can be applied to
$(4-2\eps)$-dimension for complete answer, or three and higher loop
amplitudes as demonstrated in \cite{Badger:2012dv}. It is also an
interesting problem to apply these general analysis to real
processes.

\section*{Acknowledgement}

We would like to thank  Simon Badger, Hjalte Frellesvig, Yang Zhang
for many explanations and discussions and early participant of this
project. B.F would like to thank the hospitality of Niels Bohr
International Academy and Discovery Center. This work is supported,
in part, by fund from Qiu-Shi and Chinese NSF funding under contract
No.11031005, No.11135006, No.11125523.


\appendix

\section{Some mathematical backgrounds}

In this section, we present several mathematical facts that may be
useful for determining branch structure of non-linear on-shell
equations.  First let us consider the quadratic equation of two
variables defined by equation
\bea A x^2 +B xy +Cy^2+Dx+Ey+F=0~,~~~\eea
with $A,B,C$ not all zero. This equation is usually called conic
section. In general, the conic is an irreducible one-dimensional
variety, however, when the determinant $\Delta$ of following
$3\times 3$ matrix
\bea \Delta={\rm det}\left(
       \begin{array}{ccc}
         A & B/2 & D/2 \\
         B/2 & C & E/2 \\
         D/2 & E/2 & F \\
       \end{array}
     \right)={-(B^2-4AC)F+BDE-CD^2-AE^2\over 4}~~~~\label{conic-Det}\eea
is zero, the conic splits to two branches.

Next let us consider the roots of a polynomial
\bea f(z)=\sum_{i=0}^{n}a_{i}z^i~.~~~\eea
One can use discriminant to determine whether if this polynomial has
repeated roots or not. If discriminant equals to zero, then there
are repeated roots. The simplest example is quadratic equation
$a_2z^2+a_1z+a_0=0$ whose discriminant is $D=a_1^2-4a_0a_2$. If
$D=0$, then this equation has double roots, and the polynomial can
be written as a perfect square of one factor. The discriminant of
$n=2,3$ can be found in many other references, and using it we can
tell the properties of roots just from coefficients of variable. A
special interesting example is the quartic function
\bea f(z)=Az^4+Bz^3+Cz^2+Dz+E~,~~~\eea
which can be factorized as
\bea f(z)\sim
(z-z^{(+,+)})(z-z^{(+,-)})(z-z^{(-,+)})(z-z^{(-,+)})~,~~~\eea
where $z^{\pm,\pm}$ are four roots. We want to know if it can be
expressed as perfect square terms such as
\bea f(z)=(az+b)^2(cz+d)^2~.~~~\eea
In other words, we want to know if there are repeated roots or not.
Defining
\bea &&\mathcal{A}=-{3B^2\over 8A^2}+{C\over
A}~~,~~\mathcal{B}={B^3\over 8A^3}-{BC\over 2A^2}+{D\over
A}~~,~~\mathcal{C}=-{3B^4\over 256A^4}+{CB^2\over 16A^3}-{BD\over
4A^2}+{E\over A}~,~~~\eea
then when $\mathcal{B}=0$, the quartic equation has following
solution
\bea z=-{B\over
4A}\pm_s\sqrt{{-\mathcal{A}\pm_t\sqrt{\mathcal{A}^2-4\mathcal{C}}\over
2}}~,~~~\eea
where $\pm_s$ and $\pm_t$ can take plus and minus sign
independently. If the coefficients further satisfy
$\mathcal{A}^2-4\mathcal{C}=0$, then $x^{(s,+)}=x^{(s,-)}$, and
$f(x)$ can be expressed as products of two perfect squares.

Using above results, we can check whether  variety defined by
following equations is reducible
\bea &&A x^2+Bxy+Cy^2+D x+Ey+F=0~,~~~\\
&& a(\tau)x+b(\tau)y+c(\tau)=0~,~~~\eea
where $a(\tau),b(\tau),c(\tau)$ are linear functions  of (possible free
parameter) $\tau$. After  solving the linear equation of
$x,y$ and substituting the result into quadratic equation we get
\bea (a^2 C-a b B+ b^2A) y^2+(a^2 E-a b D-a c B+2 b c A)y+(a^2 F-a c
D+ c^2A) =0~,~~~\eea
where the coefficients $A'(\tau)=(a^2 C-a b B+ b^2A)$,
$B'(\tau)=(a^2 E-a b D-a c B+2 b c A)$ and $C'(\tau)=(a^2 F-a c D+
c^2A)$ are now quadratic functions  of $\tau$. The solution of $y$
is given by
\bea y={-B'(\tau)\pm\sqrt{B'(\tau)^2-4A'(\tau)C'(\tau)}\over
2A'(\tau)}~,~~~\label{quartic-sol}\eea
thus $y$ is a rational function of $\tau$ when and only when terms
inside the square root is perfect square, {\sl i.e.}, the quartic
function $(B'(\tau)^2-4A'(\tau)C'(\tau))$ of $\tau$ is a perfect
square.


\end{document}